%% file: main.tex
\documentclass{article}
\usepackage{arxiv}
\usepackage[utf8]{inputenc} 
\usepackage[T1]{fontenc}    

\usepackage{graphicx}
\usepackage{xcolor}
\usepackage{booktabs} 
\usepackage{amsmath}
\usepackage{amssymb}
\usepackage{makecell}
\usepackage{subcaption}
\usepackage{wrapfig}
\usepackage{xspace}
\usepackage{enumitem}
\usepackage[normalem]{ulem}
\usepackage{multirow}
\usepackage{forest}
\usepackage{nicematrix}
\usepackage{url}
\usepackage{hyperref}
\newenvironment{icompact}{
  \begin{list}{$\bullet$}{
    \parsep 0pt plus 1pt
    \partopsep 0pt plus 1pt
    \topsep 2pt plus 2pt minus 1pt
    \itemsep 0pt plus 1pt
    \parskip 0pt plus 2pt
    \leftmargin 0.13in}}
  {\normalsize\end{list}}

\newcommand{\bench}{ImageDetectBench\xspace}

\usepackage[
backend=biber,
style=nature,
]{biblatex}
\newcommand{\myparatight}[1]{\noindent\textbf{#1}:}
\begin{document}

\title{\Large \bf AI-generated Image Detection: Passive or Watermark?}

\author{
    Moyang Guo$^{*,1}$, Yuepeng Hu$^{*,1}$, Zhengyuan Jiang$^{*,1}$, Zeyu Li$^{*,1}$, Amir Sadovnik$^{2}$, Arka Daw$^{2}$, Neil Gong$^{1}$ \\
    $^{1}$Duke University \\
    \texttt{\{moyang.guo, yuepeng.hu, zhengyuan.jiang, zeyu.li030, neil.gong\}@duke.edu}\\~\\
    $^{2}$Oak Ridge National Laboratory \\ \texttt{\{sadovnika, dawa\}@ornl.gov}
}

\renewcommand{\thefootnote}{\fnsymbol{footnote}}
\footnotetext[1]{These authors contributed equally and are ordered alphabetically by last name.}
\date{}

\maketitle
\input{0.abstract}
\input{1.introduction}

\input{2.taxonomy_of_detector}
\input{3.taxonomy_of_perturbation}

\input{4.dataset_collection}

\input{5.benchmarking}
\input{6.conclusion}

\printbibliography
\input{7.appendix}

\end{document}

%% file: 0.abstract.tex
\begin{abstract}
While text-to-image models offer numerous benefits, they also pose significant societal risks. Detecting AI-generated images is crucial for mitigating these risks. Detection methods can be broadly categorized into \emph{passive} and \emph{watermark-based} approaches: passive detectors rely on artifacts present in AI-generated images, whereas watermark-based detectors proactively embed watermarks into such images. A key question is which type of detector performs better in terms of effectiveness, robustness, and efficiency. However, the current literature lacks a comprehensive understanding of this issue. In this work, we aim to bridge that gap by developing \emph{\bench}, the first comprehensive benchmark to compare the effectiveness, robustness, and efficiency of passive and watermark-based detectors. Our benchmark includes four datasets, each containing a mix of AI-generated and non-AI-generated images. We evaluate five  passive detectors and four watermark-based detectors against eight types of common perturbations and three types of adversarial perturbations. Our benchmark results reveal several interesting findings. For instance, watermark-based detectors consistently outperform passive detectors, both in the presence and absence of perturbations. Based on these insights, we provide recommendations for detecting AI-generated images, e.g., when both types of detectors are applicable, watermark-based detectors should be the preferred choice. Our code and data are publicly available at \url{https://github.com/moyangkuo/ImageDetectBench.git}.
\end{abstract}

%% file: 1.introduction.tex
\section{Introduction}

Text-to-image models~\cite{rombach2022high,podellsdxl,saharia2022photorealistic,li2024hunyuan} can generate highly realistic images from simple text descriptions called \emph{prompts}, offering substantial benefits across multiple domains. In content creation, they enable individuals with limited graphic design skills or resources to produce high-quality images; in education, they provide customized imagery to aid learning and engagement; and in marketing and design, they empower creators to generate on-demand, tailored visuals, reducing time and cost barriers. As a result, these capabilities have attracted significant interest from both academic researchers and industry developers.

However, alongside these benefits, text-to-image models also introduce severe societal risks. By enabling anyone to create realistic images, these models can be misused to spread misinformation and cause public harms. For instance, AI-generated images of prominent figures, like those depicting fictitious events involving public figures such as Joe Biden~\cite{biden-fake} or Donald Trump~\cite{trump-fake}, have been used to manipulate public opinion or disseminate false information. Such instances 
raise ethical and regulatory concerns about the responsible use of these models. Addressing these risks is crucial, especially as AI-generated images become more indistinguishable from real ones.

\begin{figure*}[!t]
   \centering
    {\includegraphics[width= \textwidth]{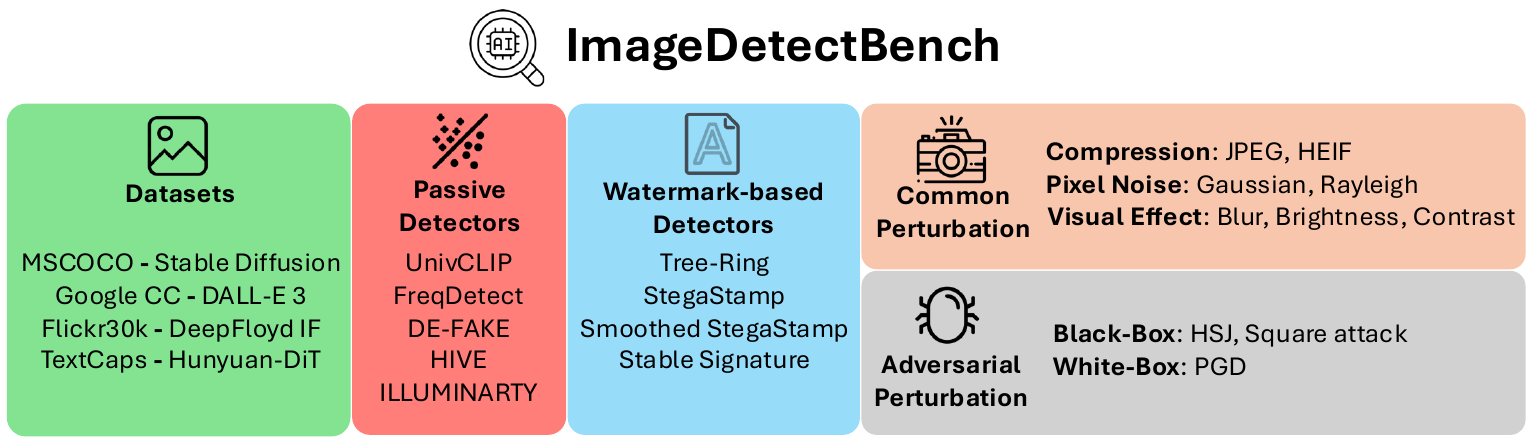}}
    \caption{Overview of our {\bench}.}
\label{framework}
\vspace{-4mm}
\end{figure*}

Detecting AI-generated images has become a crucial technology for mitigating these societal risks. At its core, a detector functions as a binary classifier, taking an image as input and determining whether it is AI-generated or non-AI-generated. Based on their design, detectors can be categorized into two types: \emph{passive} and \emph{watermark-based}.
Passive detectors~\cite{joslin2024double,sha2023fake,ojha2023towards,ricker2022towards,sha2024zerofake,alam2024ugad,hive-ai,illu-ai} rely on identifying subtle artifacts inherent to AI-generated images that may differ from  non-AI-generated ones. They leverage standard deep learning classifiers to classify images based on features extracted from their pixel values. For instance, companies like hive.ai~\cite{hive-ai} and illuminarty.ai~\cite{illu-ai} offer passive detection services to identify AI-generated images. 
By contrast, watermark-based detectors~\cite{tancik2020stegastamp,jiang2024certifiably,wen2024tree,fernandez2023stable,kim2024wouaf,HiDDeN,yang2024gaussian,jiang2024watermark} proactively embed a human-imperceptible ``artifact'' (called watermark) into images during their generation. An image is detected as AI-generated if a similar watermark can be extracted from it.  Watermark has recently gained substantial attention in both policy and industry; for example, the Executive Order on Trustworthy Artificial Intelligence from the White House~\cite{executive-order} recommends watermarking AI-generated content. Several major technology companies--such as OpenAI~\cite{dalle-c2pa}, Google~\cite{google-synthid}, Meta~\cite{fernandez2023stable}, and Microsoft~\cite{ms-watermark}--have also deployed watermarks.

While both detectors are actively deployed, there remains a key question: Which type of detector is more effective, robust, and efficient under real-world conditions? Understanding the comparative strengths and weaknesses of these two types of detectors is crucial for advancing detection technology and reducing the potential harms associated with AI-generated images. However, existing research primarily focuses on evaluating the effectiveness and robustness of either passive~\cite{ha2024organic,saberirobustness,meng2024ava,abdullah2024analysis} or watermark-based detectors~\cite{Evading-Watermark,jiang2024certifiably,hu2024transfer,an2024benchmarking,jiang2024watermark,hu2024stable}, leaving a comparative analysis under the same threat models largely unexplored.

\myparatight{Our work} We aim to bridge this gap by proposing \bench, the \emph{first} comprehensive benchmark for the comparative evaluation of passive and watermark-based detectors across diverse datasets and image perturbations. Figure~\ref{framework} provides an overview of our \bench. 

\textbf{\emph{-\quad Datasets}}. \bench includes four diverse datasets, each with distinct image characteristics. Every dataset contains an equal number of non-AI-generated and AI-generated images. We collected the non-AI-generated images from various established image benchmark datasets, while creating the AI-generated images using four text-to-image models. To reduce the influence of image content on the distinction between AI-generated and non-AI-generated images, we used the captions of the non-AI-generated images as prompts to create the AI-generated images.

\textbf{\emph{-\quad Perturbations}}.  An attacker may deceive a detector through either a \emph{removal} or \emph{forgery} attack. In a removal attack, the goal is to make the detector misclassify an AI-generated image as non-AI-generated, while in a forgery attack, a non-AI-generated image is misclassified as AI-generated. To assess the robustness of both passive and watermark-based detectors against these threats, \bench includes a variety of image perturbations. These consist of eight \emph{common perturbations}~\cite{hendrycks2018benchmarking,boyat2015review,wallace1992jpeg,jacobsen2004update,gedraite2011investigation} and three \emph{adversarial perturbations} (two black-box~\cite{chen2020hopskipjumpattack,andriushchenko2020square} and one white-box~\cite{madry2017towards}). Common perturbations encompass widely used image processing operations, such as JPEG compression, which may be applied by regular users as well as attackers. In contrast, adversarial perturbations are intentionally crafted to deceive detectors.

\textbf{\emph{-\quad Systematic benchmarking, findings, and recommendations.}} Using our datasets and perturbations, we systematically benchmark the comparative effectiveness, robustness, and efficiency of five state-of-the-art passive detectors~\cite{sha2023fake,ojha2023towards,ricker2022towards,hive-ai,illu-ai} and four watermark-based detectors~\cite{tancik2020stegastamp,jiang2024certifiably,wen2024tree,fernandez2023stable} under consistent threat models. Our comprehensive evaluation reveals several key findings. Notably, watermark-based detectors consistently outperform passive detectors across all scenarios, including those with no perturbations, common perturbations, and adversarial perturbations. Moreover, watermark-based detectors demonstrate remarkable insensitivity to variations in image distributions, whereas passive detectors are significantly impacted. Lastly, the most robust watermark-based detector is two orders of magnitude faster than the most robust passive detector. Based on these findings, we offer several recommendations. For instance, when both detector types are applicable, watermark-based detectors should be prioritized for their superior effectiveness, robustness, and efficiency.

To summarize, our key contributions are as follows:
\begin{itemize}
    \item 

We propose \bench, the \emph{first} benchmark designed to systematically compare the effectiveness, robustness, and efficiency of passive and watermark-based AI-generated image detectors.

    \item  \bench incorporates four diverse datasets, eleven types of perturbations (eight common perturbations and three adversarial perturbations), and a total of nine detectors: five passive and four watermark-based. 

    \item Based on our comprehensive benchmark study, we present key findings and offer recommendations for AI-generated image detection.
\end{itemize}

%% file: 2.taxonomy_of_detector.tex
\section{Taxonomy of Detectors}
\begin{figure*}[t!]
\centering
\begin{forest}
for tree={
    align=center,
    edge={-},
    parent anchor=south,
    child anchor=north,
    l sep=20pt,
    s sep=10pt,
    anchor=center,
    calign=center,
    inner sep=4pt,
    draw,
    rounded corners,
}
[Detector
    [Passive
        [Image-based
            [Spatial-domain-based \\ ~\cite{ojha2023towards,joslin2024double}]
            [Frequency-domain-based \\ ~\cite{ricker2022towards,alam2024ugad}]
        ]
        [{(Image, caption)-based} \\ ~\cite{sha2024zerofake,sha2023fake}]
    ]
    [Watermark-based
        [Pre-generation \\ ~\cite{wen2024tree,yang2024gaussian}]
        [Post-generation \\ ~\cite{HiDDeN,tancik2020stegastamp,jiang2024certifiably}]
        [In-generation \\ ~\cite{fernandez2023stable,kim2024wouaf}]
    ]
]
\end{forest}
\caption{Our taxonomy of detectors for AI-generated image detection.}
\label{taxonomy-detector}
\end{figure*}
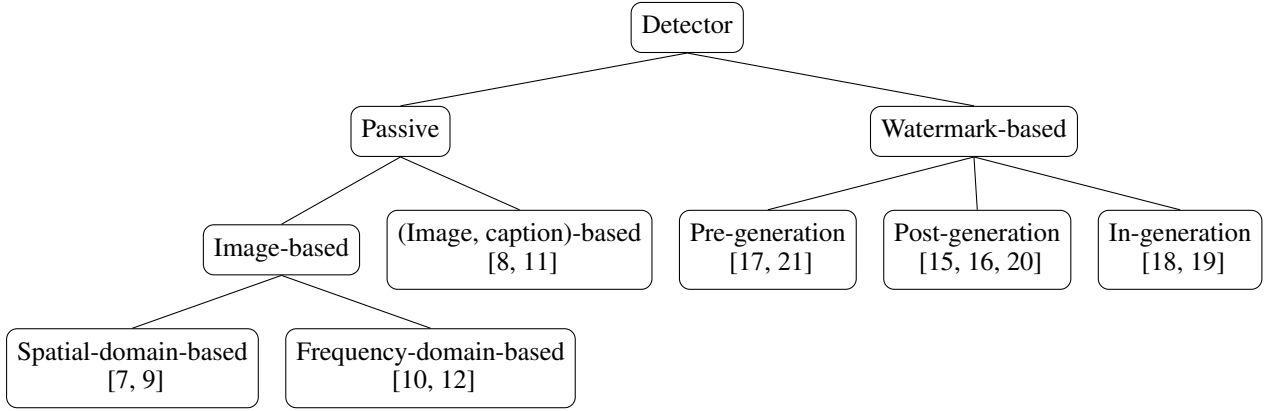

\begin{figure*}
   \centering
    {\includegraphics[width=1 \textwidth]{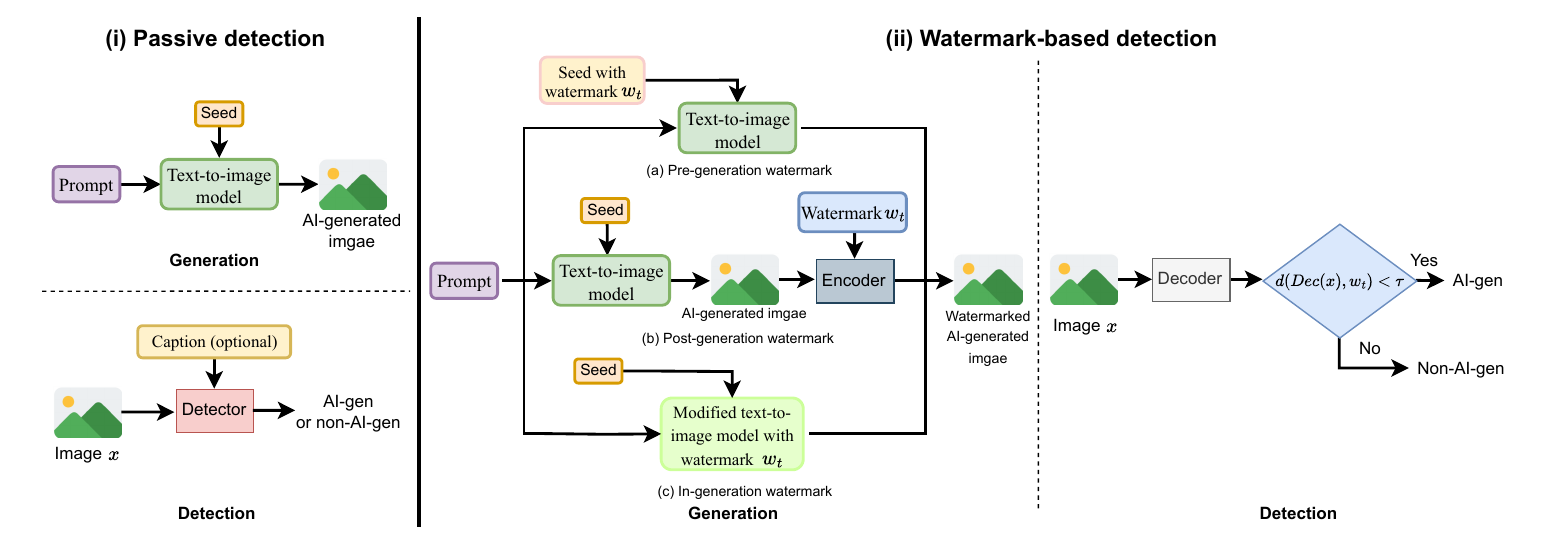}}
    \caption{Illustration of passive and watermark-based detection of AI-generated images. }
\label{illustration-passive-watermark}
\end{figure*}

A detector aims to classify images as either AI-generated or non-AI-generated. As illustrated in Figure~\ref{taxonomy-detector}, detection methods can be divided into two categories: \emph{passive}~\cite{sha2024zerofake,sha2023fake,ojha2023towards,ricker2022towards,alam2024ugad,joslin2024double,hive-ai,illu-ai} and \emph{watermark-based}~\cite{tancik2020stegastamp,jiang2024certifiably,wen2024tree,fernandez2023stable,kim2024wouaf,HiDDeN,yang2024gaussian}. Passive detectors identify AI-generated images by analyzing artifacts present within the images, whereas watermark-based detectors proactively embed a watermark into an AI-generated image, enabling later detection. Figure~\ref{illustration-passive-watermark} illustrates the difference between passive and watermark-based detection. We denote a detector as $D$, which takes an image $x$ as input and outputs either 1 or 0, where $D(x) = 1$ indicates $x$ is AI-generated and $D(x) = 0$ indicates $x$ is non-AI-generated. 
We note that some passive detectors~\cite{sha2024zerofake,sha2023fake} also incorporate an associated caption of the image $x$. In this work, we mainly focus on detectors from academic research since they are public but we also evaluate commercial detectors~\cite{hive-ai,illu-ai} in Section~\ref{sec:commercial}.

\subsection{Passive Detectors}
Passive detection methods train binary classifiers by following a standard machine learning pipeline. The initial step is to collect a training dataset comprising both AI-generated and non-AI-generated images. Some passive detection methods~\cite{sha2024zerofake,sha2023fake} also require a caption for each image in the training dataset. During training, features of images (and captions, if applicable) are extracted and used to train a binary classifier through supervised learning. During testing, given an image (and optionally its caption), features are extracted and fed into the binary classifier to identify whether this image is AI-generated or not. Different passive detectors essentially use different features and/or binary classifiers.
Based on whether captions are utilized as input, passive detectors can be further classified into two subcategories: \emph{image-based} and \emph{(image, caption)-based}.

\myparatight{Image-based detectors}
Image-based detectors~\cite{ojha2023towards,ricker2022towards,alam2024ugad,joslin2024double} rely solely on images as input and are typically categorized as \emph{spatial-domain-based} or \emph{frequency-domain-based}. Spatial-domain-based detectors~\cite{ojha2023towards,joslin2024double}  extract features in the spatial domain for detection. In contrast, frequency-domain-based detectors~\cite{ricker2022towards,alam2024ugad}  transform an image into the frequency domain, leveraging frequency-domain features to identify AI-generated images. In the following, we pick one representative detector from each category to discuss more details.

\begin{icompact}
    \item {\bf UnivCLIP~\cite{ojha2023towards}.} 
    UnivCLIP is a spatial-domain-based detector that leverages high-resolution pretrained image encoders, such as CLIP’s Vision Transformer (ViT), to extract spatial features from images. The features extracted by the CLIP encoder are passed through a single linear layer with sigmoid activation for classification.
        
    \item {\bf Frequency Domain Detection (FreqDetect)~\cite{ricker2022towards}.}
    FreqDetect is a frequency-domain-based detector. This detector extracts frequency features using transforms such as the Discrete Cosine Transform (DCT) and classifies them with logistic regression. Specifically, FreqDetect first converts an image to grayscale. The DCT then transforms the grayscale pixel values into a matrix of frequency coefficients. This matrix is flattened to serve as the feature vector for classification.
\end{icompact}

\myparatight{(Image, caption)-based detectors}
These detectors~\cite{sha2024zerofake,sha2023fake} take both an image and a caption as input, leveraging the combined information from textual and visual data. The key assumption is that AI-generated images are likely to closely align with the content described in their associated prompts/captions, whereas non-AI-generated images often contain additional details that may not be explicitly described by a  caption. In our experiments, we use DE-FAKE~\cite{sha2023fake} as an example detector in this subcategory.

\begin{icompact}
    \item {\bf DE-FAKE~\cite{sha2023fake}.}  DE-FAKE uses CLIP’s image and text encoders to extract features from an image and its caption, respectively. By concatenating these features, DE-FAKE creates a unified feature vector representing each image-caption pair. The classifier in DE-FAKE is a three-layer multilayer perceptron. During training, a set of (non-AI-generated image, caption) pairs from standard benchmark datasets and (AI-generated image, prompt) pairs are used to train the classifier, where the prompt used to generate an AI image is treated as its caption. During testing, if a caption is unavailable for a given image, it is generated by a captioning model, such as BLIP~\cite{li2022blip}.
    
\end{icompact}

\subsection{Watermark-based Detectors}
In contrast to passive detection, watermark-based detection proactively embeds watermarks in AI-generated images to facilitate their detection. In watermark-based detection, all AI-generated images are watermarked, while non-AI-generated images remain non-watermarked. The core of image watermarking methods lies in the \emph{embedding} and \emph{detection} of these watermarks. Based on the stage at which the watermark is embedded into an AI-generated image, image watermarks are categorized into three types: \emph{pre-generation}~\cite{wen2024tree,yang2024gaussian}, \emph{post-generation}~\cite{tancik2020stegastamp,jiang2024certifiably,HiDDeN}, and \emph{in-generation}~\cite{fernandez2023stable,kim2024wouaf}. Pre-generation methods modify the seed of the image generation process so that the image generated with the specific seed has a watermark. Post-generation methods add watermarks after generation by subtly modifying pixel values in the image. In-generation methods fine-tune the text-to-image model such that it generates images with the watermark intrinsically embedded.

Although these three types differ in watermark embedding, all require a watermark decoder during detection, denoted as $Dec$. We define the ground-truth watermark as $w_t$. During detection, the watermark decoder $Dec$ extracts  watermark $w$ from a given image $x$, i.e., $w=Dec(x)$. If the distance between the ground-truth watermark $w_t$ and the decoded watermark $w$ is below a threshold $\tau$, i.e., $d(w_t, w) < \tau$, the image $x$ is detected as watermarked and thus AI-generated, where $d$ is a distance metric, e.g., $\ell_0$-norm or $\ell_1$-norm. The threshold $\tau$ can be selected to ensure a desired probability of falsely detecting a non-watermarked image as watermarked based on the assumption that  $w_t$ is sampled uniformly at random. Formally, a watermark-based detector $D$ is defined as follows:
\begin{align}
    D(x) = \mathbb{I}(d(w_t, Dec(x)) \textless \tau),
\end{align}
where  $\mathbb{I}$ is the indicator function, i.e., $\mathbb{I}(d(w_t, Dec(x)) \textless \tau)$ is 1 if $d(w_t, Dec(x)) \textless \tau$ and 0 otherwise. Different watermarking methods use various ground-truth watermarks $w_t$, decoders $Dec$, and/or  distance metrics $d$. 
\myparatight{Pre-generation watermark} Given a prompt and a seed, a typical text-to-image model iteratively denoises the seed as an image. Pre-generation watermarks~\cite{wen2024tree,yang2024gaussian} embed a watermark into an AI-generated image via encoding a signal into the seed.
We use the state-of-the-art pre-generation watermarking method, Tree-Ring~\cite{wen2024tree}, in our experiments.

\begin{icompact} 
    \item {\bf Tree-Ring~\cite{wen2024tree}.} 
    When generating an image using a text-to-image model, Tree-Ring selects a seed $x_T$ whose Fourier transform contains a specific pattern, which serves as the ground-truth watermark, i.e., $w_t=FT(x_T)$, where $FT$ means Fourier transform.
    For detection, given an image $x$, a seed is reconstructed from it using inverse DDIM sampling~\cite{dhariwal2021diffusion}. Then, a watermark is extracted from the reconstructed seed. The decoded watermark can be expressed as $Dec(x)=FT(iDDIM(x))$, where $iDDIM$ is the inverse DDIM process. 
    If the $\ell_1$-norm distance between the decoded watermark and the ground-truth one is below the threshold $\tau$, the image $x$ is classified as watermarked and thus AI-generated. 
\end{icompact}

\myparatight{Post-generation watermark}
In a post-generation watermarking method~\cite{tancik2020stegastamp,jiang2024certifiably,HiDDeN}, the ground-truth watermark $w_t$ is a bitstring. These methods employ a watermark encoder, $Enc$, to embed the watermark into an image  by subtly altering the image's pixel values, producing a watermarked image. 
 In our experiments, we consider the state-of-the-art post-generation watermarking methods: StegaStamp~\cite{tancik2020stegastamp} and its smoothed version~\cite{jiang2024certifiably}.

\begin{icompact}
    \item {\bf StegaStamp~\cite{tancik2020stegastamp}.}
    StegaStamp uses neural networks as the watermark encoder $Enc$ and the decoder $Dec$. The two neural networks can be jointly trained using an image dataset. For detection, given an image $x$, if the $\ell_0$-norm distance (i.e., the number of mismatched bits) between the ground-truth watermark $w_t$ and the decoded watermark $Dec(x)$ is smaller than a threshold $\tau$, the image is classified as AI-generated.

    \item {\bf Smoothed StegaStamp~\cite{jiang2024certifiably}.}
    Smoothed StegaStamp is an enhanced version of StegaStamp, which provides certified robustness guarantees against bounded perturbations.
    During detection, given an image $x$, $N$ random Gaussian noises
    are added to it to construct $N$ noisy images; the watermark decoder $Dec$ extracts a watermark from each noisy image; and the $\ell_0$-norm distance between each decoded watermark and the ground-truth watermark $w_t$ is calculated. Finally, the image $x$ is classified as AI-generated if the median of these $N$ $\ell_0$-norm distances is smaller than the threshold $\tau$. Smoothed StegaStamp guarantees that the detection result is unaffected by any $\ell_2$-norm bounded perturbation added to $x$. 
\end{icompact}

\myparatight{In-generation watermark}
In-generation watermarks~\cite{fernandez2023stable,kim2024wouaf} modify the parameters of a text-to-image model, ensuring that the generated images contain a watermark. In our experiments, we use Stable Signature\cite{fernandez2023stable}, state-of-the-art in-generation watermark developed by Meta.

\begin{icompact} 
    \item {\bf Stable Signature~\cite{fernandez2023stable}.}
    Stable Signature is based on the post-generation watermarking method HiDDeN~\cite{HiDDeN}. Given a HiDDeN watermarking decoder $Dec$ and a ground-truth watermark $w_t$, Stable Signature fine-tunes the text-to-image model so that the watermark $w_t$  can be extracted from any generated image $\hat{x}$, i.e., $Dec(\hat{x})\approx w_t$.
    For detection, an image $x$ is classified as AI-generated if $\ell_0(w_t, Dec(x)) \textless \tau$.
\end{icompact}

%% file: 3.taxonomy_of_perturbation.tex
\section{Taxonomy of Perturbations}
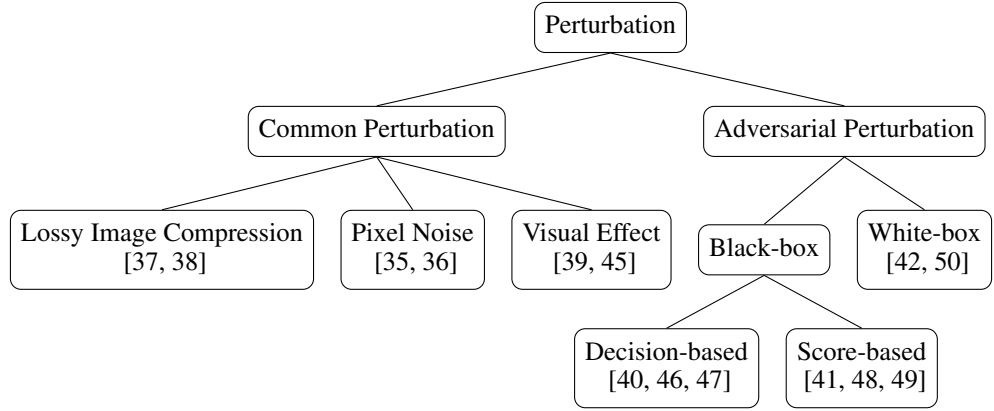
\begin{figure*}[t!]
\centering
\begin{forest}
for tree={
    align=center,
    edge={-},
    parent anchor=south,
    child anchor=north,
    l sep=20pt,
    s sep=10pt,
    anchor=center,
    calign=center,
    inner sep=4pt,
    draw,
    rounded corners,
}
[Perturbation
    [Common Perturbation
        [Lossy Image Compression \\ ~\cite{wallace1992jpeg,jacobsen2004update}]
        [Pixel Noise \\ ~\cite{boyat2015review,hendrycks2018benchmarking}]
        [Visual Effect \\ ~\cite{gedraite2011investigation,zhu2013estimating}]
    ]
    [Adversarial Perturbation
        [Black-box
            [Decision-based \\ ~\cite{chen2020hopskipjumpattack,hong2023certifiable,brendel2018decision}]
            [Score-based \\ ~\cite{andriushchenko2020square,ilyas2018prior,ilyas2018black}]
        ]
        [White-box \\ ~\cite{madry2017towards,goodfellow2014explaining}]
    ]
]
\end{forest}
\caption{Our taxonomy of perturbations to deceive a detector.}
 \label{taxonomy-perturbation}
\end{figure*}

\subsection{Removal and Forgery Attacks}

To deceive an image detector, an attacker can introduce perturbations to manipulate an AI-generated image to appear non-AI-generated (\emph{removal attack}), or, conversely, make a non-AI-generated image appear AI-generated (\emph{forgery attack}). In a removal attack, the attacker adds perturbation $\delta$ to an AI-generated image $x_a$ so that the detector $D$ mistakes it for a non-AI-generated image, i.e., $D(x_a+\delta) = 0$. In a forgery attack, the attacker adds perturbation $\delta$ to a non-AI-generated image $x_h$ so that the detector $D$ mistakes it for an AI-generated image, i.e., $D(x_h+\delta) = 1$. 

To execute these attacks, the attacker can apply either a \emph{common perturbation} or  \emph{adversarial perturbation} to alter the AI-generated or non-AI-generated images. Common perturbation involves using standard image processing techniques, whereas adversarial perturbation introduces carefully crafted noise to an image. Figure~\ref{taxonomy-perturbation} shows the detailed taxonomy of perturbations that may be introduced to deceive a detector. It is important to note that common perturbations can also be applied by normal users for benign purposes, and may not always correspond to an attack.

\subsection{Common Perturbation} \label{sec:commonpert}
These perturbations arise from common image processing operations. We categorize them into three types: lossy image compression~\cite{wallace1992jpeg,jacobsen2004update}, pixel noise~\cite{boyat2015review,hendrycks2018benchmarking}, and visual effect~\cite{gedraite2011investigation,zhu2013estimating}. Each type of perturbation has a parameter, which controls the amount of perturbation added to an image. 

\myparatight{Lossy image compression} 
Lossy image compression aims to reduce image file sizes but introduces compression artifacts that can degrade image quality. We incorporate two widely used image compression methods~\cite{wallace1992jpeg,jacobsen2004update}. 
\begin{icompact}
    \item {\bf JPEG Compression~\cite{wallace1992jpeg}.}
    This method uses a quality factor $q$ to balance compression ratio and image quality. A smaller $q$ results in more aggressive compression.

    \item {\bf High Efficiency Image File Format (HEIF)~\cite{jacobsen2004update}.}
     HEIF also uses a quality factor $q$ to balance image quality and file size. A smaller $c$ results in more aggressive compression.

\end{icompact}

\myparatight{Pixel noise}
Pixel noise perturbation introduces random noise at the pixel level. 
\begin{icompact}
    \item {\bf Gaussian noise.}
    Gaussian noise is characterized by a Gaussian distribution with zero mean and standard deviation $\sigma$, which controls the noise intensity. 
    \item {\bf Rayleigh noise.}
    Rayleigh noise is characterized by Rayleigh distribution with a scale parameter $\sigma$, which also determines the intensity of the noise. 
\end{icompact}

\myparatight{Visual effect}
Visual effect introduces carefully designed modifications to pixel values.
\begin{icompact}
    \item {\bf Gaussian blur.}
    Gaussian blur reduces image sharpness by averaging pixel values within a local neighborhood, following a Gaussian distribution. The blur radius $r$ determines the extent of the smoothing effect.

    \item {\bf Brightness adjustment.}
    This method adjusts the brightness of an image by adding a specified brightness value $b$ to the pixel intensities. 

    \item {\bf Contrast adjustment.}
    Contrast adjustment scales the difference between each pixel and the mean pixel value. Pixel intensities are centered by subtracting the mean, multiplied by a contrast factor $c$, and re-centered. A factor $c > 1$ increases contrast, making an image more vivid, while $0 < c < 1$ reduces contrast, flattening details.
    
    \item {\bf Elastic blur.}
    Elastic blur applies a spatial distortion to an image by shifting pixel locations according to a smooth displacement field. The degree of distortion is controlled by a parameter $\alpha$, which defines the intensity of the transformation. Larger $\alpha$ causes more pronounced warping, while smaller $\alpha$ leads to subtler distortions.
\end{icompact}

\subsection{Adversarial Perturbation}
\label{sec:adversarial_perturbation}
Adversarial perturbations are intentionally crafted by attackers to execute removal or forgery attacks aimed at deceiving a detector. Specifically, given an image $x$, an adversarial perturbation $\delta$ is found to deceive the detector $D$, i.e., $D(x+\delta)\neq D(x)$, where $x$ is an AI-generated image in removal attacks and a non-AI-generated image in forgery attacks. 
These adversarial perturbations can be identified by adapting adversarial examples~\cite{goodfellow2014explaining} to detectors. Specifically, we treat a detector as a binary classifier and apply adversarial examples to it. Based on different levels of attackers' background knowledge, we consider two attack settings: \emph{black-box} and \emph{white-box}.

\myparatight{Black-box attacks}
In the black-box setting, we assume the attacker can query the detector $D$ using various images. Given an image $x$ and a query budget, these attacks iteratively query $D$ to optimize  $\delta$ until  deceiving the detector or reaching the query budget. Based on the detector's response, these attacks can be further categorized into \emph{decision-based}~\cite{chen2020hopskipjumpattack,hong2023certifiable,brendel2018decision} and \emph{score-based}~\cite{andriushchenko2020square,ilyas2018black,ilyas2018prior}. In decision-based attacks, the detector simply indicates whether the queried image is AI-generated or not. In score-based attacks, the detector not only provides this detection result but also returns additional information, such as classification probability (for passive detectors) or distance between the decoded watermark and the ground-truth one (for watermark-based detectors). In our experiments, we use the HopSkipJump attack~\cite{chen2020hopskipjumpattack}, a decision-based attack, and the Square attack~\cite{andriushchenko2020square}, a score-based attack. 
\begin{icompact}
    \item {\bf HopSkipJump (HSJ) attack~\cite{chen2020hopskipjumpattack}.}
    Given an image $x$, this method begins with a large perturbation $\delta$ that successfully deceives the detector $D$, i.e., $D(x+\delta)\neq D(x)$.  Such initial perturbation $\delta$ can be obtained via applying severe common perturbations onto the image $x$~\cite{Evading-Watermark}. The perturbation $\delta$ is then iteratively optimized using gradients estimated from the detector's responses. Specifically, the HopSkipJump attack defines a loss function to quantify the attack's effectiveness and uses Monte Carlo estimation~\cite{mooney1997monte} to approximate its gradient. This gradient is then used to search for the minimal perturbation capable of deceiving the detector. 

    \item {\bf Square attack~\cite{andriushchenko2020square}.}
    This method employs randomly generated noise to iteratively perturb randomly selected square regions of an image $x$. Specifically, in each iteration, the attack applies noise to the selected square regions if the noise reduces the classification probability of the correct class (for passive detectors) or the bitwise accuracy (for watermark-based detectors). 
\end{icompact}

\myparatight{White-box attacks}
In the white-box setting, we assume that the attacker has full knowledge of the detector $D$, including its parameters and architecture. Given an image $x$ and a perturbation budget $r$, the attacker’s objective is to identify a perturbation $\delta$ within the budget that is most likely to deceive the detector $D$. The process of finding this perturbation $\delta$ can be formulated as an optimization problem, typically solved using gradient-based methods~\cite{madry2017towards,goodfellow2014explaining}.

For passive detectors, the attacker optimizes $\delta$ such that the detector’s prediction is opposite to the original prediction. The optimization problem is defined as follows~\cite{madry2017towards}:
\begin{equation}
\begin{aligned}
\max _\delta \  CE(x+\delta, D(x);D) \quad s.t. \  \| \delta \|_\infty \leq r,
\end{aligned}
\label{optim-wb-passive}
\end{equation}
where $CE$ denotes the cross-entropy loss of $D$ for $x+\delta$ when treating $D(x)$ as its label, and $\| \cdot \|_\infty$ indicates $\ell_\infty$-norm. For watermark-based detectors, since the decoder $Dec$ does not output a class probability, the typical white-box adversarial example methods cannot be directly applied. Following Jiang et al.~\cite{Evading-Watermark}, the attacker optimizes a perturbation $\delta$ to either maximize the distance between the ground-truth watermark ${w}_t$ and decoded watermark $Dec(x + \delta)$ for a removal attack or minimize it for a forgery attack. Specifically, for a removal attack, the optimization problem can be formulated as follows:
\begin{equation}
\begin{aligned}
\max  _\delta \  l_2({w}_t, Dec(x_a+\delta)) \quad s.t. \  \| \delta \|_\infty \leq r,
\end{aligned}
\label{optim-wb-wm-removal}
\end{equation}
where $x_a$ is a watermarked, AI-generated image. 
For a forgery attack, the optimization problem is as follows:
\begin{equation}
\begin{aligned}
\min  _\delta \  l_2({w}_t, Dec(x_h+\delta)) \quad s.t. \  \| \delta \|_\infty \leq r,
\end{aligned}
\label{optim-wb-wm-forgery}
\end{equation}
where $x_h$ is a non-AI-generated image. To solve the optimization problems for both passive and watermark-based detectors, we employ \emph{projected gradient descent (PGD)}~\cite{madry2017towards} to optimize $\delta$. Specifically, we optimize $\delta$ under an $\ell_{\infty}$-norm constraint defined by the perturbation budget $r$. If $\delta$ exceeds this budget, it is projected back onto the $\ell_{\infty}$-norm ball.

%% file: 4.dataset_collection.tex
\begin{table*}[!t]
\centering
\caption{Summary of our datasets. Each dataset includes 11,000 non-AI-generated and 11,000 AI-generated images, with separate columns for training and testing splits.}
\label{tab:dataset-summary} 
\renewcommand{\arraystretch}{1.2}
\begin{NiceTabular}{|c|c|c|c|c|c|c|}
\hline
           \Block{2-1}{\textbf{Dataset}} & \multicolumn{3}{c|}{\textbf{Non-AI-generated}} & \multicolumn{3}{c|}{\textbf{AI-generated}} \\
           \cline{2-7}
           & \textbf{Data source} & \textbf{\#Training} & \textbf{\#Testing} & \textbf{Model} & \textbf{\#Training} & \textbf{\#Testing} \\
\hline
\textbf{MS} &     MSCOCO         &      10,000     & 1,000 & Stable Diffusion       & 10,000 & 1,000                 \\
\hline
\textbf{GD}     &     Google CC        &      10,000     & 1,000 & DALL-E 3               & 10,000 & 1,000                 \\
\hline
\textbf{FD}  &      Flickr30k        &      10,000     & 1,000 & DeepFloyd IF           & 10,000 & 1,000                 \\
\hline
\textbf{TH}    &       TextCaps       &      10,000     & 1,000 & Hunyuan-DiT           & 10,000 & 1,000                 \\
\hline
\end{NiceTabular}
\end{table*}

\section{Collecting Datasets} 

Our goal is to collect diverse image datasets with varied content generated by different text-to-image models. To achieve this, we compiled four datasets, each containing both non-AI-generated and AI-generated images. For each dataset, we followed these steps: beginning with a public dataset of non-AI-generated images with captions, we sampled 11,000 images. Then, we used the corresponding captions as prompts to generate 11,000 AI-generated images through a text-to-image model. Using the same captions as prompts helps minimize the influence of image content on the distinction between AI-generated and non-AI-generated images. We pair each dataset of non-AI-generated images with a different text-to-image model, though the pairings are not based on any specific pattern. Figure~\ref{fig:example_images_MS}-\ref{fig:example_images_TH} in Appendix show examples of images from each dataset. Table~\ref{tab:dataset-summary} provides an overview of the four datasets, which we elaborate on in the following.

\myparatight{MSCOCO -- Stable Diffusion (MS)} 
MSCOCO~\cite{lin2014microsoft} is a dataset containing 320,000 non-AI-generated images with captions, featuring common scenes intended for object detection, segmentation, and captioning tasks. We sampled 11,000 non-AI-generated images from MSCOCO. Each image includes one or more captions; for each, we selected the first caption as a prompt to generate an AI-generated image using Stable Diffusion v2.1~\cite{rombach2022high}. These prompts have an average of 10.2 tokens, with a standard deviation of 2.4 tokens, and a median of 10 tokens.

\myparatight{Google CC -- DALL-E 3 (GD)} 
The Google CC dataset~\cite{sharma2018conceptual} contains more than 3,300,000 non-AI-generated images sourced from various websites, where the image  captions are extracted and refined from the associated HTML tags. We sampled 11,000 non-AI-generated images from Google CC. Furthermore, for each non-AI-generated image, we treat its caption as a prompt to generate an AI-generated image using DALL-E 3 (version 2024/02/01). These prompts have an average of 6.9 tokens, with a standard deviation of 1.5 tokens, and a median of 6 tokens.

\myparatight{Flickr30k -- DeepFloyd IF (FD)} The Flickr30k dataset~\cite{plummer2015flickr30k} contains more than 30,000 non-AI-generated images, each paired with multiple captions. We sampled 11,000 non-AI-generated images from Flickr30k. Similar to our treatment of MSCOCO, for each image, we use its first caption as a prompt to generate an AI-generated image through  DeepFloyd IF~\cite{saharia2022photorealistic}. 
We use the first two modules of DeepFloyd IF (``IF-I-XL-v1.0'' and ``IF-II-L-v1.0'') in its pipeline.  These prompts have an average of 17.3 tokens, with a standard deviation of 5.5 tokens, and a median of 16 tokens.

\myparatight{TextCaps -- Hunyuan-DiT (TH)} The TextCaps dataset~\cite{sidorov2020textcaps} includes 39,408 non-AI-generated images and there are 5 captions for each image. We sampled 11,000 non-AI-generated images from TextCaps. Furthermore, for each image,  we use its first caption as a prompt to generate an AI-generated image using Hunyuan-DiT~\cite{li2024hunyuan}. These prompts have an average of 12.6 tokens, with a standard deviation of 3.7 tokens, and a median of 12 tokens.

\myparatight{Training and testing splits} For each dataset, we sample 10,000 non-AI-generated and 10,000 AI-generated images for the training set, reserving the remaining 1,000 non-AI-generated and 1,000 AI-generated images for the testing set.   
For passive detectors, we train them on the training set of each dataset and test them on the corresponding testing set. For watermark-based detectors, we train the watermark encoders and decoders using a dataset separate from the four primary datasets, with further details provided in Section~\ref{train_watermark_detectors}. It is worth noting that our training setup gives advantages to passive detectors.

\section{Training Detectors} 
\label{sec:training_detectors}

\subsection{Passive Detectors}
We evaluate three passive detectors: UnivCLIP~\cite{ojha2023towards}, FreqDetect~\cite{ricker2022towards}, and DE-FAKE~\cite{sha2023fake}. We use the publicly available code for each detector. Unless specified otherwise, for each of our four datasets,  the training set is used to train these detectors, and their performance is evaluated using the corresponding testing set. Detailed training procedures for each detector are outlined below.

\myparatight{UnivCLIP} 
 UnivCLIP employs the ViT-B/32 CLIP model~\cite{radford2021learning} to extract spatial features from images, with a linear layer serving as the classifier. We train the classifier using the Adam optimizer, with a batch size of 500 for 800 iterations. The initial learning rate is set at $3 \times 10^{-4}$ and decays to $1 \times 10^{-6}$ following a cosine decay schedule. To enhance robustness, we use \emph{training with perturbations}, where images are perturbed during training. These perturbations include JPEG compression, Gaussian noise, Gaussian blur, brightness, and contrast adjustments. Each image undergoes a perturbation picked from the five choices uniformly at random in each iteration.

\myparatight{FreqDetect} 
  FreqDetect uses the DCT frequency coefficients as image features, with logistic regression serving as the classifier. The logistic regression classifier is trained using the L-BFGS solver over 1,000 iterations with an $\ell_2$-norm penalty. We also apply training with perturbations to enhance robustness.

\myparatight{DE-FAKE}
 DE-FAKE uses the ViT-B/32 CLIP model~\cite{radford2021learning} to extract features from both images and captions, with a three-layer multilayer perceptron serving as the classifier based on these features. Since DE-FAKE takes image-caption pairs as input, the training set includes both images and their corresponding captions. During training, the classifier is optimized using the Adam optimizer with a batch size of 500 for 800 iterations. The learning rate is initially set to $3 \times 10^{-4}$ and decays to $1 \times 10^{-6}$ following a cosine decay schedule. Additionally, training with perturbations is used to enhance robustness. For testing, we use BLIP~\cite{li2022blip} to generate a caption for each image and classify the resulting image-caption pair with the trained classifier.

\subsection{Watermark-based Detectors}\label{train_watermark_detectors}
In our experiments, we use four watermark-based detectors: Tree-Ring~\cite{wen2024tree},  StegaStamp~\cite{tancik2020stegastamp},  Smoothed StegaStamp~\cite{jiang2024certifiably}, and Stable Signature~\cite{fernandez2023stable}. Their respective training details (if any) are provided below.

\myparatight{Tree-Ring} Tree-Ring does not require training, and we use the default watermark embedding and detection settings provided by Wen et al.~\cite{wen2024tree}. Specifically, a circle pattern with a radius of 10 serves as the ground-truth watermark $w_t$. For detection, Tree-Ring utilizes the inverse DDIM process to reconstruct a seed for a given image. This reconstruction process requires a prompt, as is typical for text-to-image models. In our experiments, we follow Wen et al.~\cite{wen2024tree} to use an empty string as the prompt. Note that Tree-Ring is tailored to Stable Diffusion, and thus we will only use it to watermark AI-generated images in our MS dataset.

\myparatight{StegaStamp} For StegaStamp, we train the watermark encoder and decoder on 10,000 non-AI-generated images sampled from the MSCOCO dataset~\cite{lin2014microsoft}. Importantly, these images do not overlap with those  in our MS dataset, which is to show the generalization ability across datasets. 

    Following Tancik et al.~\cite{tancik2020stegastamp}, we employ a U-Net architecture as the watermark encoder $Enc$ and a spatial transformer network~\cite{jaderberg2015spatial} as the watermark decoder $Dec$. The watermark is a bitstring with 32 bits. The training process involves two phases: \emph{standard training} and \emph{training with perturbations}. In the standard training phase, the watermark encoder and decoder are jointly trained such that 1) the watermark decoded from a watermarked image is similar to the true watermark, i.e., $Dec(Enc(x,w))\approx w$ for any watermark $w$, where $Enc(x,w)$ indicates a watermarked image with watermark $w$; and 2) a watermarked image (i.e., $Enc(x,w)$) looks visually the same as the image $x$.
    
    Training with perturbations differs by applying perturbations to the watermarked images during training.  This encourages the watermark encoder to embed a robust pixel pattern into an image as a watermark. Specifically, the watermark encoder and decoder are jointly trained such that the watermark decoded from a perturbed watermarked image is similar to the true watermark, i.e., $Dec(P(Enc(x,w)))\approx w$ for $w$, where $P$ is a perturbation operation. We apply the same five types of perturbations as training the passive detectors. For each training image in each iteration, we randomly sample one perturbation from these five options and apply it to the corresponding watermarked image.
    
    We train the watermark encoder and decoder for 3,000 iterations with a batch size of 32. The first 1,000 iterations use standard training, while the remaining 2,000 iterations employ training with perturbations to enhance robustness. After training, a random 32-bit bitstring is sampled as $w_t$, which is embedded into AI-generated images. 

\myparatight{Smoothed StegaStamp} Smoothed StegaStamp uses the trained StegaStamp model, so it does not require extra training. Specifically, Smoothed StegaStamp applies the regression smoothing~\cite{jiang2024certifiably} to the trained StegaStamp model. During detection, given an image $x$, Smoothed StegaStamp samples 100 Gaussian noise vectors with zero mean and a standard deviation of 0.1, adds them to the image to construct 100 noisy images. For each noisy image, we use the StegaStamp's watermark decoder $Dec$ to decode a watermark and calculate its $\ell_0$-norm distance with the ground-truth watermark $w_t$. We then take the median across the 100 $\ell_0$-norm distances as the final $\ell_0$-norm distance, and classify the image $x$ as AI-generated if the final $\ell_0$-norm distance is smaller than a threshold $\tau$.

\myparatight{Stable Signature}   For Stable Signature, we use the publicly available text-to-image model~\cite{fernandez2023stable}. In this model, the ground-truth  watermark $w_t$ is a 48-bit bitstring. This model is adversarially trained with JPEG compression, random cropping, and random resizing. The watermark decoder $Dec$ follows the architecture of HiDDeN~\cite{HiDDeN}. We note that the publicly available  text-to-image model fine-tuned by Stable Signature is Stable Diffusion, and thus in experiments, we will only apply Stable Signature to watermark AI-generated images in our MS dataset. 

\myparatight{Detection threshold $\tau$} Passive detectors rely on standard classifiers with widely used detection thresholds. In contrast, via properly setting the threshold $\tau$, watermark-based detectors can ensure a desired probability of falsely detecting a non-AI-generated image as AI-generated, assuming the ground-truth watermark is sampled uniformly at random. Following previous studies~\cite{Evading-Watermark}, we select \(\tau\) in our experiments to ensure this probability is less than 0.0001. Specifically, for StegaStamp and Smoothed StegaStamp, \(\tau\) is set to 0.8125; for Stable Signature, \(\tau\) is set to 0.7708; and for Tree-Ring, \(\tau\) is set to 70.1875. It is important to note that this probability differs from the false positive rate (FPR) reported in our experiments. FPR refers to the fraction of non-AI-generated images falsely detected as AI-generated for a given ground-truth watermark.

%% file: 5.benchmarking.tex
\section{Benchmarking Results}

\subsection{Evaluation Metrics}
\vspace{-2mm}
\myparatight{FNR, FPR, and ACC} We evaluate the performance of the detectors using  \emph{false negative rate (FNR)}, \emph{false positive rate (FPR)}, and \emph{accuracy (ACC)}. FNR is the fraction of AI-generated images incorrectly detected as non-AI-generated; FPR is the fraction of non-AI-generated images incorrectly detected as AI-generated; and  ACC is the overall fraction of correct detection. Unless otherwise mentioned, for each of our four datasets, we calculate FNR, FPR, and ACC of a detector using the testing set comprising 1,000 AI-generated and 1,000 non-AI-generated images. Under removal (or forgery) attacks, we add common or adversarial perturbations to AI-generated (or non-AI-generated) images.  A smaller FNR (or FPR) for the perturbed AI-generated (or non-AI-generated) images indicates that the detector is more robust against removal (or forgery) attacks. 

\myparatight{LPIPS}  We will apply common and adversarial perturbations to the images to evaluate the robustness of detectors. In those experiments, we further assess the visual similarity between an image and its perturbed version using the \emph{Learned Perceptual Image Patch Similarity (LPIPS)}~\cite{zhang2018unreasonable}. LPIPS is a widely used metric that quantifies the visual similarity between two images by comparing their deep feature representations. In our experiments, the feature representation of an image is calculated using the pre-trained VGG neural network~\cite{karen2014very}. A lower LPIPS score indicates that an image and its perturbed version are more visually similar.

\myparatight{Running time} 
We also evaluate the efficiency of the detectors by measuring their runtime per image detection. Given the high volume of images processed in real-world scenarios, detection may need to be performed frequently and at scale. For instance, social media platforms may label AI-generated images before publication. According to \cite{twitter}, X (formerly Twitter) publishes approximately 3,000 images per second. This immense volume highlights the necessity for detectors to operate efficiently.

\begin{table*}[t!]
\centering
\caption{FNRs, FPRs, and ACCs of various detectors on our four datasets. We cannot get FNRs (and thus ACCs) for Tree-Ring and Stable Signature on  GD, FD and TH datasets since their implementations are only applicable for Stable Diffusion model.}
\renewcommand{\arraystretch}{1.2}
\resizebox{16.5cm}{!}{
\begin{NiceTabular}{|c|c|c|c|c|c|c|c|c|c|c|c|c|c|}[hvlines]
\Block{2-2}{\textbf{Detector}} & & \multicolumn{3}{c|}{\textbf{MS}} & \multicolumn{3}{c|}{\textbf{GD}} & \multicolumn{3}{c|}{\textbf{FD}} & \multicolumn{3}{c|}{\textbf{TH}} \\ \cline{3-14}
& & \textbf{FNR} & \textbf{FPR} & \textbf{ACC} & \textbf{FNR} & \textbf{FPR} & \textbf{ACC} & \textbf{FNR} & \textbf{FPR} & \textbf{ACC} & \textbf{FNR} & \textbf{FPR} & \textbf{ACC} \\ \hline

\Block{3-1}{Passive} & UnivCLIP & 0.05 & 0.07 & 0.94 & 0.01 & 0.06 & 0.96 & 0.01 & 0.01 & 0.99 & 0.02 & 0.09 & 0.94 \\ \cline{2-14}
& FreqDetect & 0.26 & 0.15 & 0.80 & 0.18 & 0.28 & 0.77 & 0.08 & 0.06 & 0.93 & 0.18 & 0.30 & 0.76 \\ \cline{2-14}
& DE-FAKE & 0.05 & 0.06 & 0.94 & 0.01 & 0.05 & 0.97 & 0.01 & 0.01 & 0.99 & 0.03 & 0.06 & 0.96 \\ \hline \hline

\Block{4-1}{Watermark-based} & Tree-Ring & 0.00 & 0.00 & 1.00 & - & 0.00 & - & - & 0.00 & - & - & 0.00 & - \\ \cline{2-14}
& StegaStamp & 0.00 & 0.00 & 1.00 & 0.00 & 0.00 & 1.00 & 0.00 & 0.00 & 1.00 & 0.00 & 0.00 & 1.00 \\ \cline{2-14}
& Smoothed StegaStamp & 0.00 & 0.00 & 1.00 & 0.00 & 0.00 & 1.00 & 0.00 & 0.00 & 1.00 & 0.00 & 0.00 & 1.00 \\ \cline{2-14}
& Stable Signature & 0.01 & 0.00 & 0.99 & - & 0.00 & - & - & 0.00 & - & - & 0.00 & - \\ \hline

\end{NiceTabular}
}
\label{tab:no_pert}
\end{table*}

\subsection{No Perturbations}
\vspace{-2mm}
\myparatight{In-distribution performance}  
Table~\ref{tab:no_pert} presents the FNRs, FPRs, and ACCs of various detectors for each dataset when no perturbations are added to the testing images. It is important to note that all passive detectors are trained and evaluated respectively on the training and testing set of each dataset, i.e., the testing and training images follow the same distribution. In contrast, watermark-based detectors are trained using a dataset that is different from the four datasets (see Section~\ref{train_watermark_detectors} for more details). This setting gives advantages to  passive detectors. 
We also note that the public implementations of Tree-Ring and Stable Signature are only applicable to watermark images generated by Stable Diffusion. Therefore, we can only calculate FNR (and thus ACC) on the MS dataset. However, we can calculate their FPRs on all the four datasets. 

\begin{figure}[!t]
\centering
\renewcommand*{\arraystretch}{0}
\begin{tabular}{ccc}
\begin{subfigure}{.33\linewidth}
  \centering
  \includegraphics[width=\linewidth]{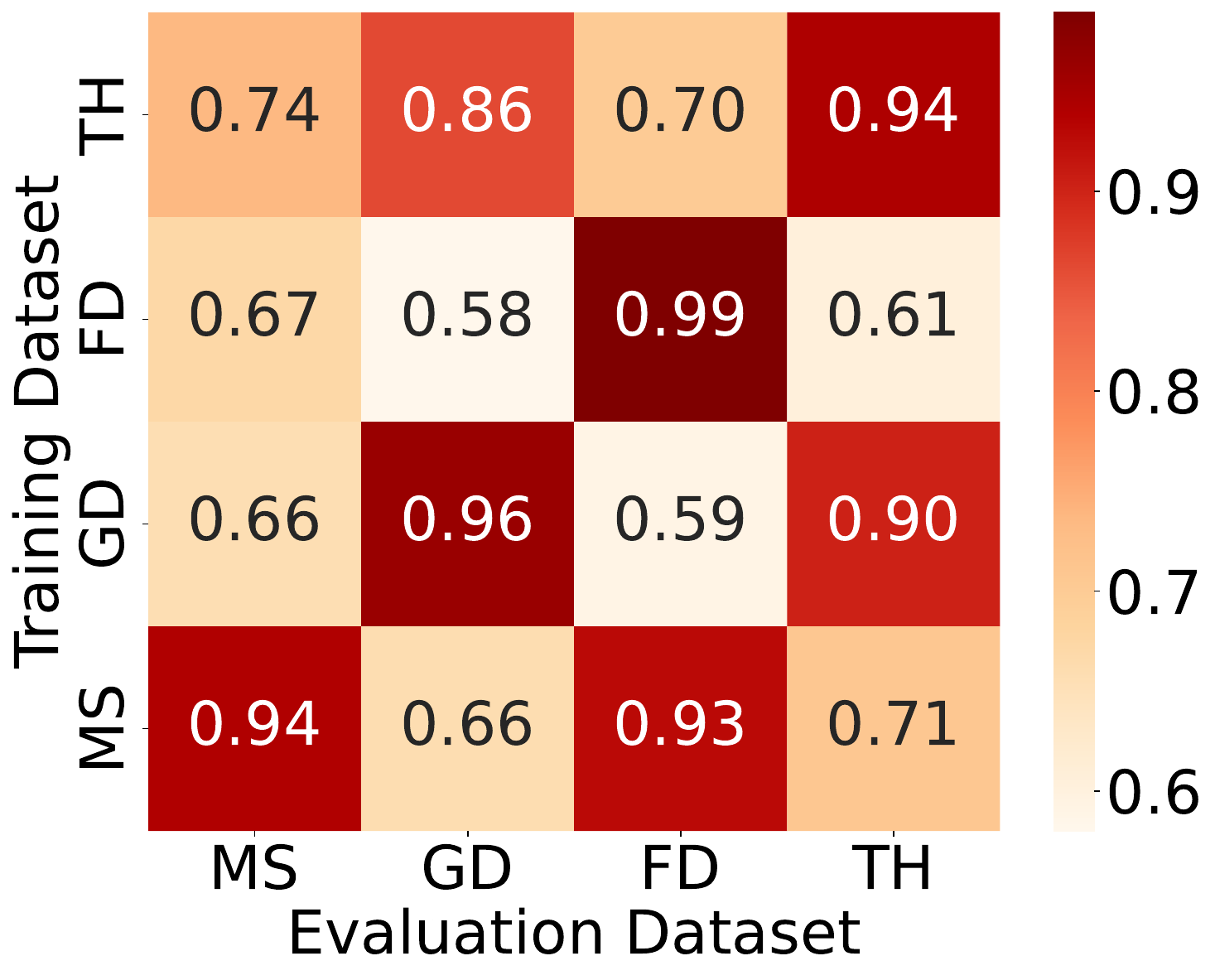}
  \caption{UnivCLIP}
\end{subfigure} 
\begin{subfigure}{.33\linewidth}
  \centering
  \includegraphics[width=\linewidth]{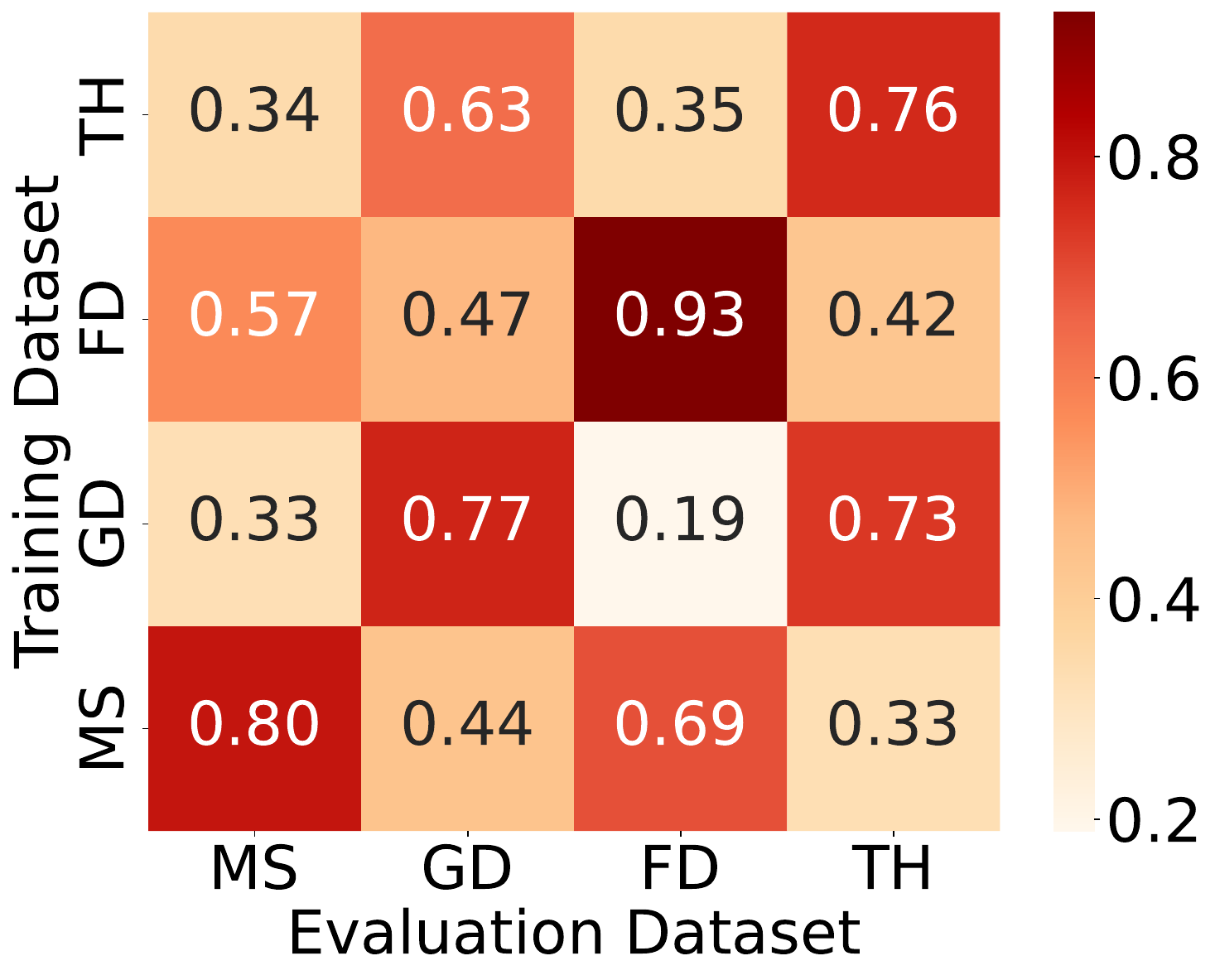}
  \caption{FreqDetect}
\end{subfigure}
\begin{subfigure}{.33\linewidth}
  \centering
  \includegraphics[width=\linewidth]{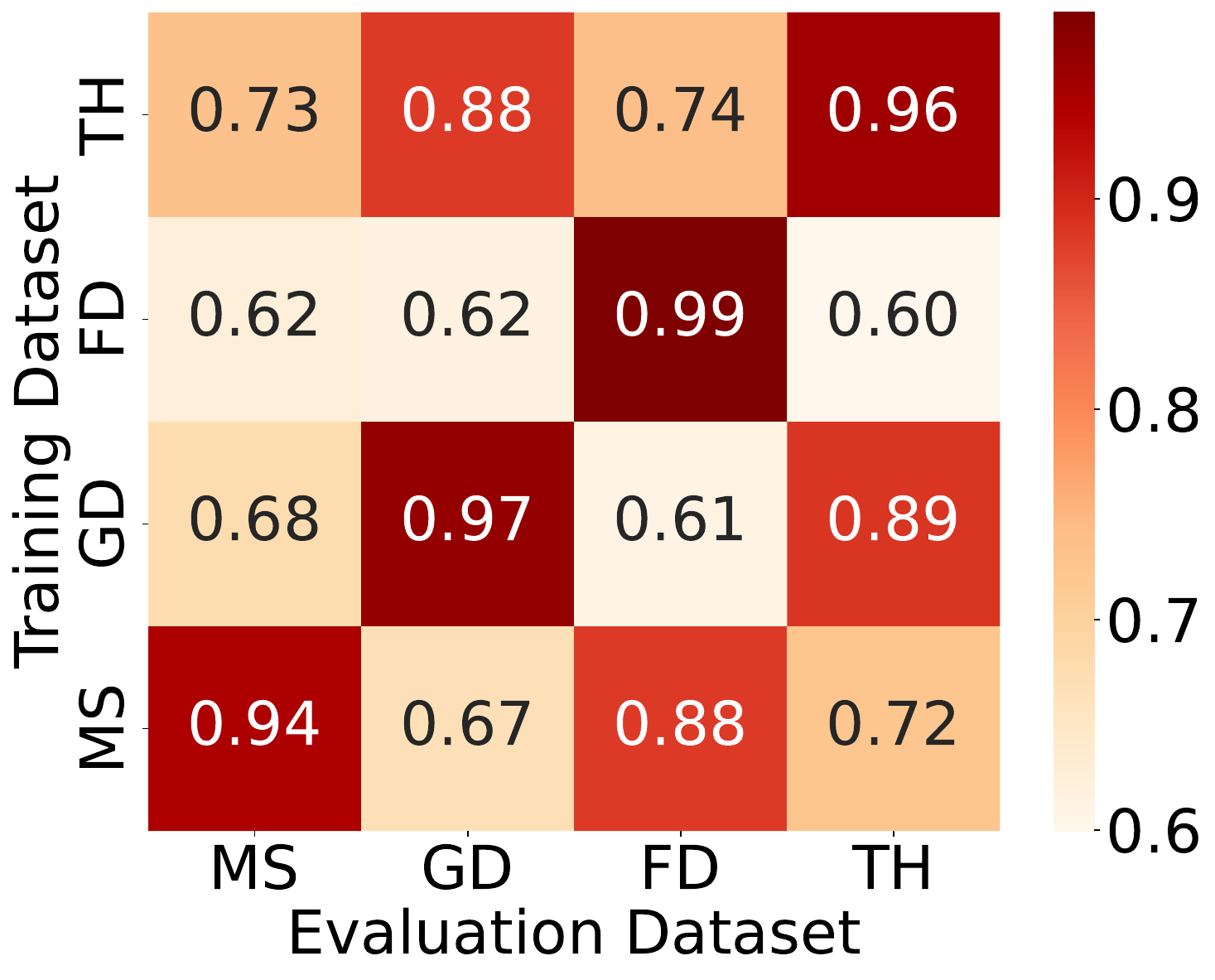}
  \caption{DE-FAKE}
\end{subfigure} \\
\end{tabular}
\caption{ACCs of passive detectors when trained and evaluated across datasets.}
\label{fig:crossdataset}
\vspace{-4mm}
\end{figure}

\begin{figure*}[!t]
    \centering
    \includegraphics[width=\linewidth]{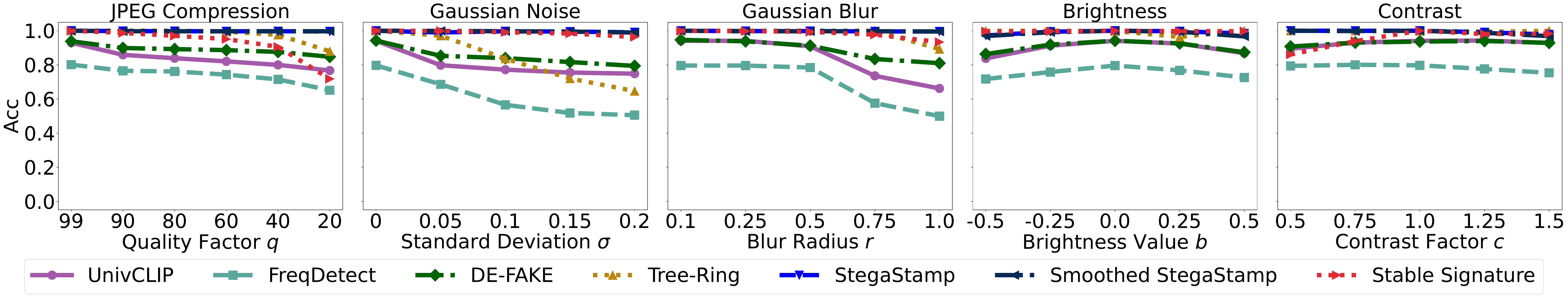}
    \caption{ACCs of various detectors on the MS dataset under 5 types of common perturbations seen during training. The FNR and FPR results are shown in Figure~\ref{fig:common_pert_fnrfpr} in Appendix.}
    \label{fig:common_pert}
    \vspace{-2mm}
\end{figure*}

First, we observe that all watermark-based detectors outperform passive detectors. Specifically, all watermark-based detectors except Stable Signature achieve an ACC of 1.00, while Stable Signature achieves an ACC of 0.99 on MS. In contrast, the best-performing passive detector achieves an average ACC of only 0.965 across the four datasets, even when trained and evaluated on data from the same distribution. This highlights both the superior detection accuracy and the image distribution insensitivity of watermark-based detectors.

Second, among the passive detectors, DE-FAKE demonstrates the highest performance, achieving an average ACC of 0.965, while the other two passive detectors achieve average ACCs of 0.958 and 0.815, respectively. DE-FAKE’s superior performance may result from its use of both images and their corresponding captions for detection, as the additional information likely enhances detection capabilities.

Third, the performance of passive detectors varies significantly across datasets. For instance, FreqDetect achieves an ACC of 0.93 on FD but only 0.76 on TH, suggesting that passive detectors may be sensitive to image distribution.

\myparatight{Out-of-distribution performance} 
In real-world scenarios, the distribution of testing data may differ from the training data. For instance, a text-to-image model may be updated, or a detector might be required to classify non-AI-generated images from a different distribution or AI-generated images from different models. Thus, we further evaluate the out-of-distribution performance of passive detectors by training on one dataset and testing on others. Since Table~\ref{tab:no_pert} already presents the out-of-distribution performance of watermark-based detectors, we focus our analysis here on passive detectors. Specifically, we train each passive detector using the training set of one dataset and evaluate it on the testing set of each of the four datasets. Figure~\ref{fig:crossdataset} shows the ACC of the three passive detectors under these cross-dataset conditions. Each grid cell shows the ACC obtained when a detector is trained on one dataset and evaluated on another.

We observe that the diagonal cells (from the bottom left to the top right), which represent detectors trained and evaluated on the same dataset, consistently outperform the off-diagonal cells. For example, UnivCLIP achieves an ACC of 0.99 on FD when training and testing on the same dataset, but only achieves 0.58 when trained on FD and tested on GD. Similarly, DE-FAKE achieves an ACC of 0.94 on MS when evaluated in-distribution, but this drops to 0.67 when tested on GD. Overall, the average out-of-distribution ACC across the off-diagonal cells is 29\% lower than the average in-distribution ACC, indicating a substantial performance drop when tested on unseen distributions. This finding suggests that passive detectors struggle with out-of-distribution generalization. A likely reason is that passive detectors primarily rely on memorizing content-specific features, making them less adaptable to new, unseen image characteristics present in different datasets.

\myparatight{Visual quality of watermarked images} To assess the visual quality of watermarked images, we calculate the LPIPS between each AI-generated image and its watermarked counterpart. The results are presented in Figure~\ref{fig:encoded_lpips} in Appendix. Our results show that watermarks preserve the visual quality of AI-generated images.  

\begin{figure}[!t]
    \centering
    \includegraphics[width=0.8\linewidth]{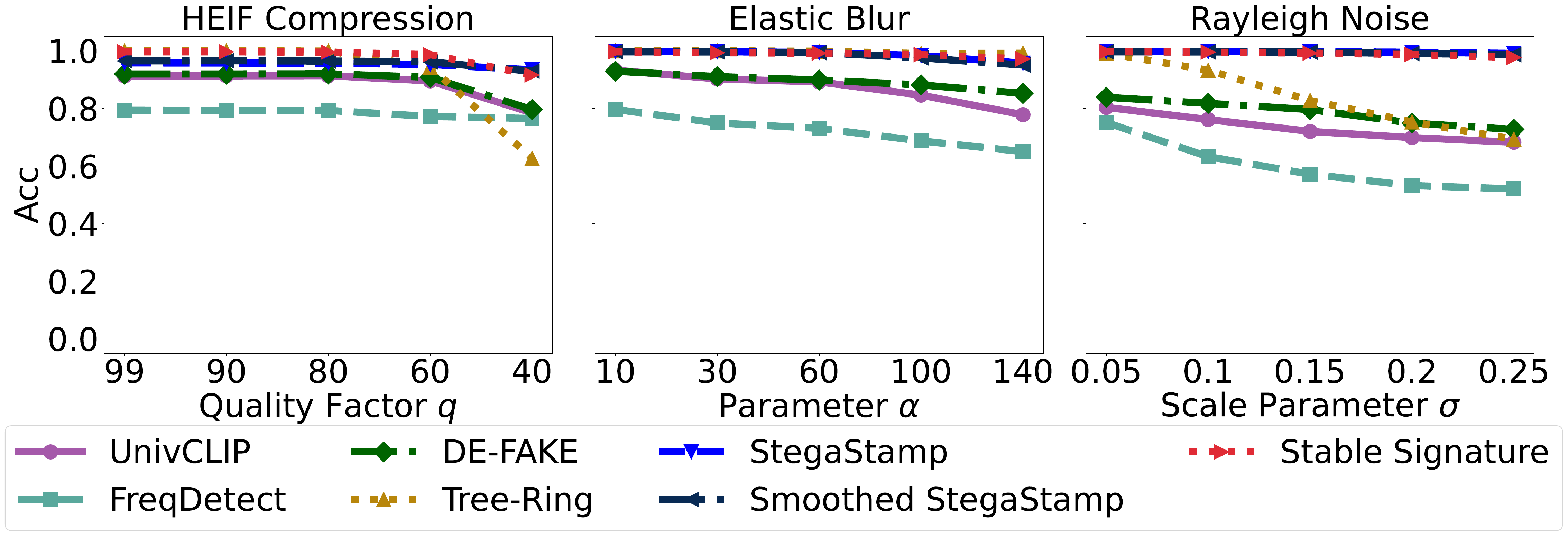}
    
    \caption{ACCs of various detectors under 3 types of common perturbations unseen during training. The FNR and FPR results are shown in Figure~\ref{fig:unseen_common_pert_fnrfpr} in Appendix.}
    \label{fig:unseen_common_pert}
    \vspace{-4mm}
\end{figure}

\subsection{Common Perturbations}

To evaluate a detector's robustness on common perturbation, we calculate FNR, FPR, and ACC for each type of perturbation. Each perturbation type includes a parameter to control its intensity; for example, JPEG compression uses a quality factor $q$. Given a specific perturbation type and parameter value, we apply the perturbation to the 1,000 testing AI-generated  images in a dataset as a removal attack and compute the detector's FNR on these perturbed images. We also apply the same perturbation to the 1,000 testing non-AI-generated  images as a forgery attack and compute the detector's FPR. Subsequently, we calculate  ACC based on the combined 2,000 perturbed images. Additionally, for each parameter value, we measure the LPIPS between an image and its perturbed version, averaging this score across all 2,000 images. This process is repeated across different parameter values for each perturbation type to assess the detector's robustness, provided that the perturbations do not significantly degrade image quality (i.e., LPIPS remains below a certain threshold).

\myparatight{Performance under seen common perturbations} 
Figure~\ref{fig:common_pert} shows ACCs of various detectors on the MS dataset under five types of common perturbations with varying parameter values; Figure~\ref{fig:common_pert_fnrfpr} in Appendix shows the FNRs and FPRs; and Figure~\ref{fig:common_perturbations} in Appendix shows the LPIPS for each perturbation type.   All detectors except  Tree-Ring have undergone  training with perturbations on these five types of perturbations, and thus they are referred to as \emph{seen common perturbations}. Training with perturbations is not applicable to Tree-Ring because it does not require training.

First, we observe that StegaStamp and Smoothed StegaStamp consistently outperform passive detectors across all tested perturbations. For example, under JPEG compression with a quality factor of \( q = 40 \), Smoothed StegaStamp achieves an ACC of 1.00, whereas the top-performing passive detector, DE-FAKE, only reaches 0.88. Other watermark-based detectors also demonstrate superior performance over passive detectors in most cases. Specifically, common perturbations fail to conduct successful forgery attacks against any watermark-based detectors (i.e., FPRs remain close to 0). While such common perturbations also fail to conduct successful removal attacks against StegaStamp and Smoothed StegaStamp (i.e., FNRs remain close to 0), some of them (e.g., JPEG compression with a quality factor $q=20$) can perform successful removal attacks on Tree-Ring and Stable Signature. In contrast, passive detectors are vulnerable to both removal and forgery attacks under common perturbations. This superior robustness of watermark-based detectors is likely due to the stability provided by the embedded watermarks.

Second, we observe that Smoothed StegaStamp stands out as the most robust watermark-based detector. It consistently achieves slightly lower  FNRs and FPRs, as well as higher  ACCs compared to StegaStamp. Moreover, it significantly outperforms Tree-Ring and Stable Signature in some scenarios, particularly under more aggressive common perturbations. For instance, under JPEG compression with a quality factor \( q = 20 \), Smoothed StegaStamp achieves an ACC of 0.99, while Tree-Ring and Stable Signature reach ACCs of 0.88 and 0.72, respectively. 
This robustness advantage likely stems from its use of the randomized smoothing technique~\cite{cohen2019certified}. Notably, Smoothed StegaStamp offers certified robustness against bounded perturbations. Specifically, its detection result for an image remains provably unaffected by any perturbation whose \(\ell_2\)-norm is within a certain bound.

Third, among passive detectors, we observe that DE-FAKE provides the best performance under seen common perturbations. While passive detectors are generally less robust, DE-FAKE consistently achieves higher ACC than UnivCLIP and FreqDetect. For example, under Gaussian blur with a radius $r = 1.0$, DE-FAKE reaches an ACC of 0.81, compared to UnivCLIP's 0.66.

\begin{wrapfigure}{r}{0.3\textwidth}
    \centering
    \includegraphics[width=0.99\linewidth]{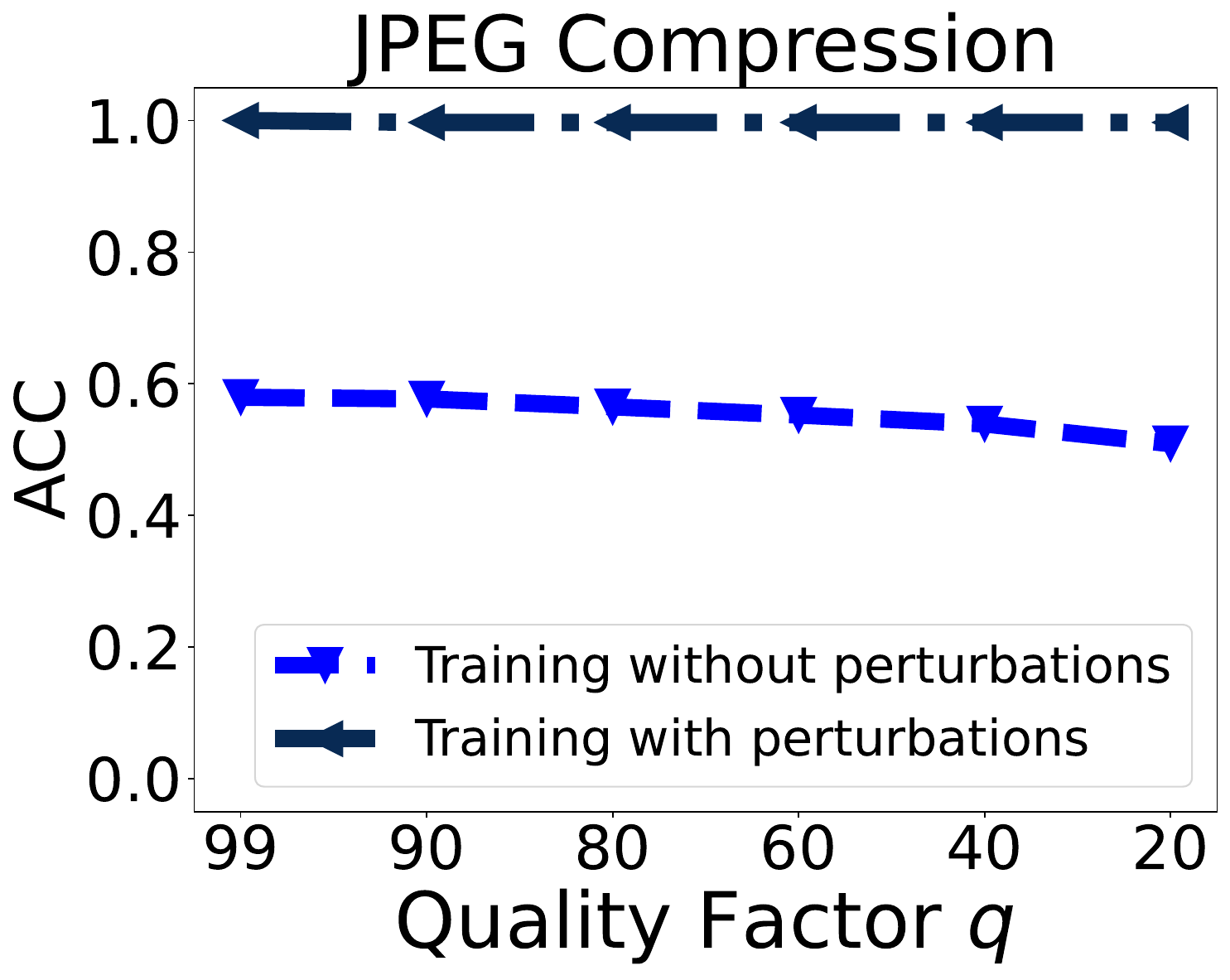}
    \caption{ACC of Smoothed StegaStamp on the MS dataset under JPEG compression with different quality factors when training with vs. without perturbations.}
    \label{fig:training_with_perturbation}
    \vspace{-1mm}
\end{wrapfigure}

\myparatight{Performance under unseen common perturbations}
In real-world scenarios, images may encounter perturbations, referred to as \emph{unseen common perturbations}, that were not included in a detector's training with perturbations. To assess the robustness of different detectors to such unseen common perturbations, we further evaluate their performance under three other types of common perturbations: HEIF, Rayleigh noise, and elastic blur. Figure~\ref{fig:unseen_common_pert} presents ACCs of various detectors on the MS dataset under these unseen common perturbations with varying parameter values, while Figure~\ref{fig:unseen_common_pert_fnrfpr} in Appendix shows the FNRs and FPRs.

Notably, the key observations regarding robustness under seen common perturbations also hold for unseen ones. For example, watermark-based detectors continue to demonstrate strong robustness even with perturbation types not included in their training. Specifically, they maintain ACCs around 0.95 when the LPIPS is at most 0.20, whereas passive detectors achieve ACCs around (with mean of) 0.85. This result indicates that watermark-based detectors can reliably identify AI-generated images even under unseen common perturbations, underscoring their robustness.

\myparatight{Training with vs. without perturbations} 
By default, we train the detectors (except Tree-Ring) using 5 types of common perturbations. We observe that detectors trained without perturbations show reduced robustness against perturbations. For example, as shown in Figure~\ref{fig:training_with_perturbation},   ACC of  Smoothed StegaStamp on the MS dataset declines under JPEG compression across various quality factors when training is conducted without perturbations. These results highlight that training with perturbations enhances robustness, as ACCs are notably higher for detectors trained with them.

\subsection{Adversarial Perturbations}

Due to the inefficiency of Tree-Ring (details in Section~\ref{sec:efficiency}), we exclude it in our evaluation on black-box attacks since they require a large number of queries to a detector for each image. Moreover,  the inverse DDIM process in Tree-Ring causes gradients to accumulate, demanding excessive GPU memory to complete white-box attacks. Thus, we also exclude Tree-Ring from the evaluation on white-box attacks due to our limited computation resources.

\begin{figure}[!t]
    \centering
    \subfloat[Removal attack]{\includegraphics[width=0.35\linewidth]{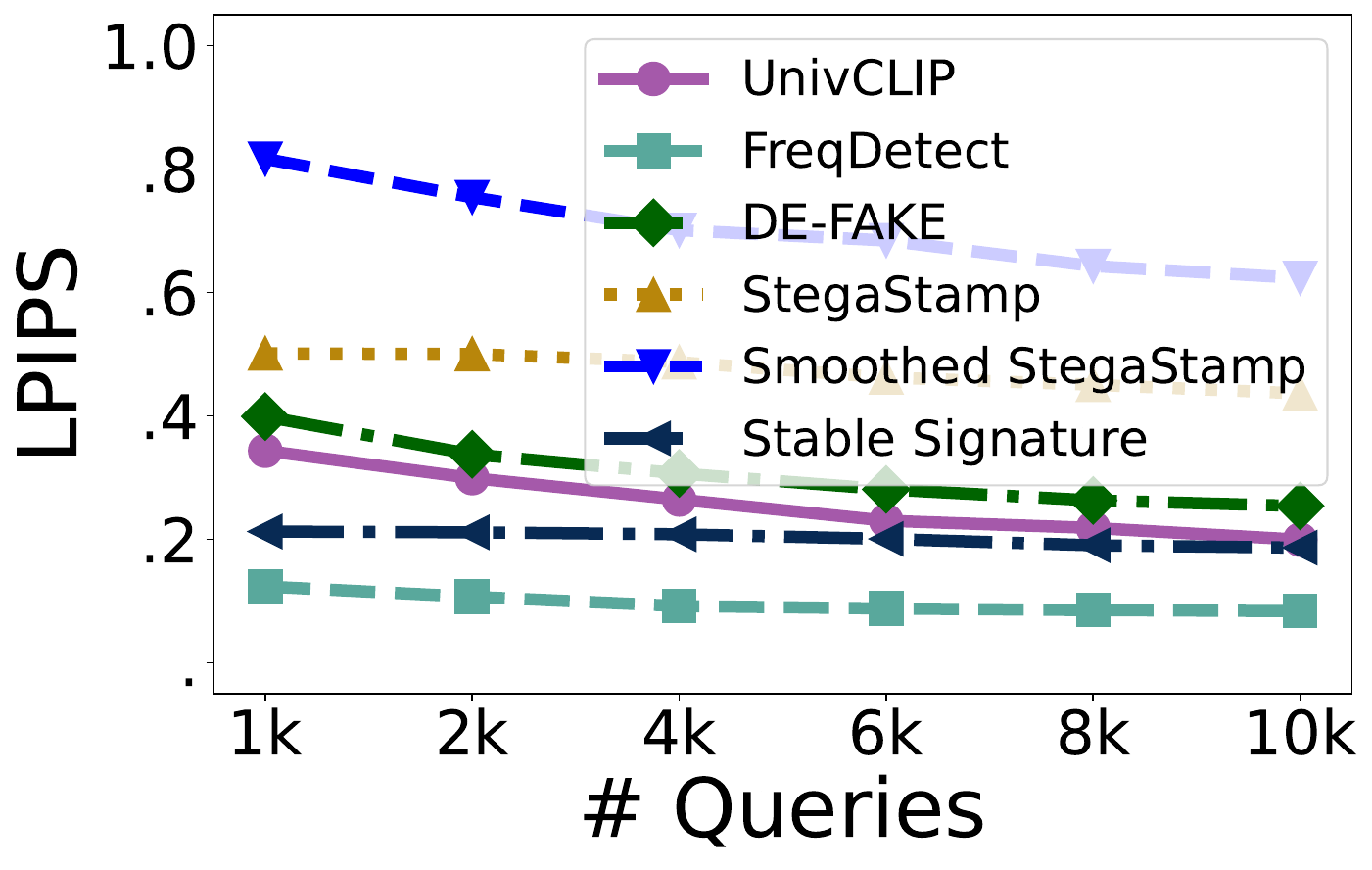}}
    \subfloat[Forgery attack]{\includegraphics[width=0.35\linewidth]{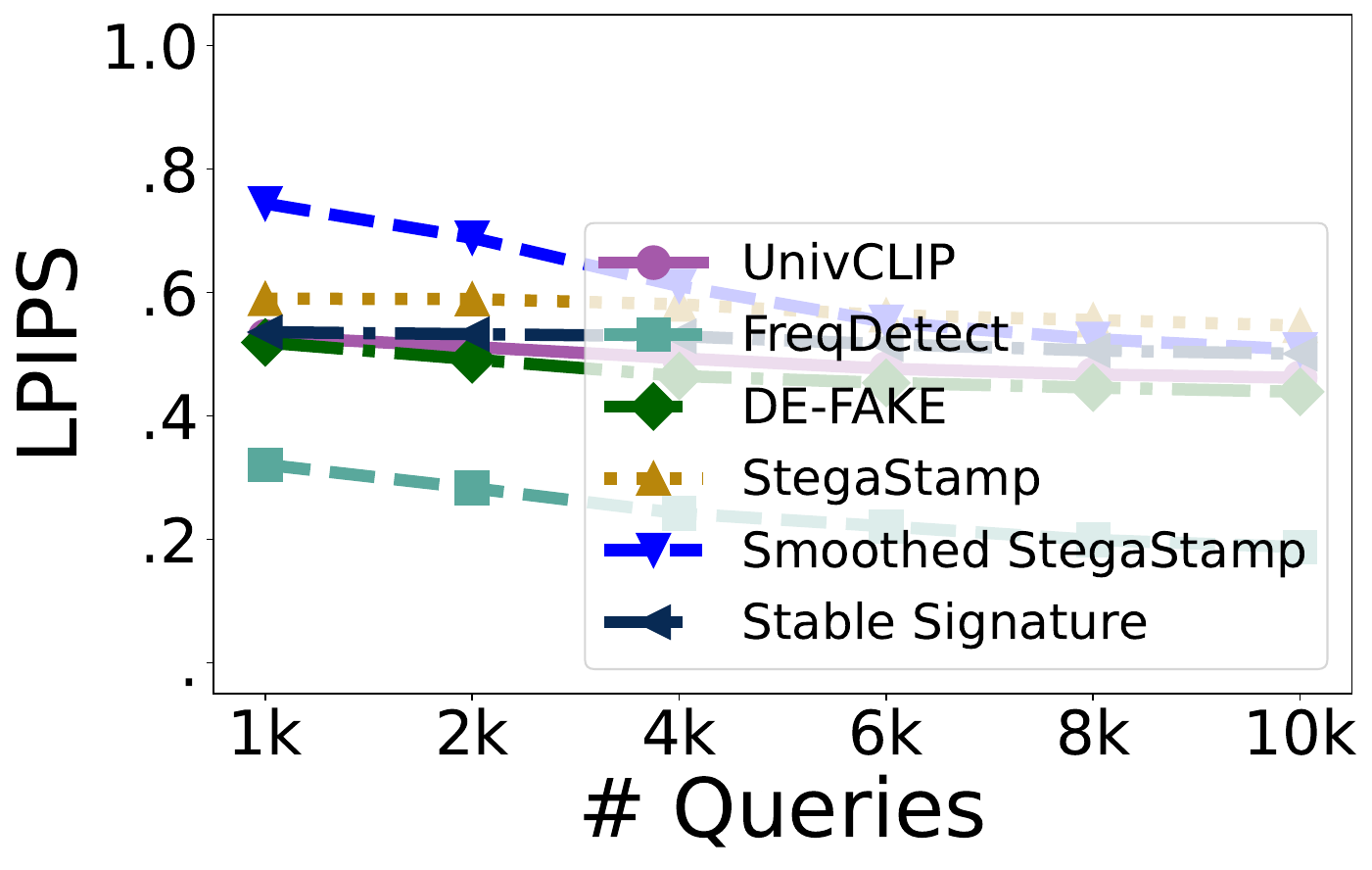}}
    \caption{LPIPS for various detectors under removal and forgery attacks involving the black-box HopSkipJump attack with varying query budget.}
    \label{fig:hsj}
\end{figure}

\myparatight{Black-box} We evaluate removal and forgery attacks using HopSkipJump and Square attacks. For removal attacks using HopSkipJump on an AI-generated image, we initialize the attack with a JPEG-compressed version of the image that is classified as non-AI-generated. For forgery attacks on a non-AI-generated image, we initialize the attack with a watermarked, AI-generated image, as common perturbations like JPEG compression cannot forge a watermark. HopSkipJump ensures deceiving the detector during the attack process, maintaining FNRs and FPRs at 1. Therefore, we use the  LPIPS between  images and their final perturbed versions to measure the robustness of the detectors. Figure~\ref{fig:hsj} presents the LPIPS values of various detectors under HopSkipJump attacks across different query budgets.

\begin{figure}[!t]
    \centering
    \subfloat[Removal attack]{\includegraphics[width=0.35\linewidth]{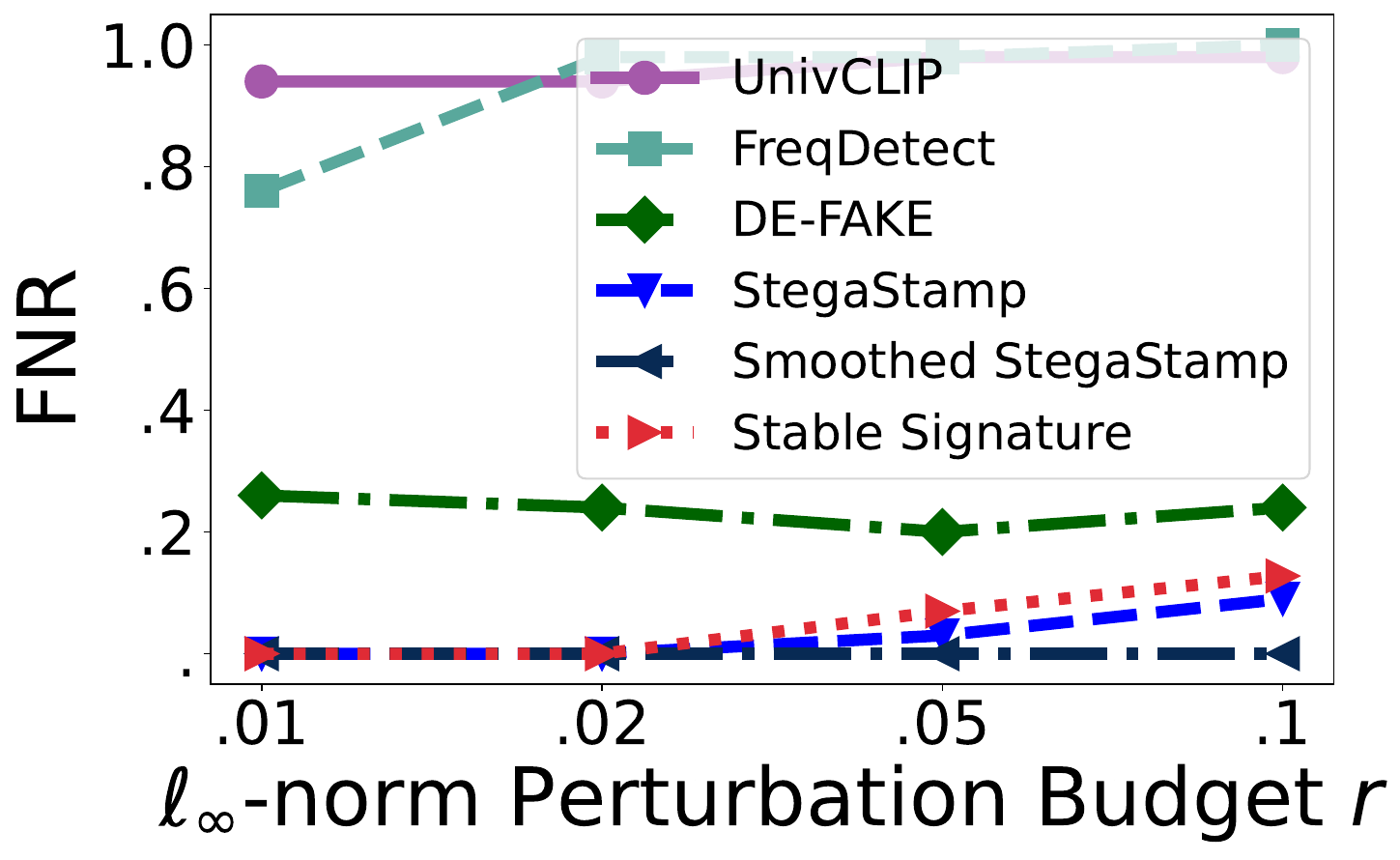}}
    \subfloat[Forgery attack]{\includegraphics[width=0.35\linewidth]{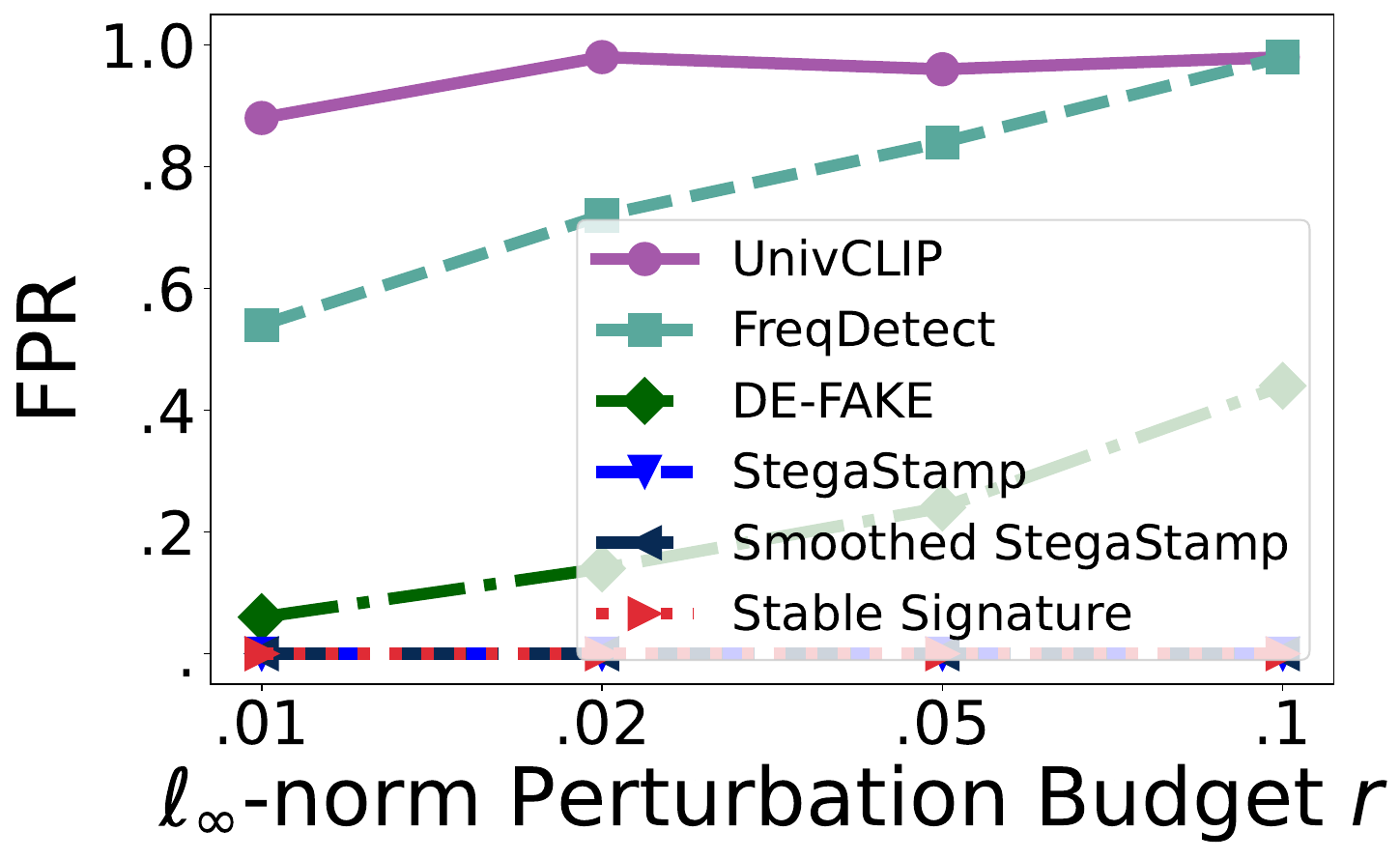}}
    \caption{FNRs and FPRs of various detectors under removal and forgery attacks involving black-box Square attack with varying $\ell_\infty$-norm perturbation budgets $r$.}
    \label{fig:square}
     \vspace{-4mm}
\end{figure}

Given an \(\ell_{\infty}\)-norm perturbation budget \(r\), the Square attack iteratively perturbs an image to deceive a detector. Unlike HopSkipJump, the Square attack does not guarantee deceiving a detector. Thus, to evaluate the robustness of a detector, we report its  FNR and  FPR under removal and forgery attacks, respectively, for varying \(\ell_{\infty}\)-norm perturbation budgets. Figure~\ref{fig:square} shows the FNRs and FPRs of various detectors under Square attacks with different perturbation budgets. For each perturbation budget and image, the Square attack is allowed to query the detector up to 10,000 times in our experiments.

We observe that watermark-based detectors, particularly Smoothed StegaStamp, demonstrate strong robustness against black-box adversarial perturbations. Notably, forgery attacks consistently fail against watermark-based detectors. For instance, forgery attacks using HopSkipJump significantly degrade the visual quality of non-AI-generated images, resulting in large LPIPS values (Figure~\ref{fig:hsj_illus} in Appendix shows several examples).  Similarly, forgery attacks using the Square attack achieve FPRs close to 0 against watermark-based detectors. In addition, Smoothed StegaStamp shows high resilience to removal attacks. Specifically, these attacks result in large LPIPS values (for HopSkipJump) or low FNRs (for Square attack). While other watermark-based detectors may eventually generate perturbed images with acceptable visual quality under high query budgets in HopSkipJump attacks, or show slight increases in FNRs under Square attacks, they generally maintain robust detection performance against these adversarial perturbations.

In contrast, passive detectors show limited robustness against black-box adversarial perturbations. Notably, DE-FAKE, the most robust passive detector, is effectively deceived by the HopSkipJump attack, resulting in low LPIPS values. Similarly, under the Square attack, DE-FAKE reaches an FNR greater than 0.2, and its FPR increases rapidly as the perturbation budget grows.

\begin{figure}[!t]
    \centering
    \subfloat[Removal attack]{\includegraphics[width=0.35\linewidth]{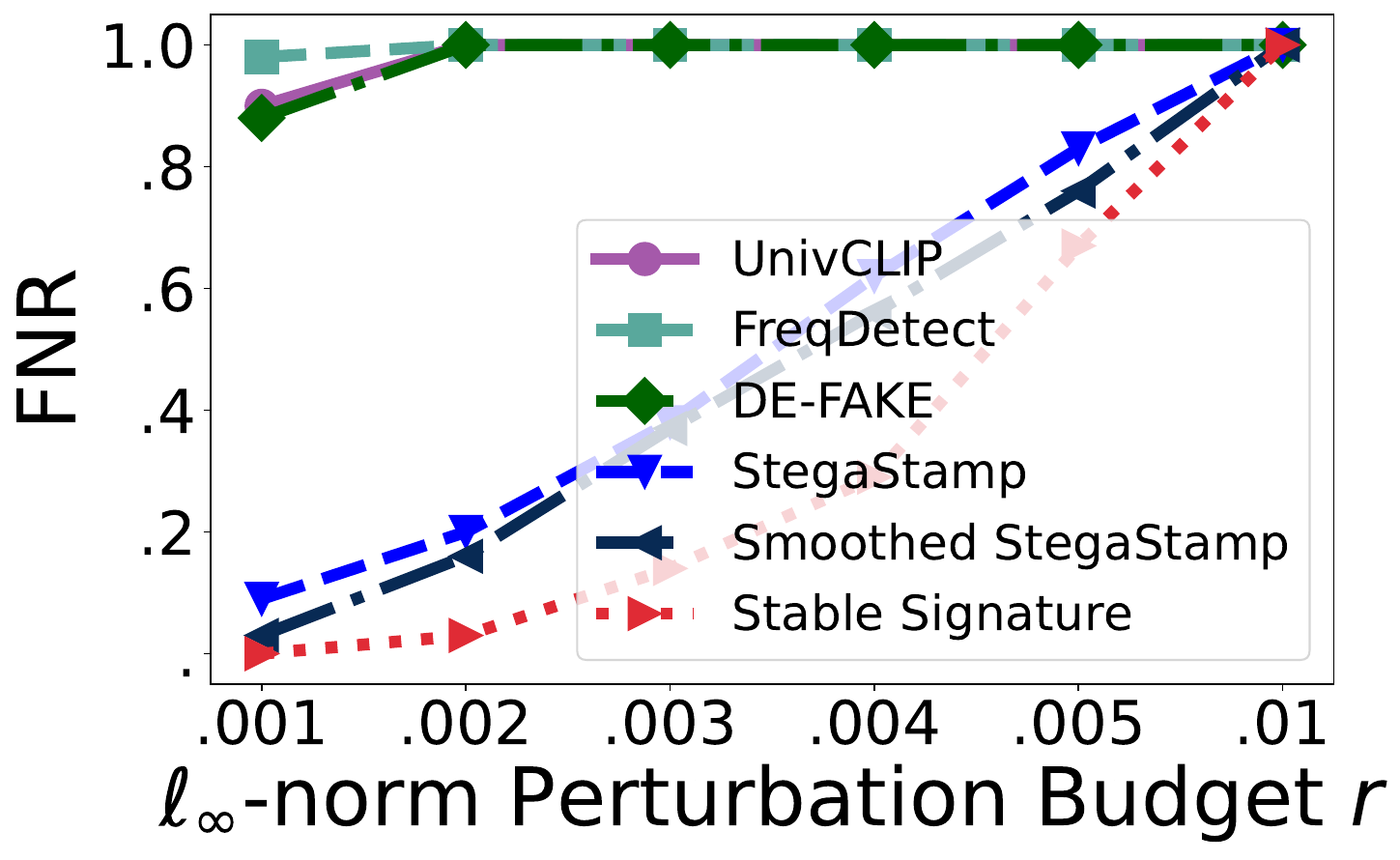}}
    \subfloat[Forgery attack]{\includegraphics[width=0.35\linewidth]{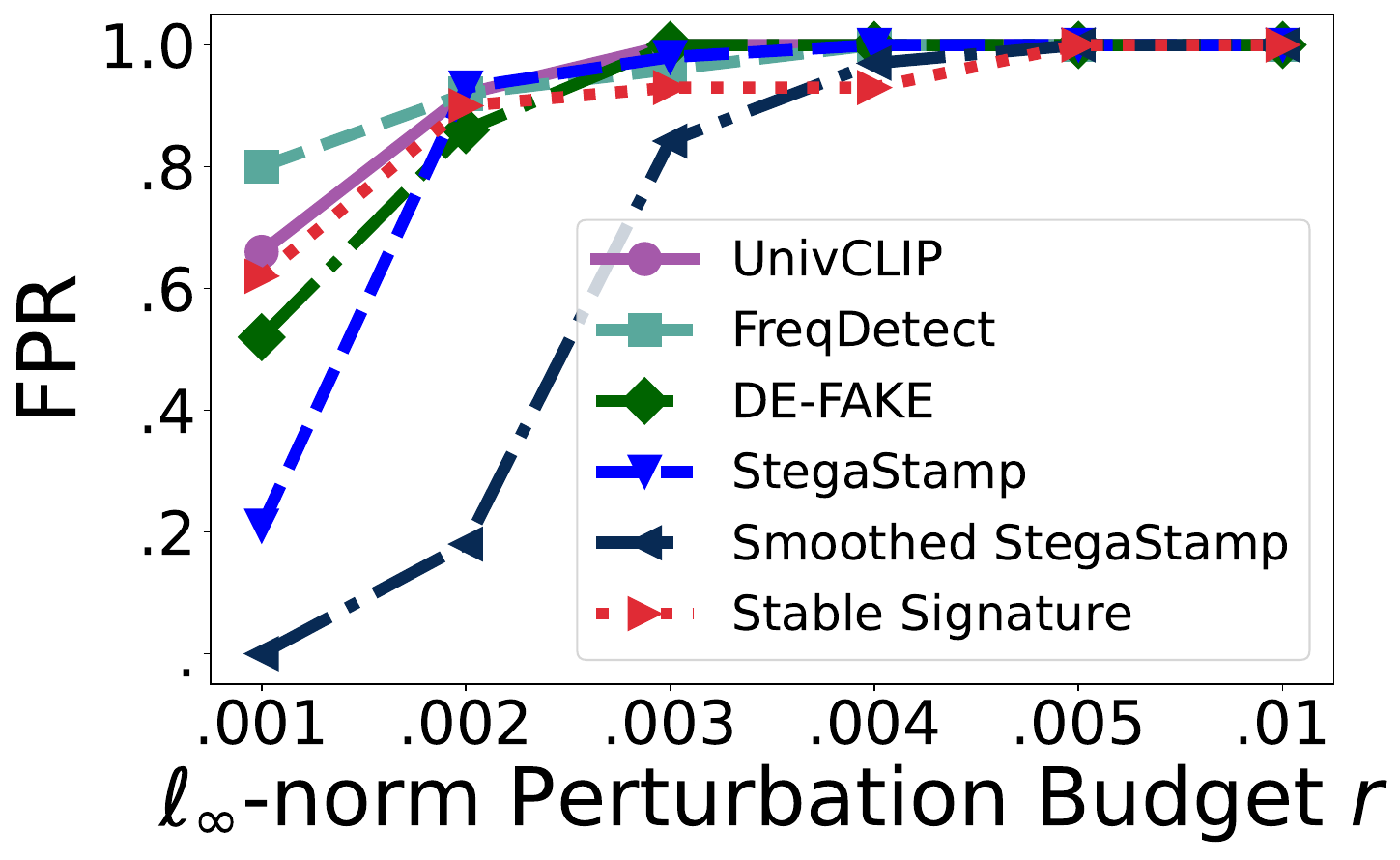}}
    \caption{FNRs and FPRs of various detectors under removal and forgery attacks involving white-box adversarial perturbations with varying $\ell_\infty$-norm perturbation budgets $r$.}
    \label{fig:pgd_linf}
\end{figure}

\myparatight{White-box}
To evaluate the worst-case robustness of passive and watermark-based detectors, we apply white-box adversarial perturbations  to execute both removal and forgery attacks. Specifically, given a detector,  we find a white-box adversarial perturbation for each testing AI-generated (or non-AI-generated) image using PGD as a removal (or forgery) attack. For Smoothed StegaStamp, because the standard white-box attack in Section~\ref{sec:adversarial_perturbation} is suboptimal, we adopt the white-box attack~\cite{jiang2024certifiably} tailored to Smoothed StegaStamp. Then, we compute the detector's FNR (or FPR) of the perturbed AI-generated (or non-AI-generated) images.  Figure~\ref{fig:pgd_linf} presents the FNRs and FPRs of various detectors on the MS dataset under white-box adversarial perturbations at different $\ell_\infty$-norm perturbation budgets $r$. 

We observe that none of the detectors are robust against white-box adversarial perturbations. Specifically, for removal attacks, the FNRs of all detectors reach 1.0 when $r$ is only 0.01. Similarly, for forgery attacks, the FPRs of all detectors reach 1.0 when $r$ is only 0.005. However, watermark-based detectors demonstrate better robustness compared to passive detectors. Specifically, it requires large perturbations to make FNRs and FPRs of watermark-based detectors reach 1, compared to passive detectors. For instance, DE-FAKE reaches an FNR and FPR of 1.0 at $r = 0.003$ for both removal and forgery attacks. In contrast, Smoothed StegaStamp reaches an FNR and FPR of 1.0 at $r = 0.01$ and $r = 0.005$ for removal and forgery attacks, respectively. StegaStamp and Smoothed StegaStamp are overall the most robust detectors, which achieve comparable FNRs and FPRs across perturbation budgets $r$. Although Stable Signature achieves comparable FNRs with   StegaStamp and Smoothed StegaStamp, its FPRs are much larger than those of  StegaStamp and Smoothed StegaStamp when $r$ is small.

\subsection{Detection Efficiency}
\label{sec:efficiency}
Table~\ref{tab:efficiency} shows the time for each detector to classify one image averaged over 100 images. The detection time is defined as the time taken from receiving an input image until outputting the binary detection result. All measurements are conducted on a single A6000 GPU with 24GB memory and an Intel(R) Xeon(R) Gold 6230 CPU on a local machine. Tree-Ring is 1-4 orders of magnitude slower than other detectors. This is because it inverts the diffusion process during detection. DE-FAKE is the second slowest detector as it uses BLIP to generate a caption for each classification.  FreqDetect is the most efficient detector due to its simplicity. UnivCLIP and StegaStamp have comparable efficiency and are the second most efficient. 

\begin{table}[!t]
\centering 
\caption{Time (seconds) taken to classify one image for passive and watermark-based detectors.}

\renewcommand{\arraystretch}{1}
\resizebox{8.5cm}{!}{
\begin{NiceTabular}{|l|c|c|}[hvlines]
\multicolumn{2}{c|}{\textbf{Detector}} &  \textbf{Time} \\  \cline{1-3}

\Block{3-1}{Passive} & UnivCLIP & $6.48 \cdot 10^{-3}$  \\ \cline{2-3} 
& FreqDetect & $6.71\cdot 10^{-4}$  \\ \cline{2-3}
& DE-FAKE & $3.24\cdot 10^{-1}$ \\ \hline \hline

\Block{4-1}{Watermark-based} & Tree-Ring & $2.17\cdot 10^0$ \\ \cline{2-3}
& StegaStamp & $6.69 \cdot 10^{-3}$  \\ \cline{2-3}
& Smoothed StegaStamp & $6.79\cdot 10^{-3}$ \\ \cline{2-3}
& Stable Signature & $1.39\cdot 10^{-2}$ \\ \hline

\end{NiceTabular}
}
\label{tab:efficiency}
\end{table}

\begin{table}[!t]
\centering
\caption{FNRs, FPRs, and ACCs of commercial passive detectors and watermark-based detectors  on 100 AI-generated images and 100 non-AI-generated images under no perturbations and Gaussian noise perturbations.}
\renewcommand{\arraystretch}{1.2}
\resizebox{8.5cm}{!}{
\begin{NiceTabular}{|c|c|c|c|c|c|c|}[hvlines]
\Block{2-1}{\textbf{Detector}} & \multicolumn{3}{c|}{\textbf{No perturbation}} & \multicolumn{3}{c|}{\textbf{Gaussian noise}} \\ \cline{2-7}
& \textbf{FNR} & \textbf{FPR} & \textbf{ACC} & \textbf{FNR} & \textbf{FPR} & \textbf{ACC} \\ \hline

HIVE & 0.00 & 0.00 & 1.00 & 0.23 & 0.00 & 0.88 \\ \cline{2-7}
ILLUMINARTY & 0.20 & 0.20 & 0.80 & 0.86 & 0.00 & 0.57 \\ \hline \hline
 StegaStamp & 0.00 & 0.00 & 1.00 & 0.00 & 0.00 & 1.00 \\ \cline{2-7}
 Smoothed StegaStamp & 0.00 & 0.00 & 1.00 & 0.00 & 0.00 & 1.00 \\ \hline
\end{NiceTabular}
}
\label{tab:clean_ms}
\end{table}

\subsection{Commercial Passive Detectors} 
\label{sec:commercial}
We evaluate two commercial detectors: HIVE~\cite{hive-ai} and ILLUMINARTY~\cite{illu-ai},  deployed by hive.ai and illuminarty.ai, respectively. Although the specific algorithms of these commercial detectors are not publicly disclosed, they are considered passive because they detect AI-generated images without relying on watermarks embedded during image generation. HIVE is the state-of-the-art commercial detector according to a recent study~\cite{ha2024organic}. Due to a daily limit of 100 queries and the need to manually upload images to the web interfaces,\footnote{They offer APIs for a fee. We reached out to HIVE for API access and pricing information but did not receive a response.} we sample 100 AI-generated images and 100 non-AI-generated images from the testing set of the MS dataset and use the commercial detectors to classify them. Additionally, we add Gaussian noise with a standard deviation of 0.2 to each image and then query the commercial detectors again to classify the perturbed images.

Table~\ref{tab:clean_ms} presents the FNR, FPR, and ACC results for the two commercial passive detectors alongside two watermark-based detectors. In the scenario of no perturbations, HIVE and the two watermark-based detectors perform very well, while ILLUMINARTY does not. Specifically, HIVE achieves an ACC of 1.0. The superior performance of HIVE is consistent with prior findings on detecting AI-generated art images reported by Ha et al.~\cite{ha2024organic}. In contrast, ILLUMINARTY shows a lower ACC of 0.8. This difference may be because HIVE's training dataset likely includes a large number of AI images generated by a diverse set of text-to-image models and non-AI-generated images sourced from the Internet~\cite{hive-nvidia}. Since our dataset also comprises images collected from the Internet or generated using publicly accessible text-to-image models, there is a possibility that HIVE has encountered similar examples during training, thus boosting its performance on these images.

When Gaussian noise is added to each image, the performance gap between the detectors widens considerably. HIVE's ACC decreases to 0.88, while ILLUMINARTY's ACC drops to 0.57, indicating their vulnerability to common perturbations. In contrast, the watermark-based detectors maintain ACCs of 1.0 even under these perturbations, showing their superior robustness.

\subsection{Summary and Recommendations}
\vspace{-2mm}
\myparatight{Summary of findings} Our comprehensive benchmark study reveals several key findings. Watermark-based detectors consistently outperform passive detectors across all conditions, including scenarios with no perturbations, common perturbations, and adversarial perturbations. Additionally, watermark-based detectors exhibit a remarkable insensitivity to variations in image distributions, in contrast to passive detectors, which are significantly affected. The state-of-the-art watermark-based detector, Smoothed StegaStamp, demonstrates strong robustness against both common and black-box perturbations. However, neither passive nor watermark-based detectors show robust performance under white-box adversarial perturbations, although watermark-based detectors still show better resilience in these cases. Finally, Smoothed StegaStamp, the best-performing watermark-based detector, is two orders of magnitude more efficient than DE-FAKE, the best-performing passive detector.

\myparatight{Recommendations}  Based on our findings, we offer several recommendations. When both passive and watermark-based detectors are applicable, watermark-based detectors—particularly Smoothed StegaStamp—should be the preferred choice due to their superior effectiveness, robustness, and efficiency. Furthermore,  training with perturbations should be adopted to enhance the detectors' robustness. 

We acknowledge that watermark-based detectors are not applicable for detecting AI images that have already been generated and propagated on the Internet before watermarks are deployed. Additionally, malicious actors may  fine-tune public text-to-image models and use them to generate images without embedding  watermarks. In these situations, passive detectors are necessary. Our study yields several recommendations for training effective passive detectors. First, training should include AI-generated images from a diverse set of models and non-AI-generated images from diverse sources to capture subtle differences in texture, color distribution, and structural characteristics typical of AI-generated images. Second, training with perturbations is beneficial, as it exposes the detectors to challenging cases and enhances robustness against perturbations that may encounter in real-world scenarios.

%% file: 6.conclusion.tex
\section{Discussion and Limitations}
\label{sec:discussion}
\vspace{-2mm}
\myparatight{Other modalities (text, audio, and video)} 
Similar studies can be conducted for other modalities, since both passive and watermark-based detectors have been developed for them. For example, in the context of AI-generated text, multiple passive detectors~\cite{mitchell2023detectgpt,liu2023detectability,pu2023deepfake} and watermark-based detectors~\cite{abdelnabi2021adversarial,kirchenbauer2023watermark,dathathri2024scalable} have been proposed.  
For AI-generated audio, various passive~\cite{li2024safeear,blue2022you} and watermark-based detectors~\cite{audioseal,timbre,audiomarkbench} have been developed as well. 
We believe our key take-away message that watermark-based detectors outperform passive detectors still hold, though it is a valuable future work to conduct a systematic benchmark study to confirm. 

\myparatight{Adversarial training} 
In our experiments, we used training with perturbations rather than adversarial training~\cite{madry2017towards}. The key difference lies in the approach to introducing perturbations during training. Adversarial training involves applying adversarial perturbations to each image, where the adversarial perturbations are specifically optimized during each training iteration to maximize the detector’s vulnerability. 
By contrast, training with perturbations involves randomly sampling a common perturbation (e.g., Gaussian noise) and applying it to each image in each training iteration. 
Although adversarial training may further improve robustness against adversarial perturbations, we opted not to use it in our experiments due to its significantly higher computational cost. Adversarial training requires additional optimization steps for each training image in each training iteration, making it less efficient and challenging to scale. Additionally, while adversarial training can improve robustness against white-box adversarial perturbations~\cite{Evading-Watermark}, i.e., a larger adversarial perturbation is needed to deceive a detector, it is still insufficient, i.e., the adversarial perturbation is still small. 

\section{Conclusion and Future Work}
Through a comprehensive benchmark study comparing the effectiveness, robustness, and efficiency of passive and watermark-based detectors, we find that watermark-based detectors consistently outperform passive detectors in all scenarios with respect to effectiveness and robustness. Moreover, the most effective and robust watermark-based detector is two orders of magnitude faster than the most effective and robust passive detector. Based on these findings, we recommend that, when both detector types are applicable, watermark-based detectors should be the preferred choice for their superior effectiveness, robustness, and efficiency. An interesting future work is to extend our \bench to other modalities such as text, audio, and video.

%% file: 7.appendix.tex
\appendix

\begin{figure}[h!]
    \centering
    \subfloat[StegaStamp]{\includegraphics[width=0.35\linewidth]{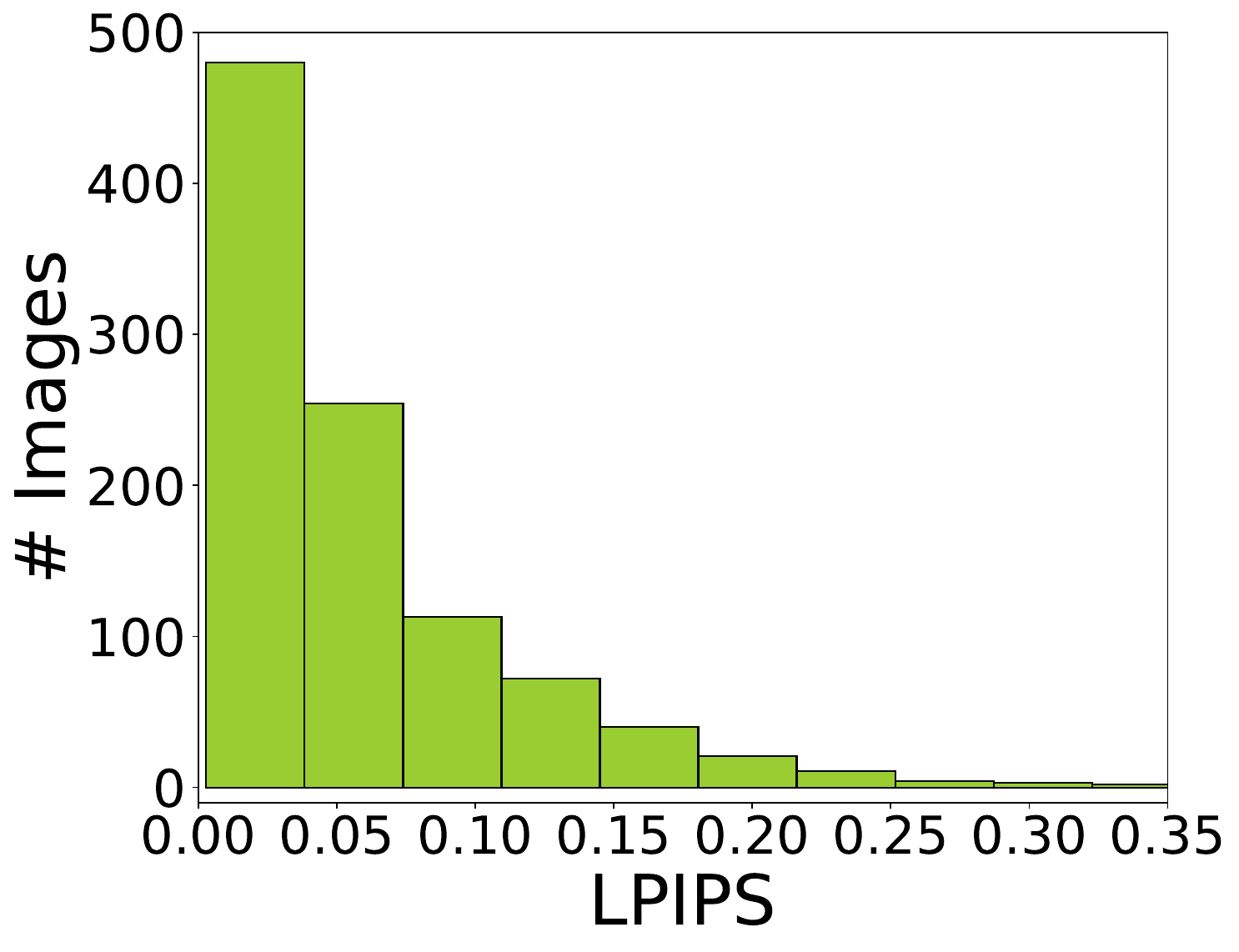}}
    \subfloat[Stable Signature]{\includegraphics[width=0.35\linewidth]{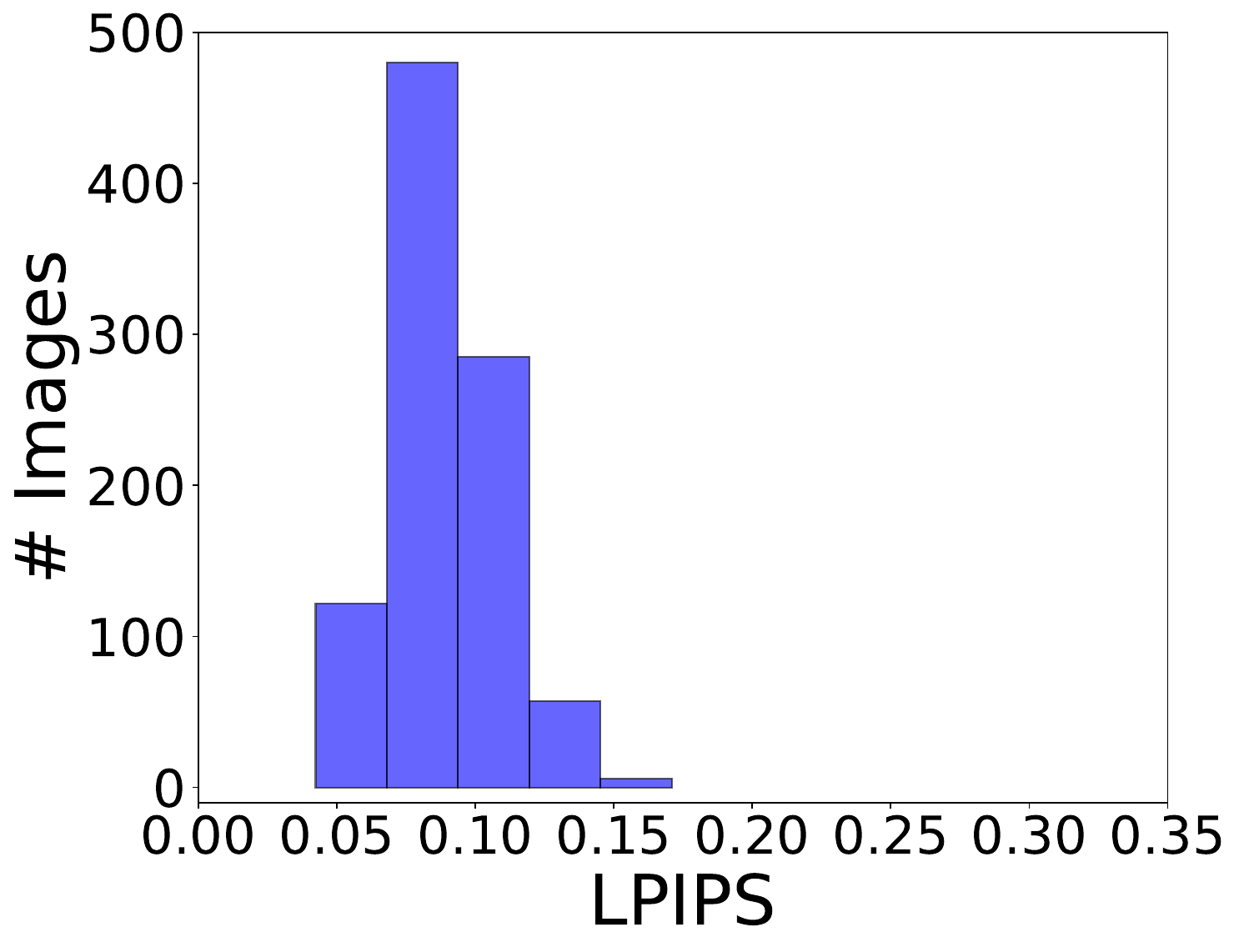}}

    \caption{Distribution of LPIPS between AI-generated images and their watermarked counterparts for (a) StegaStamp and (b) Stable Signature on the MS dataset.  Smoothed StegaStamp and StegaStamp use the same watermark encoder, and thus they have the same  LPIPS distribution. The unwatermarked AI-generated images for Stable Signature are generated using the unmodified Stable Diffusion model with the same seeds. Since Tree-Ring modifies the seeds, there are no unwatermarked versions of watermarked AI-generated images and thus LPIPS is not applicable. The watermarks preserve visual quality well since the LPIPS values are small.} 
    \label{fig:encoded_lpips}
\end{figure}

\vspace{-6mm}

\begin{figure}[h!]
    \centering
    \includegraphics[width=0.8\linewidth]{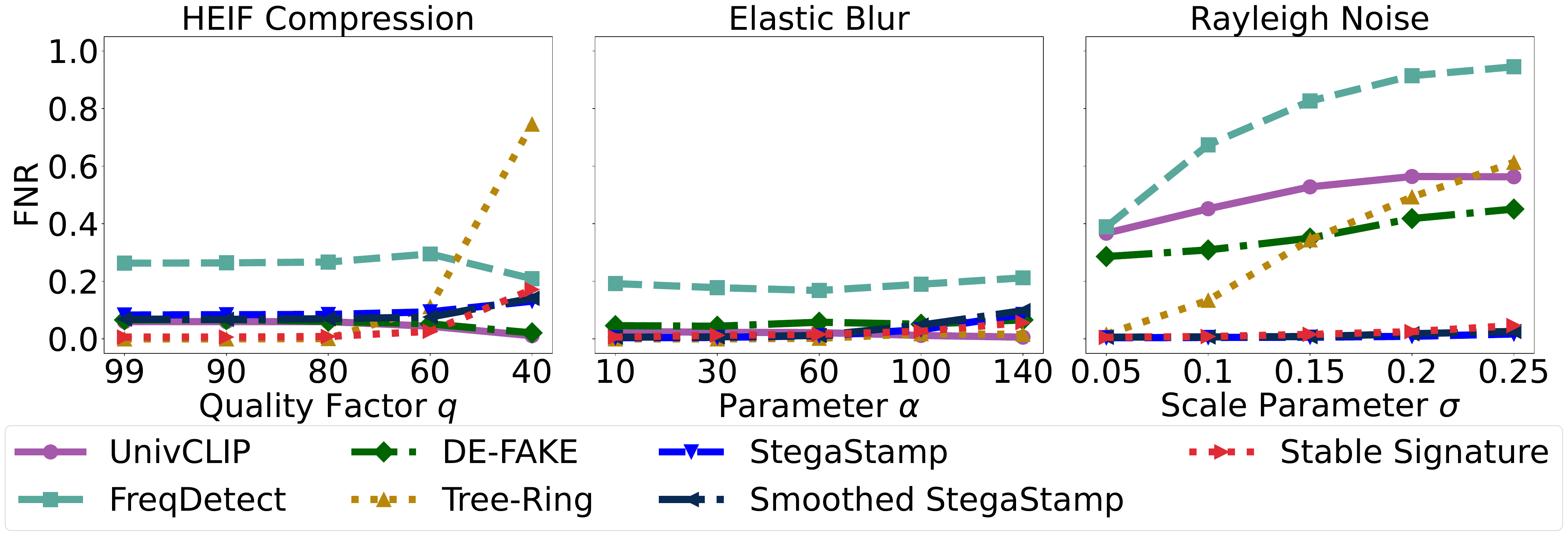}
    \includegraphics[width=0.8\linewidth]{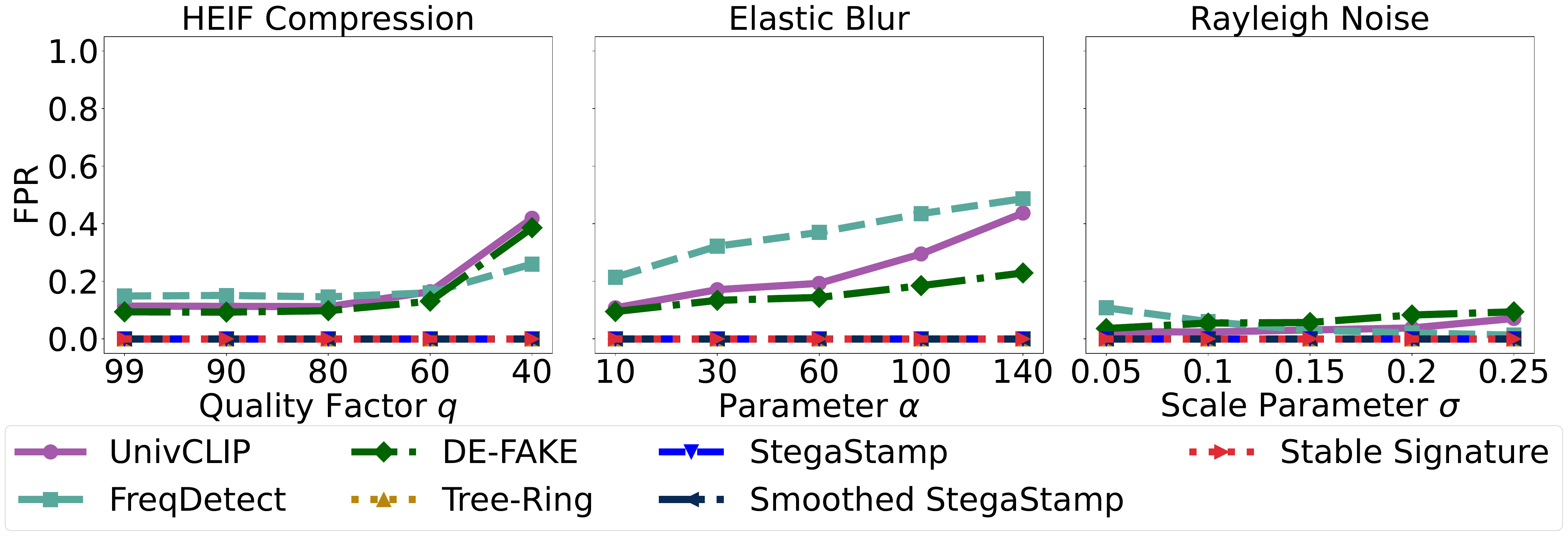}
    \caption{FNRs and FPRs of various detectors under 3 types of common perturbations unseen during training.}
    \label{fig:unseen_common_pert_fnrfpr}
\end{figure}

\begin{figure}[h!]
\centering
\renewcommand*{\arraystretch}{0}
\begin{tabular}{*{3}{@{}c}@{}}
\begin{subfigure}{.3\linewidth}
  \centering
  \includegraphics[width=\linewidth, height=\linewidth]{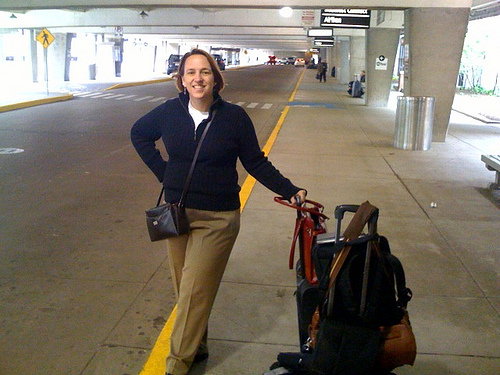}
\end{subfigure} 
\begin{subfigure}{.3\linewidth}
  \centering
  \includegraphics[width=\linewidth, height=\linewidth]{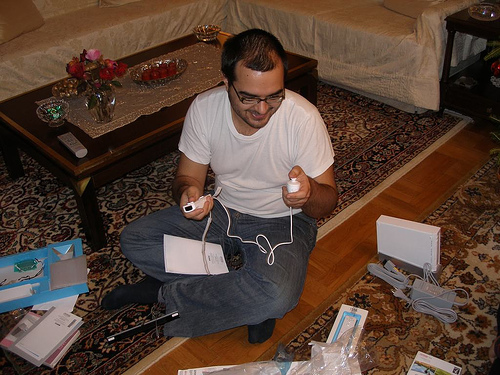}
\end{subfigure}
\begin{subfigure}{.3\linewidth}
  \centering
  \includegraphics[width=\linewidth, height=\linewidth]{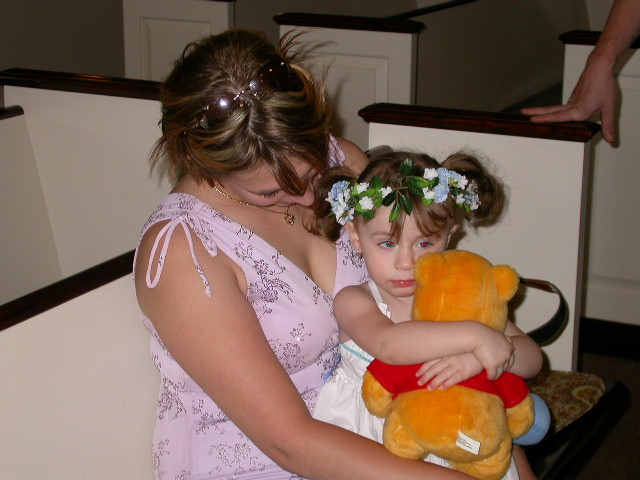}
\end{subfigure} \\

\begin{subfigure}{.3\linewidth}
  \centering
  \includegraphics[width=\linewidth, height=\linewidth]{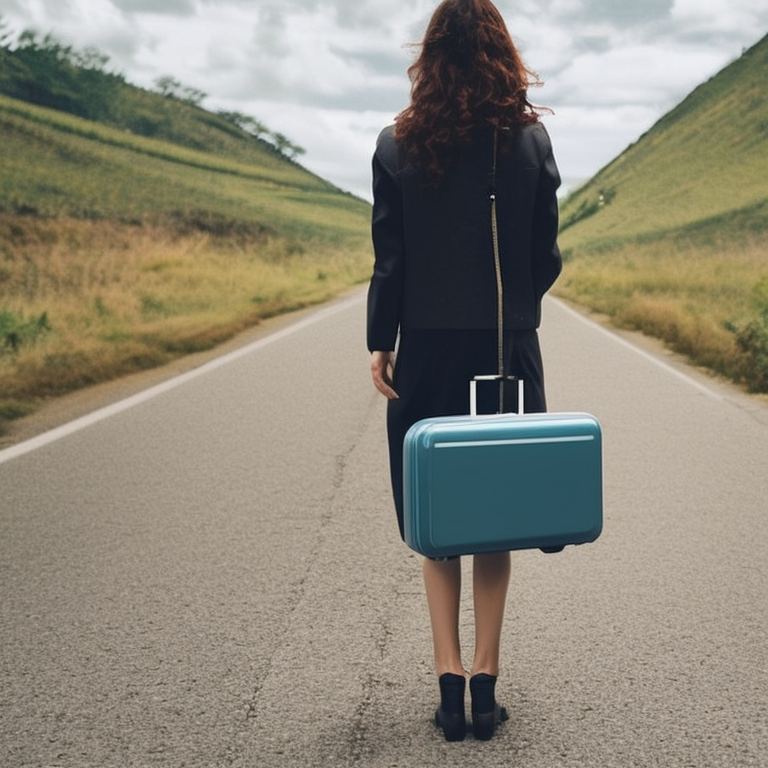}
\end{subfigure}
\begin{subfigure}{.3\linewidth}
  \centering
  \includegraphics[width=\linewidth, height=\linewidth]{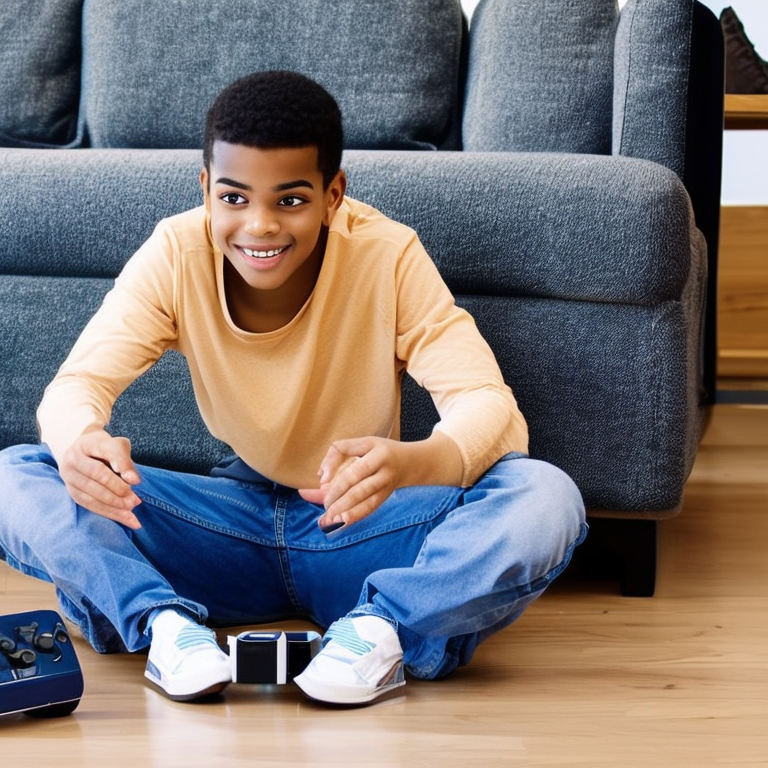}
\end{subfigure}
\begin{subfigure}{.3\linewidth}
  \centering
  \includegraphics[width=\linewidth, height=\linewidth]{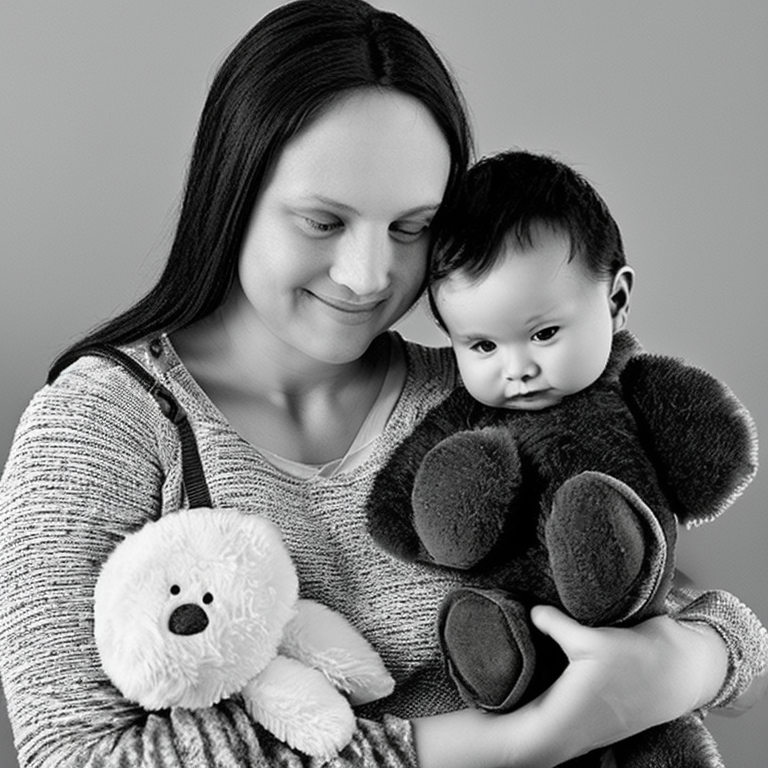}
\end{subfigure}
\end{tabular}
\caption{Examples of non-AI-generated (top row) and AI-generated (bottom row) images in the MS dataset. The captions of the non-AI-generated images are: \textbf{(a)} ``a woman standing by the road with a suitcase.'', \textbf{(b)} ``the young man sits on the floor to look at his new game system.'', and \textbf{(c)} ``a mom and a baby who is holding a teddy bear.''. The AI-generated images are generated by Stable Diffusion using the captions as prompts.}
\label{fig:example_images_MS}
\end{figure}

\begin{figure}[h!]
\centering
\renewcommand*{\arraystretch}{0}
\begin{tabular}{*{3}{@{}c}@{}}
\begin{subfigure}{.3\linewidth}
  \centering
  \includegraphics[width=\linewidth, height=\linewidth]{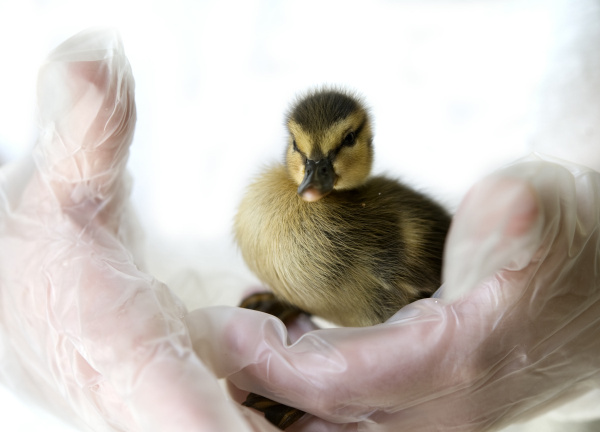}
\end{subfigure}
\begin{subfigure}{.3\linewidth}
  \centering
  \includegraphics[width=\linewidth, height=\linewidth]{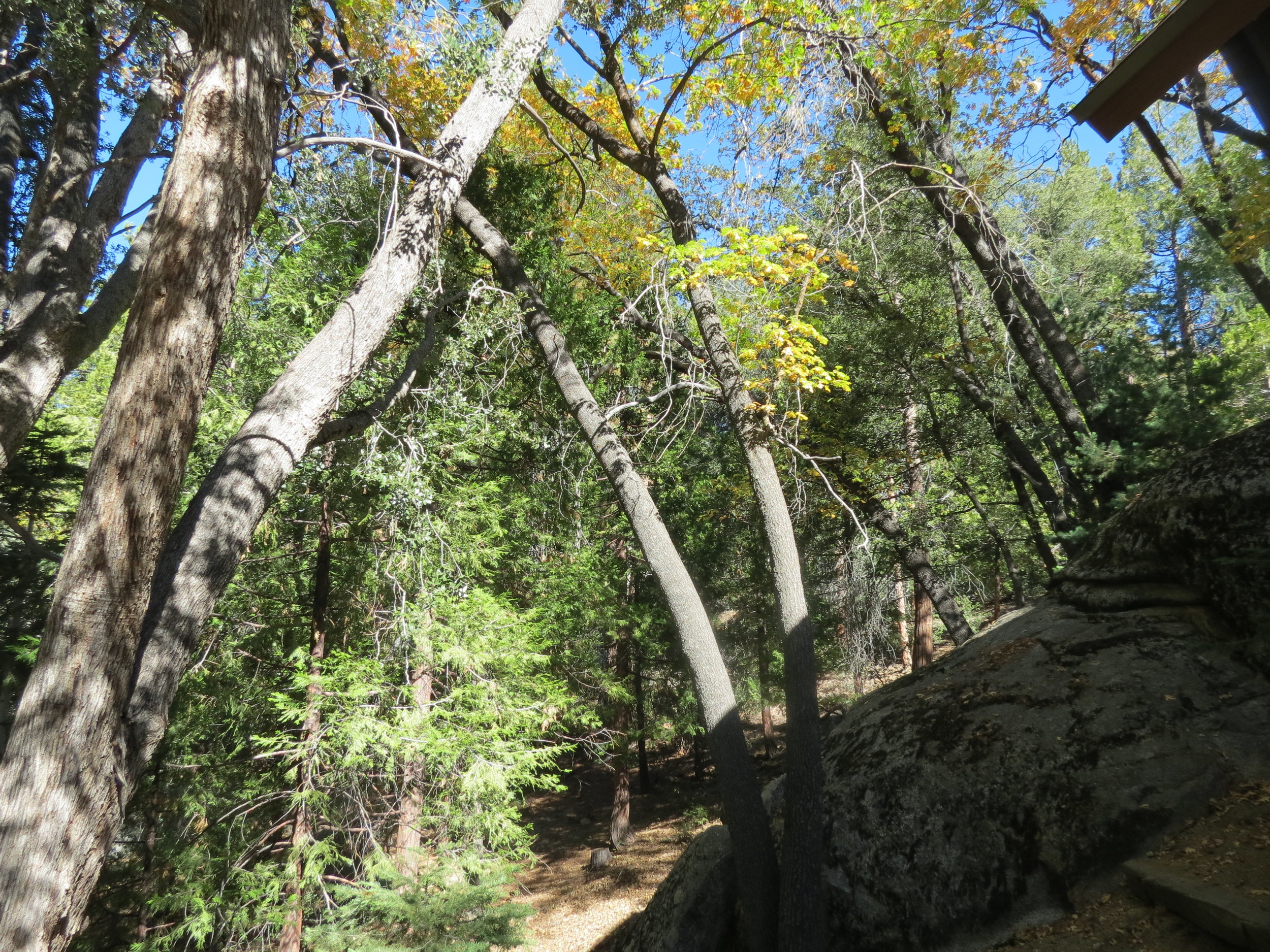}
\end{subfigure}
\begin{subfigure}{.3\linewidth}
  \centering
  \includegraphics[width=\linewidth, height=\linewidth]{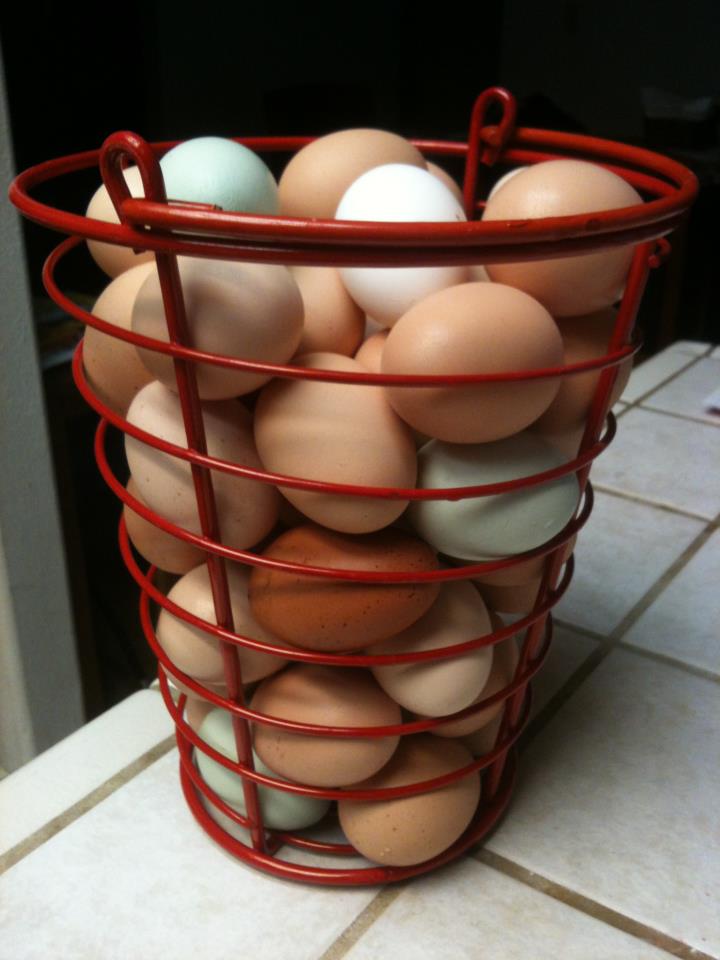}
\end{subfigure}\\

\begin{subfigure}{.3\linewidth}
  \centering
  \includegraphics[width=\linewidth, height=\linewidth]{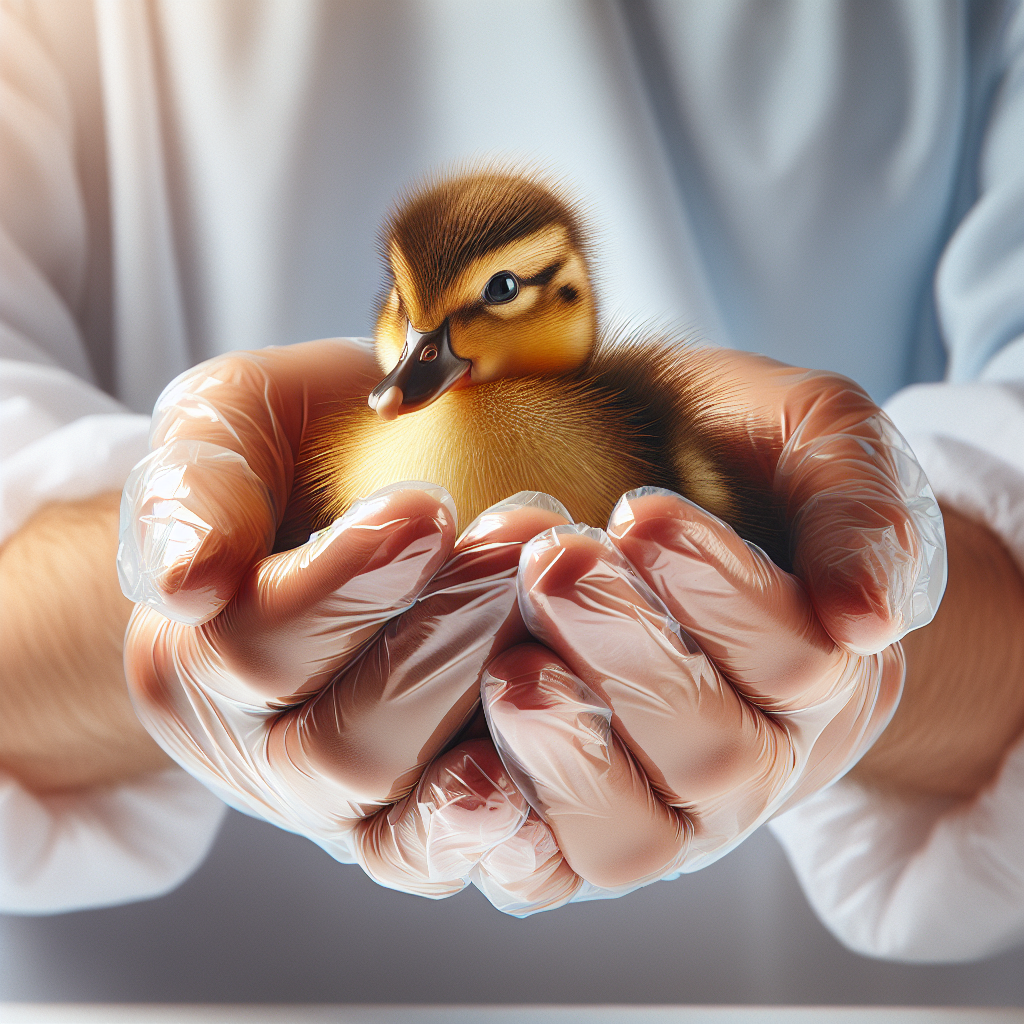}
\end{subfigure}
\begin{subfigure}{.3\linewidth}
  \centering
  \includegraphics[width=\linewidth, height=\linewidth]{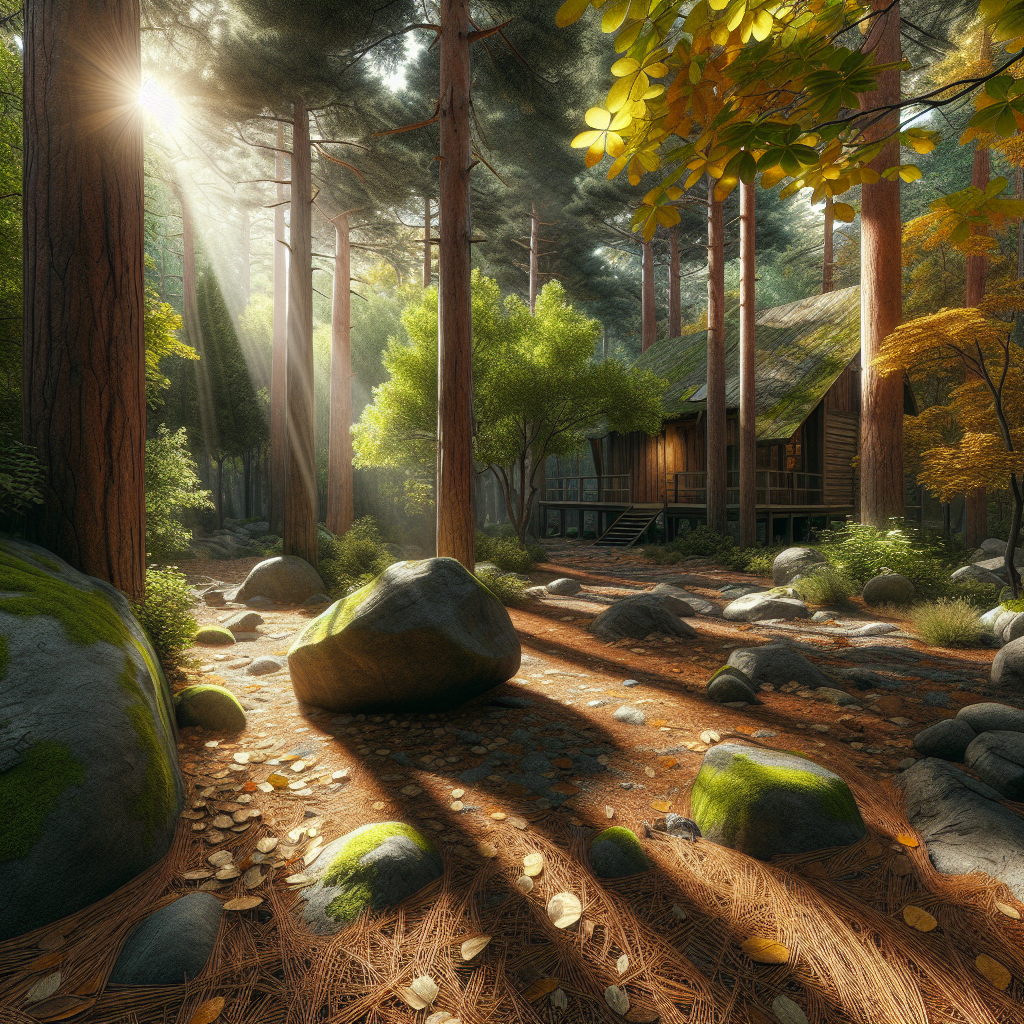}
\end{subfigure}
\begin{subfigure}{.3\linewidth}
  \centering
  \includegraphics[width=\linewidth, height=\linewidth]{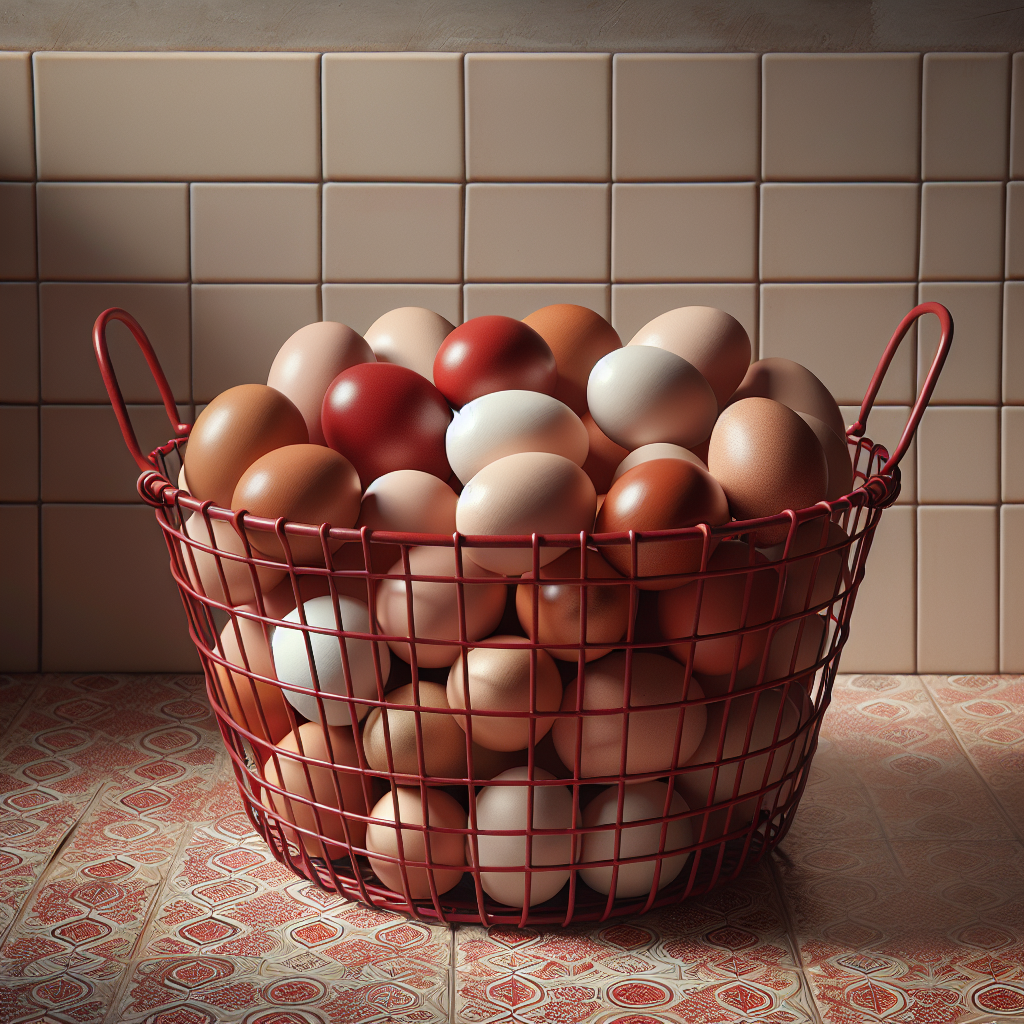}
\end{subfigure}

\end{tabular}

\caption{Examples of non-AI-generated (top row) and AI-generated (bottom row) images in the GD dataset. The {captions} of the non-AI-generated images are \textbf{(a)} ``a baby mallard duck was featured'', \textbf{(b)} ``a backyard of boulders and trees'', and \textbf{(c)} ``a basket of farm fresh eggs''. The AI-generated images are generated by DALL-E 3 using the captions as prompts.}
\label{fig:example_images_GD}
\end{figure}

\begin{figure}[h!]
\centering
\renewcommand*{\arraystretch}{0}
\begin{tabular}{*{3}{@{}c}@{}}
\begin{subfigure}{.3\linewidth}
  \centering
  \includegraphics[width=\linewidth, height=\linewidth]{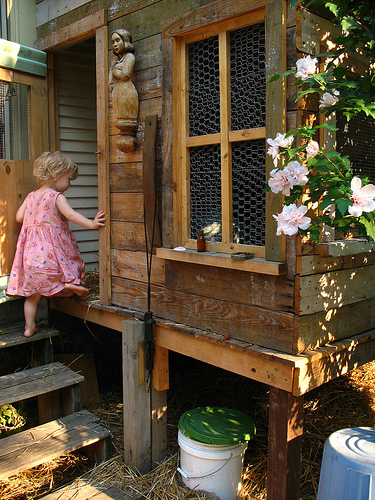}
\end{subfigure} 
\begin{subfigure}{.3\linewidth}
  \centering
  \includegraphics[width=\linewidth, height=\linewidth]{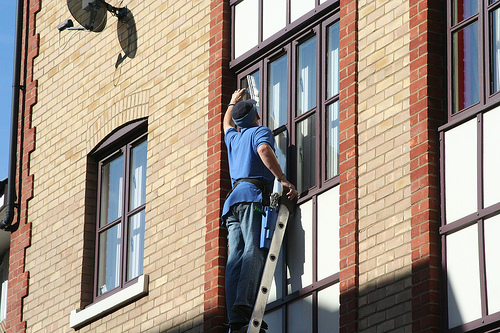}
\end{subfigure}
\begin{subfigure}{.3\linewidth}
  \centering
  \includegraphics[width=\linewidth, height=\linewidth]{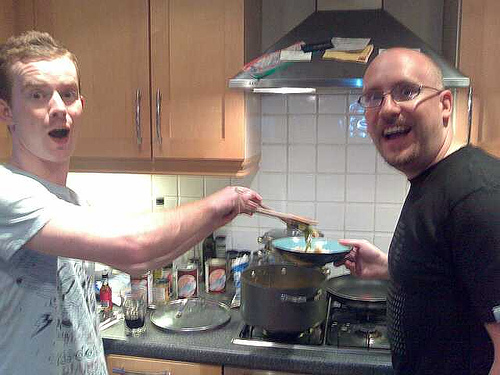}
\end{subfigure} \\

\begin{subfigure}{.3\linewidth}
  \centering
  \includegraphics[width=\linewidth]{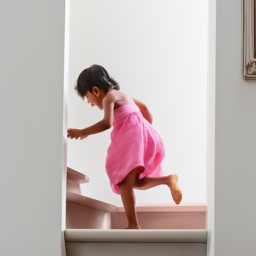}
\end{subfigure}
\begin{subfigure}{.3\linewidth}
  \centering
  \includegraphics[width=\linewidth]{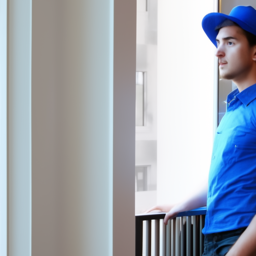}
\end{subfigure}
\begin{subfigure}{.3\linewidth}
  \centering
  \includegraphics[width=\linewidth]{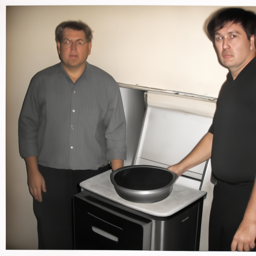}
\end{subfigure}
\end{tabular}

\caption{Examples of non-AI-generated (top row) and AI-generated (bottom row) images in the FD dataset. The {captions} of the non-AI-generated images are: \textbf{(a) }``A child in a pink dress is climbing up a set of stairs in an entry way.'', \textbf{(b)} ``Someone in a blue shirt and hat is standing on stair and leaning against a window.'', and \textbf{(c)} ``Two men, one in a gray shirt, one in a black shirt, standing near a stove.''. The AI-generated images are generated by DeepFloyd IF using the captions as prompts.}
\label{fig:example_images_FD}
\end{figure}

\begin{figure}[h!]
\centering
\renewcommand*{\arraystretch}{0}
\begin{tabular}{*{3}{@{}c}@{}}
\begin{subfigure}{.3\linewidth}
  \centering
  \includegraphics[width=\linewidth]{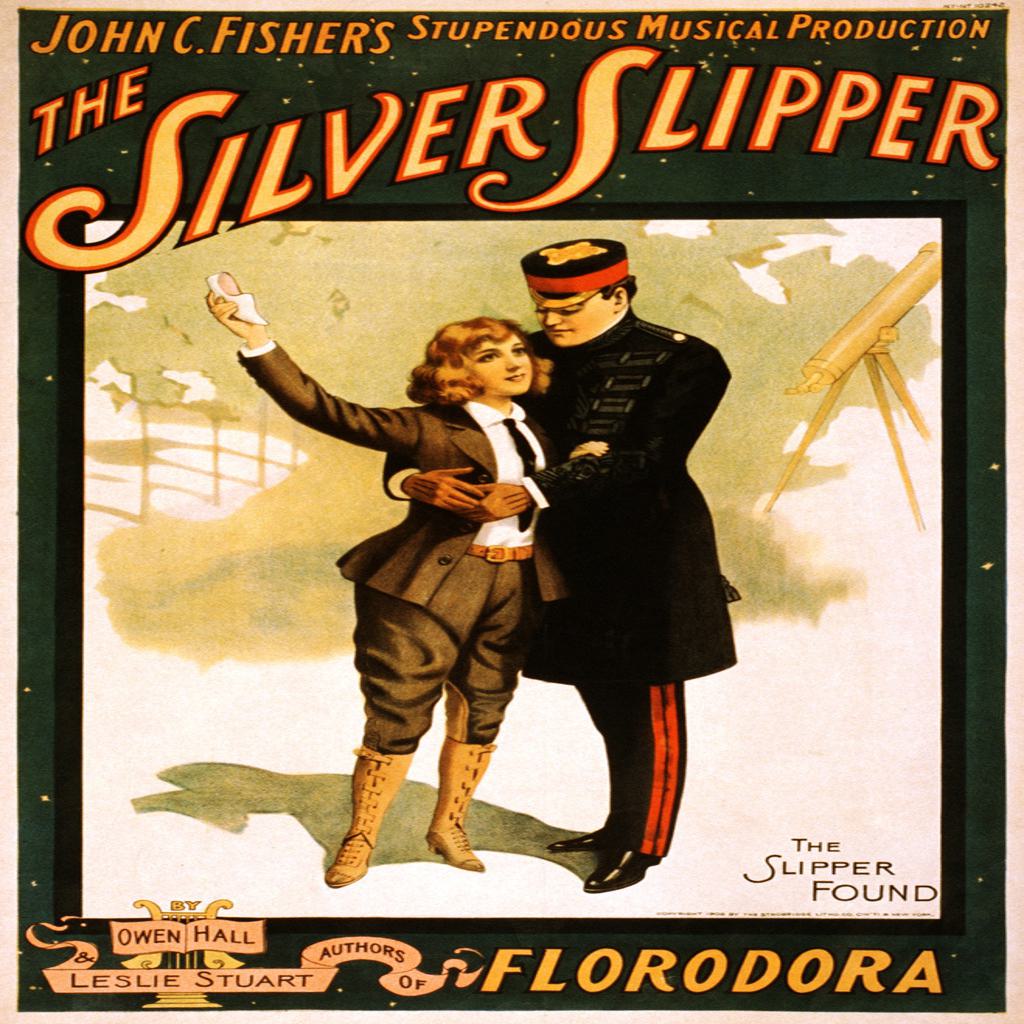}
\end{subfigure} 
\begin{subfigure}{.3\linewidth}
  \centering
  \includegraphics[width=\linewidth]{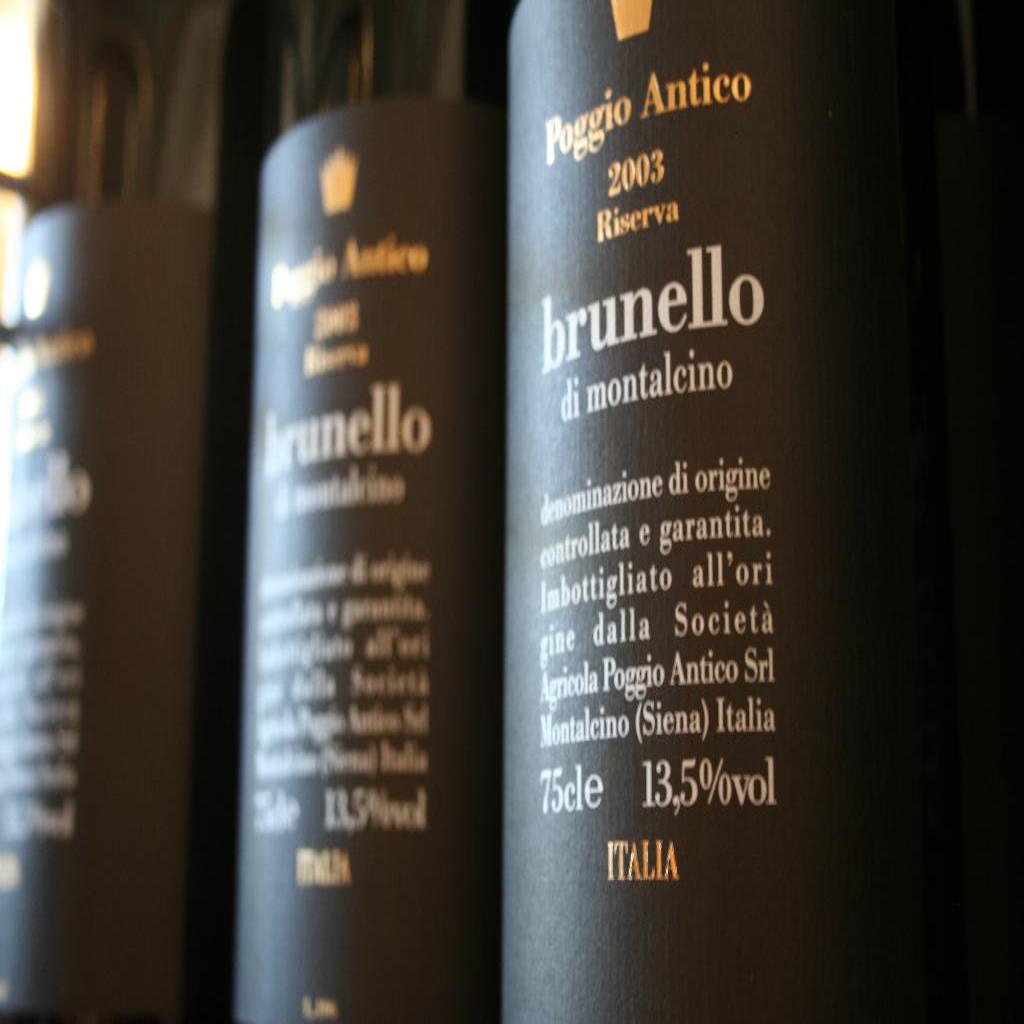}
\end{subfigure}
\begin{subfigure}{.3\linewidth}
  \centering
  \includegraphics[width=\linewidth]{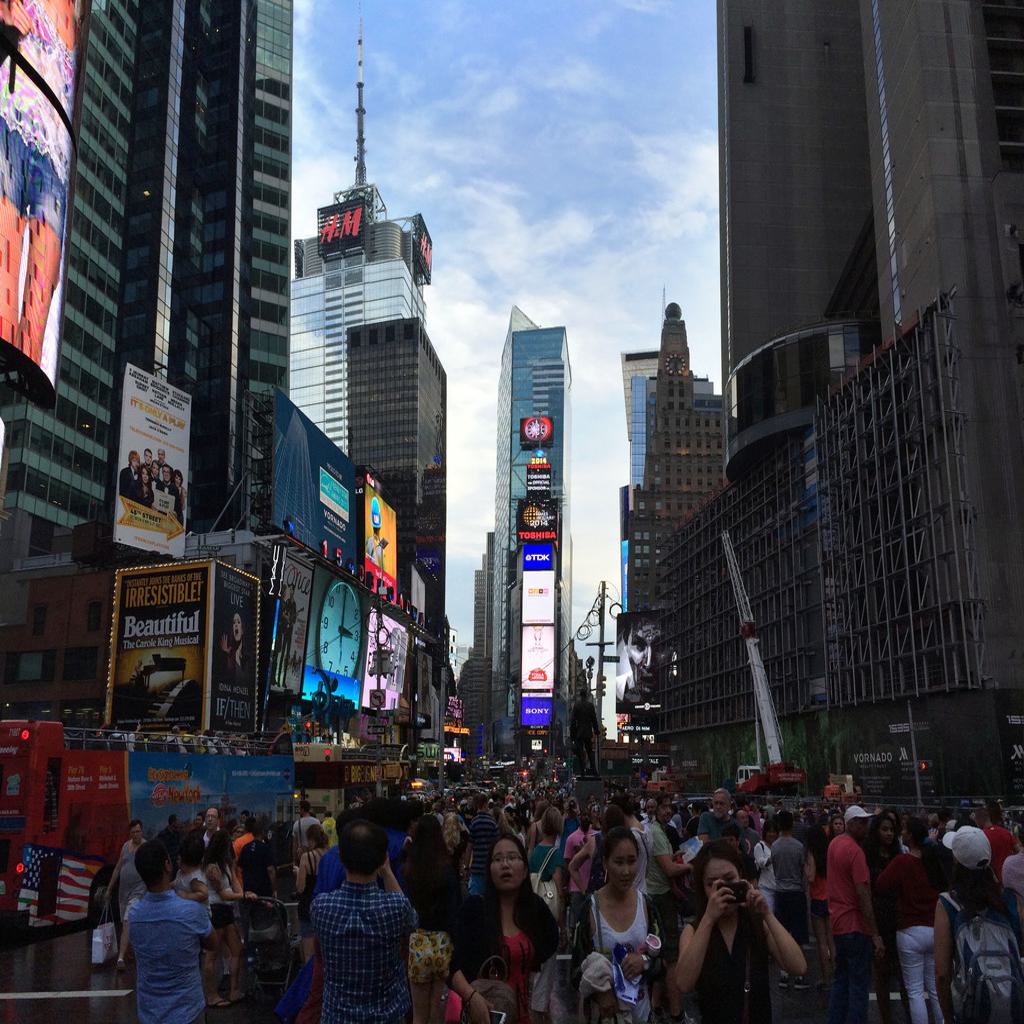}
\end{subfigure} \\

\begin{subfigure}{.3\linewidth}
  \centering
  \includegraphics[width=\linewidth]{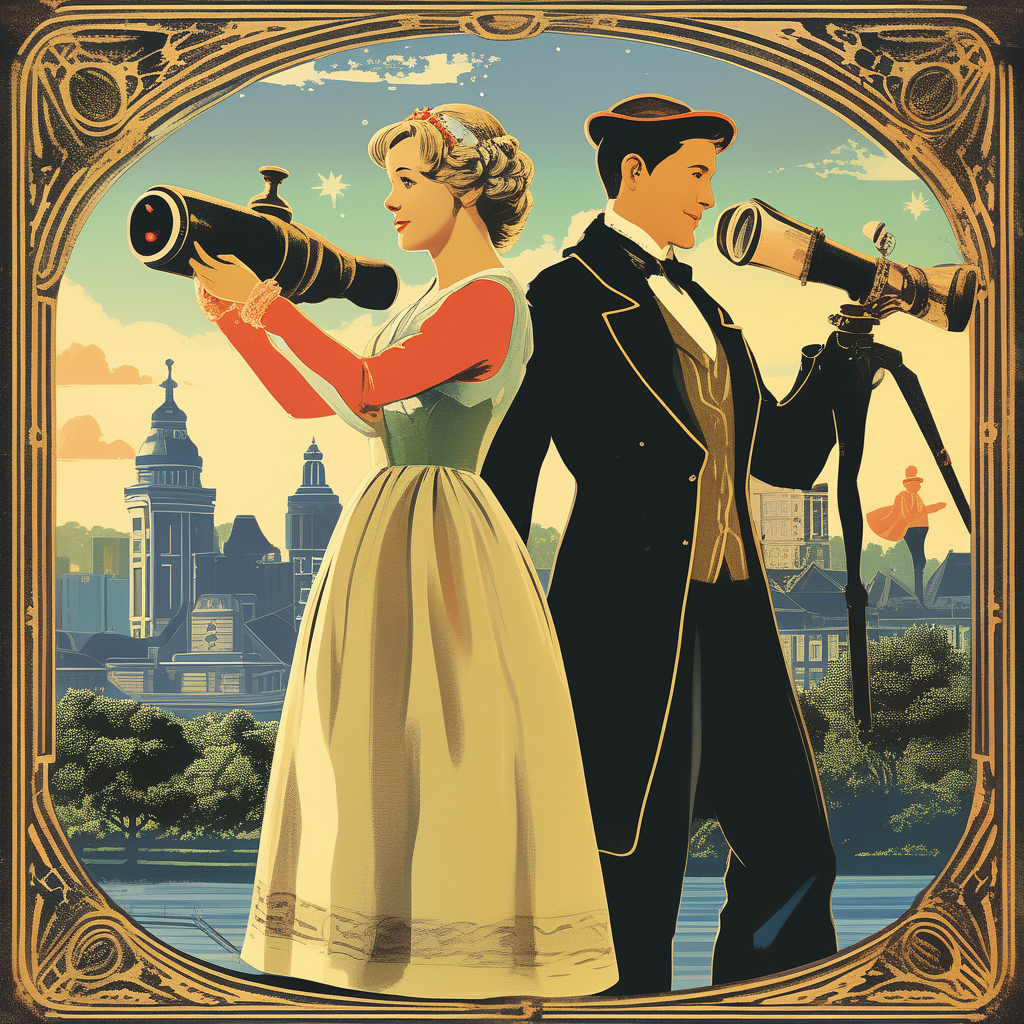}
\end{subfigure}
\begin{subfigure}{.3\linewidth}
  \centering
  \includegraphics[width=\linewidth]{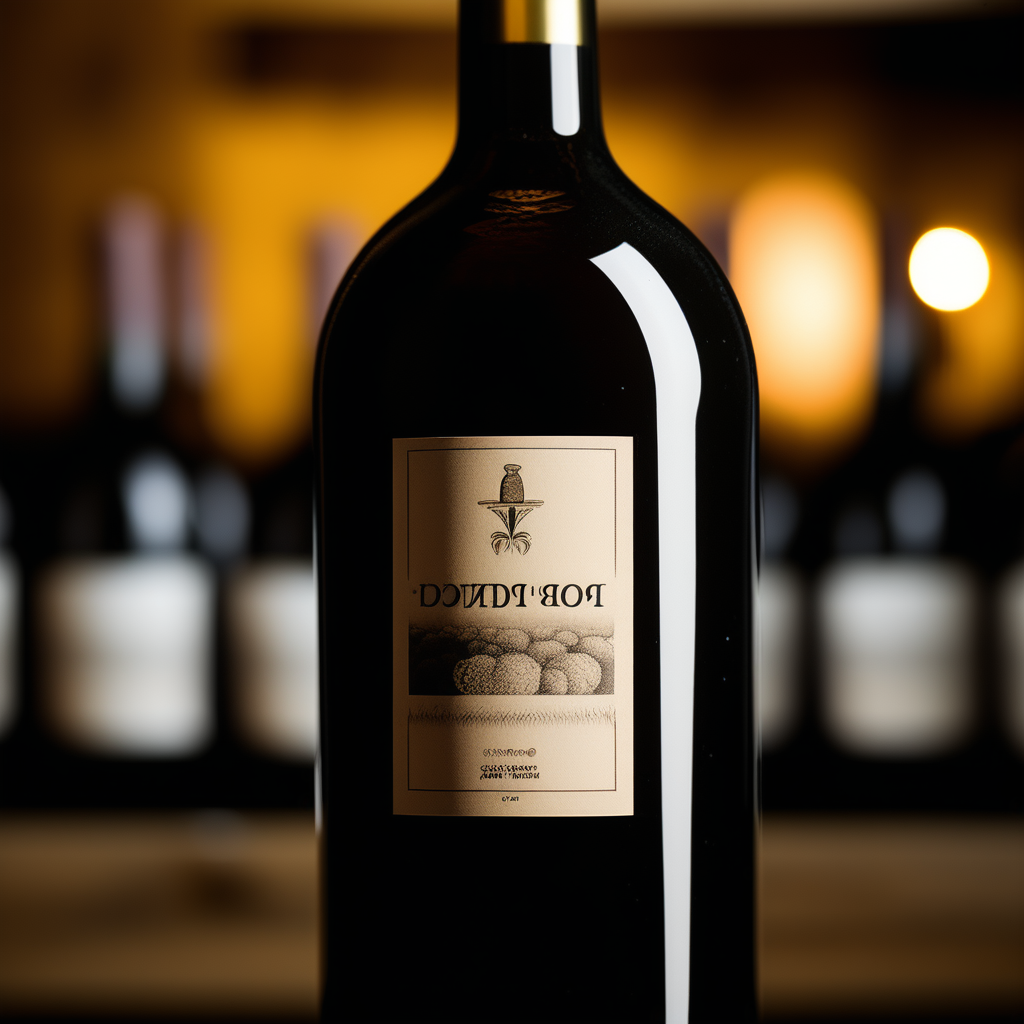}
\end{subfigure}
\begin{subfigure}{.3\linewidth}
  \centering
  \includegraphics[width=\linewidth]{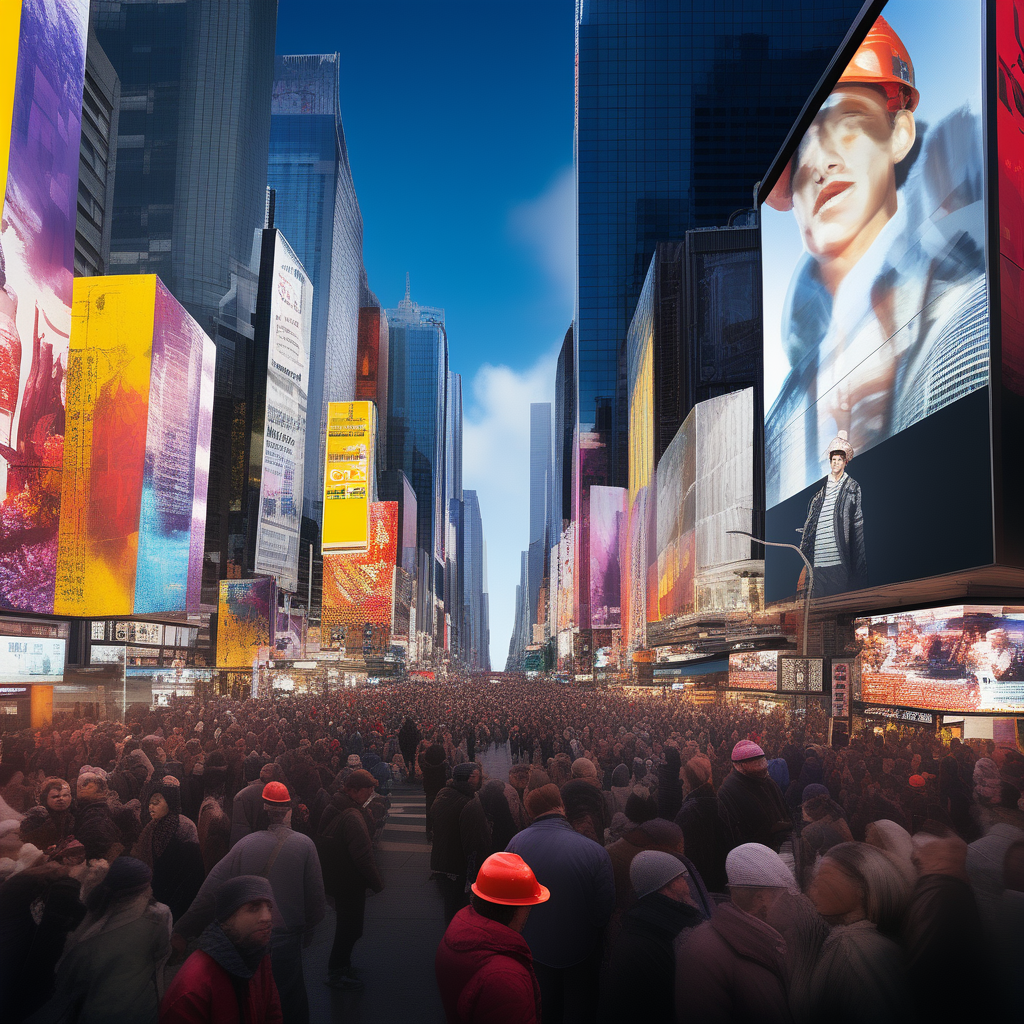}
\end{subfigure}
\end{tabular}

\caption{Examples of non-AI-generated (top row) and AI-generated (bottom row) images in the TH dataset. The {captions} of the non-AI-generated images are \textbf{(a)} ``John C Fisher's The Silver Slipper movie poster.'', \textbf{(b)} ``A row of bottles of Poggio Antico wine on a store shelf.'', and \textbf{(c)} ``The city of New York crowded with tourists with many ads on the buildings like H and M and Beautiful movie.''. The AI-generated images are generated by Hunyuan-DiT using the captions as prompts.}
\label{fig:example_images_TH}
\end{figure}

\begin{figure*}[!t]
    \centering
    \includegraphics[width=\linewidth]{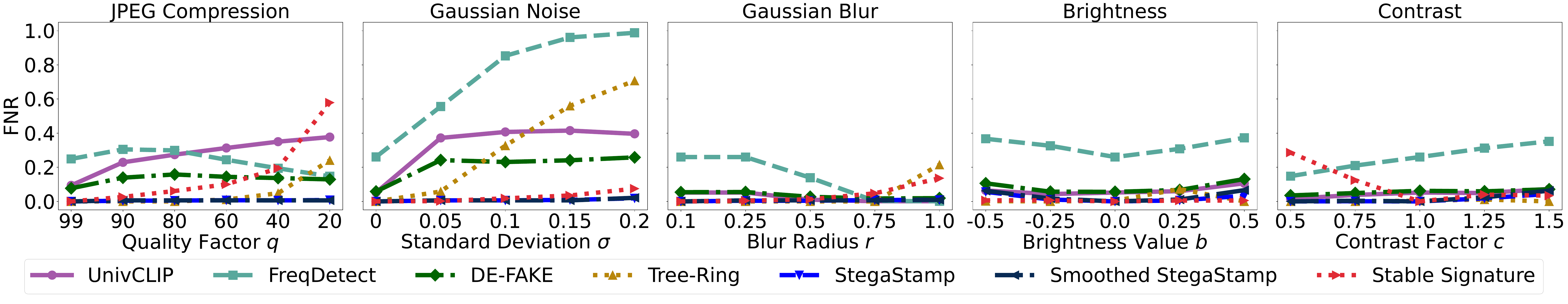}
    
    \includegraphics[width=\linewidth]{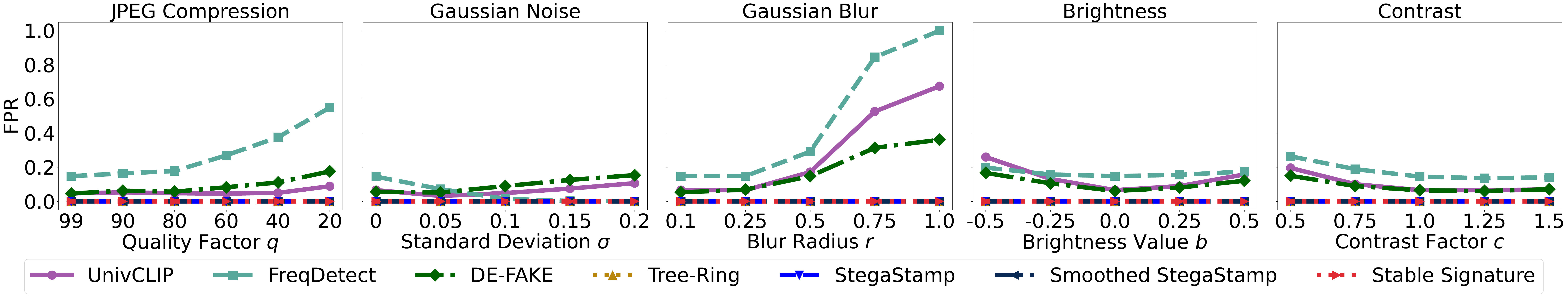}
    \caption{FNRs and FPRs of various detectors on the MS dataset under 5 types of common perturbations seen during training.}
    \label{fig:common_pert_fnrfpr}
\end{figure*}

\begin{figure*}[!t]
    \centering
    \includegraphics[width=0.24\textwidth]{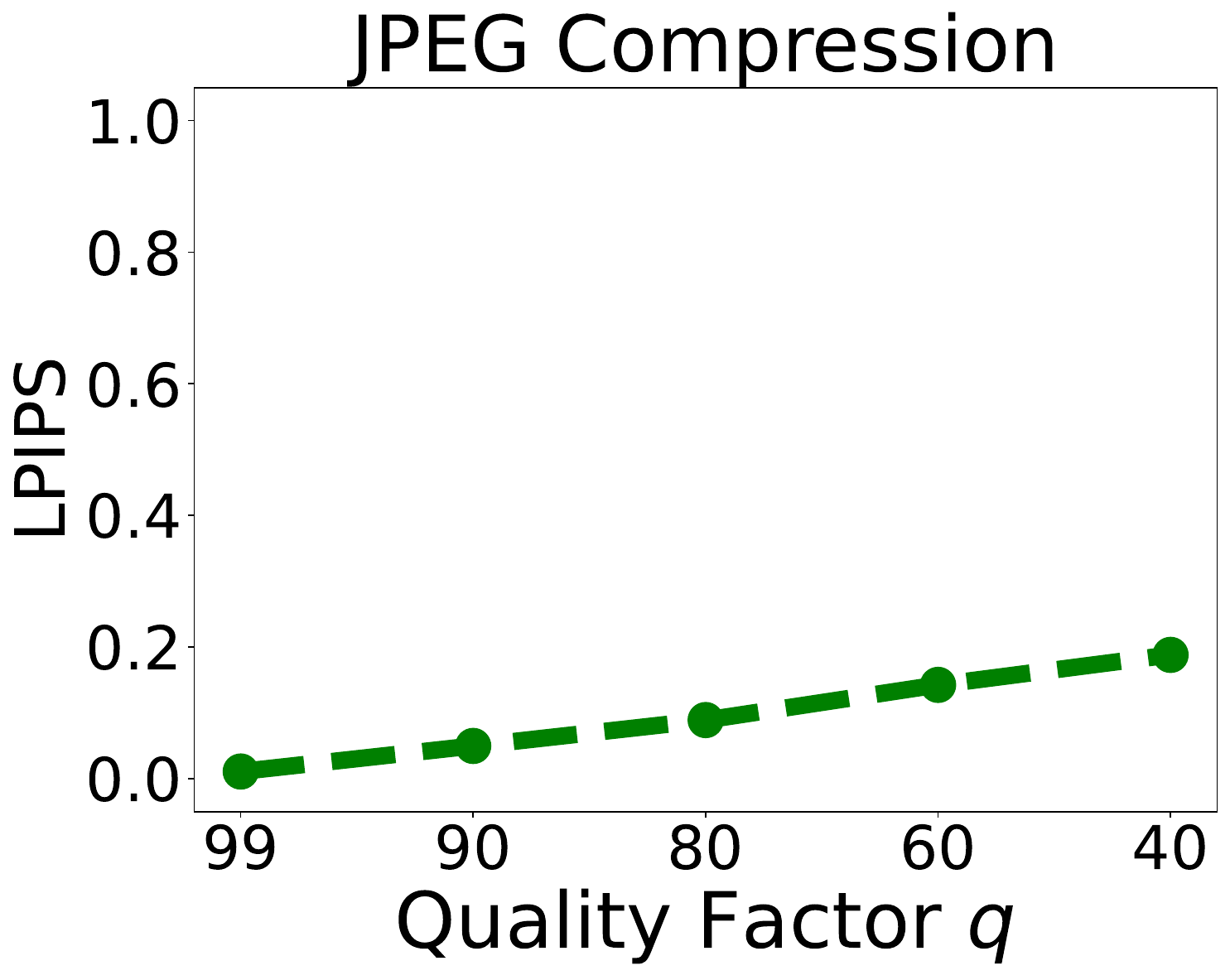}
    \includegraphics[width=0.24\textwidth]{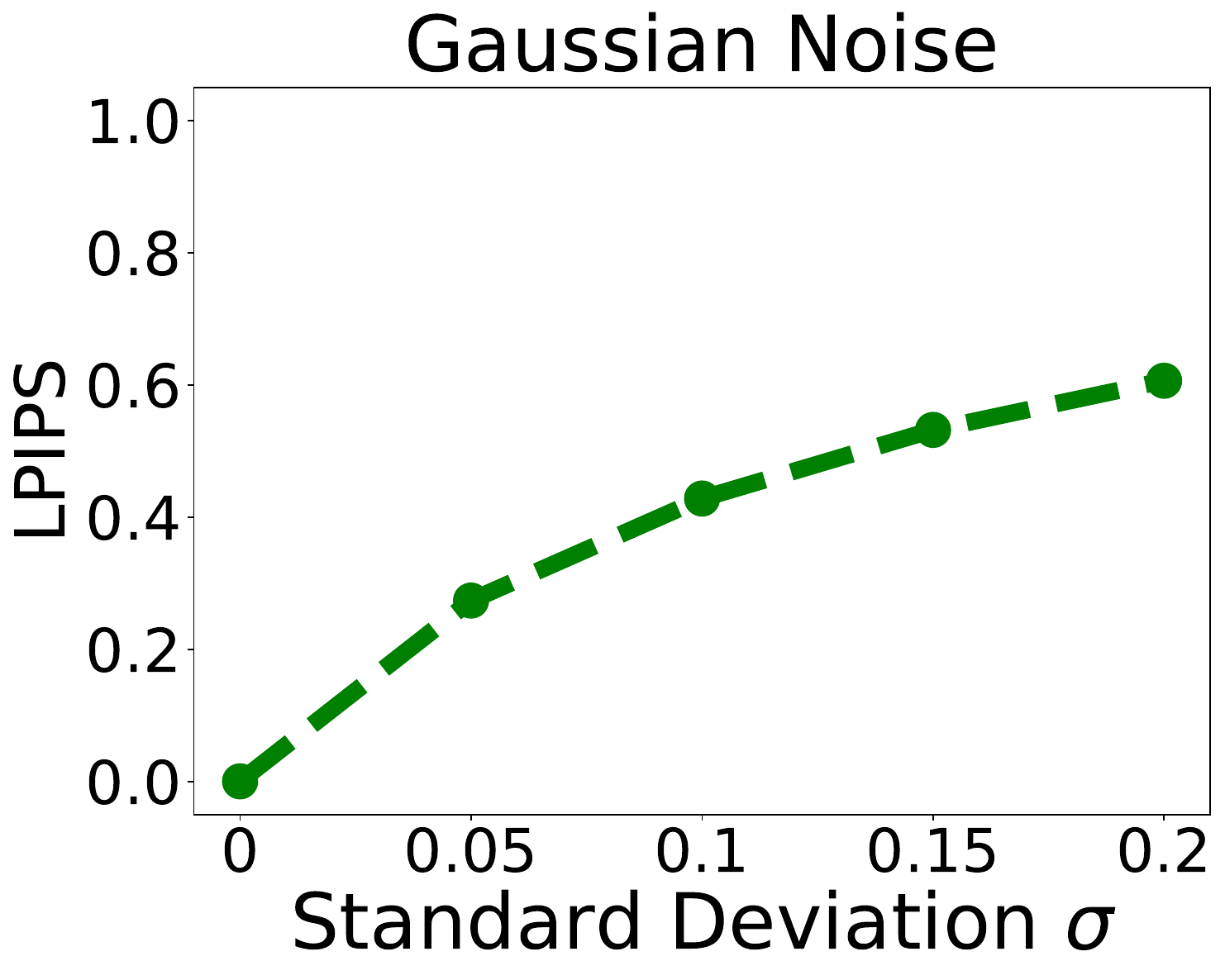}
    \includegraphics[width=0.24\textwidth]{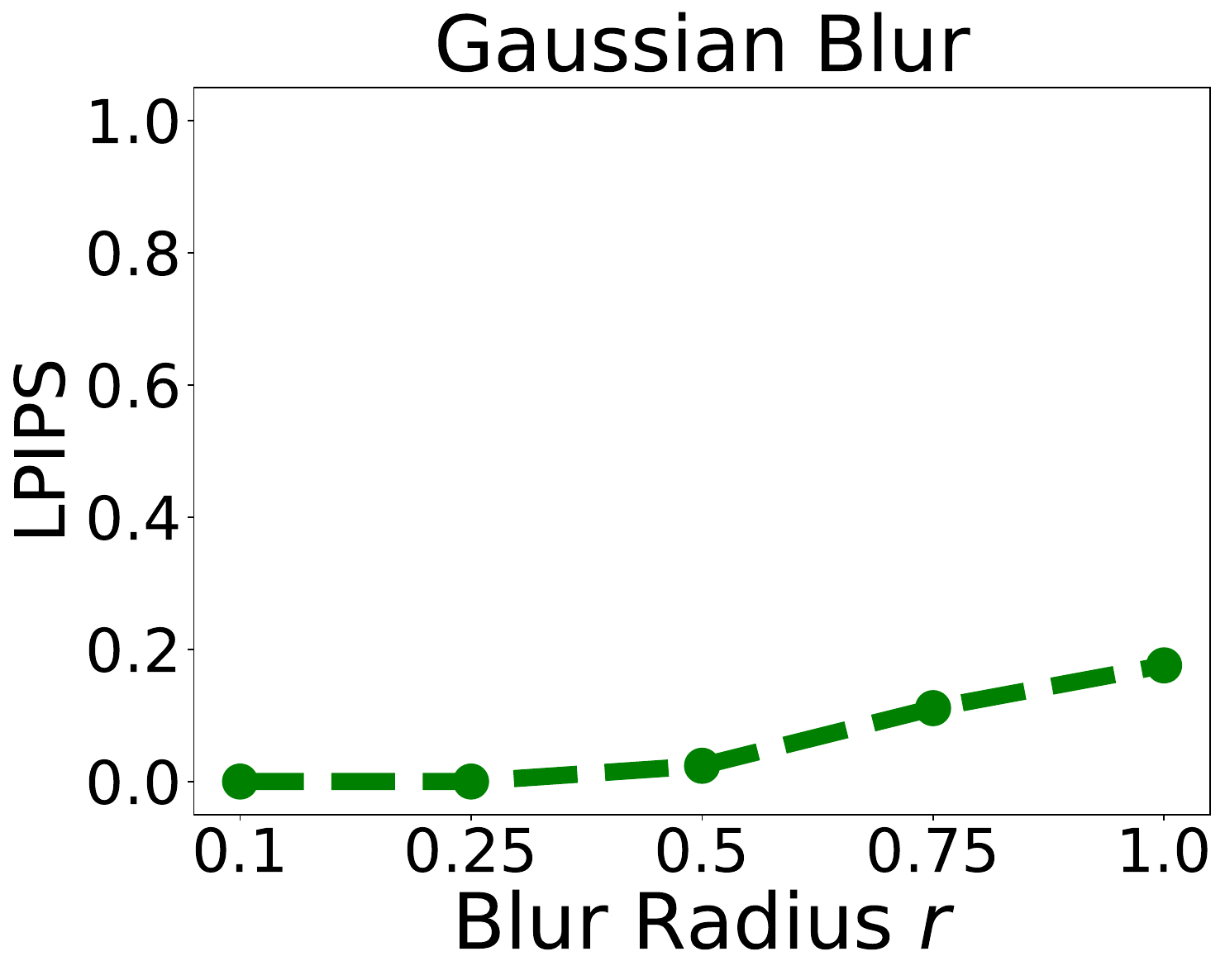}
    \includegraphics[width=0.24\textwidth]{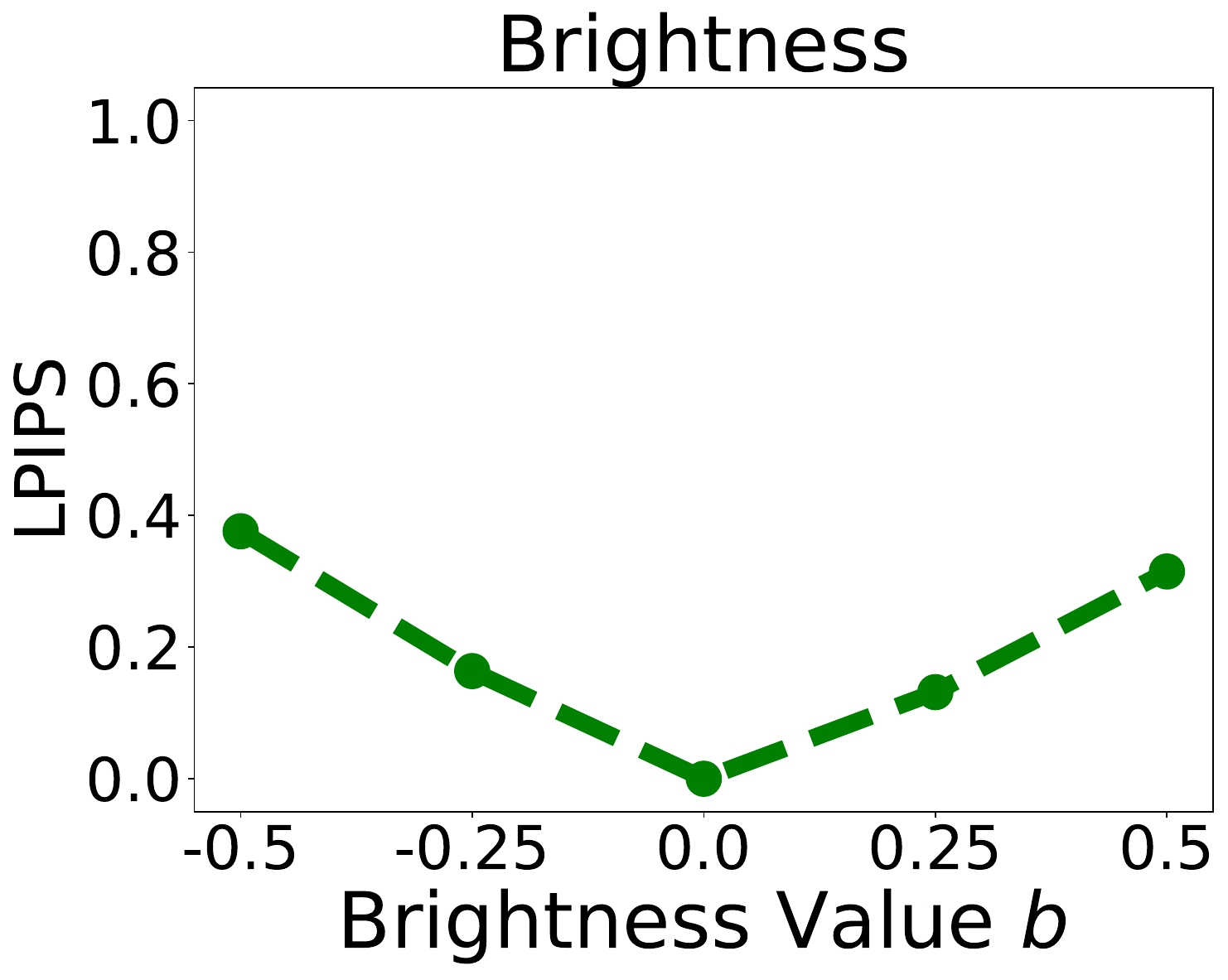}
    \vspace{4mm}

        \includegraphics[width=0.24\textwidth]{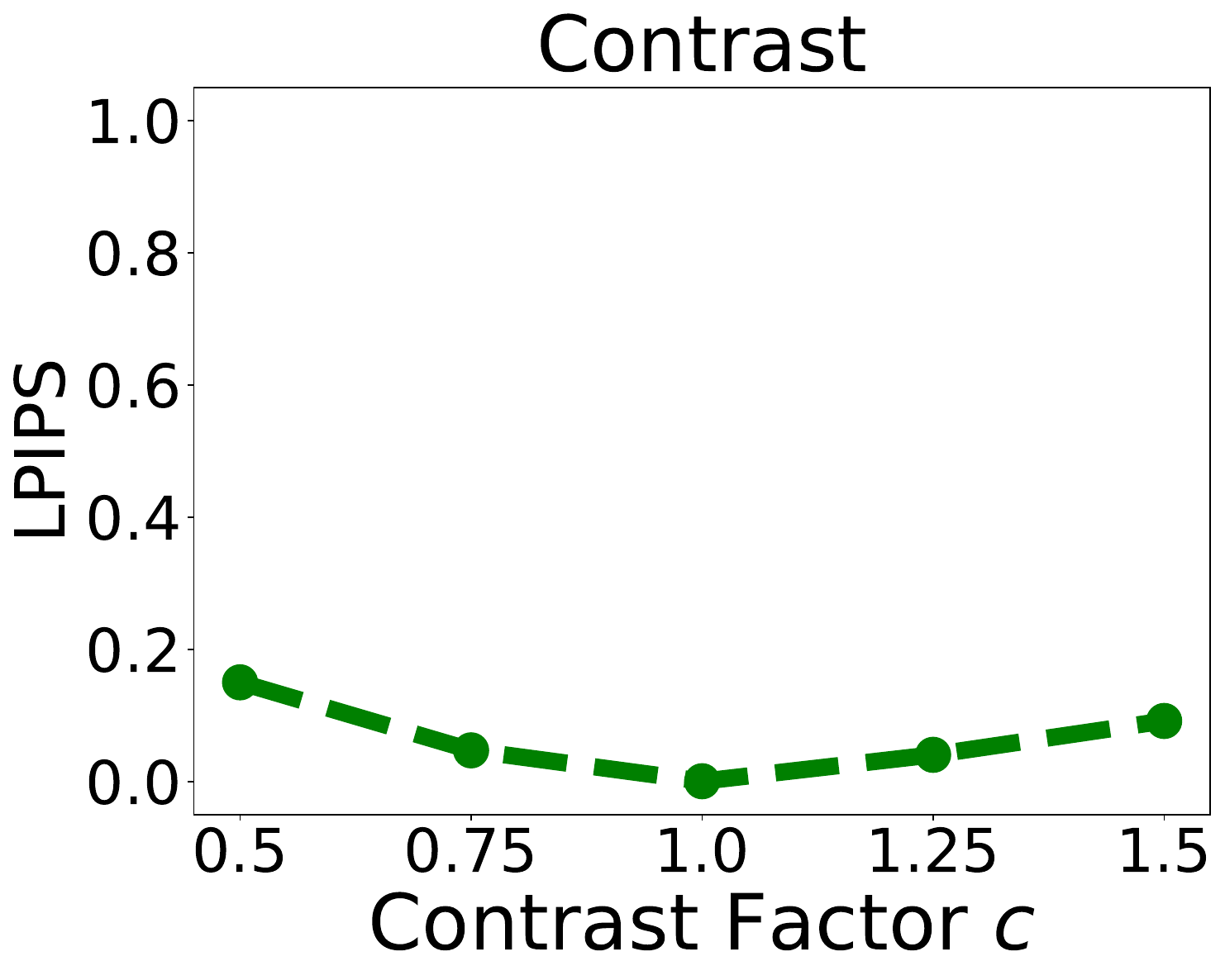}
        \includegraphics[width=0.24\textwidth]{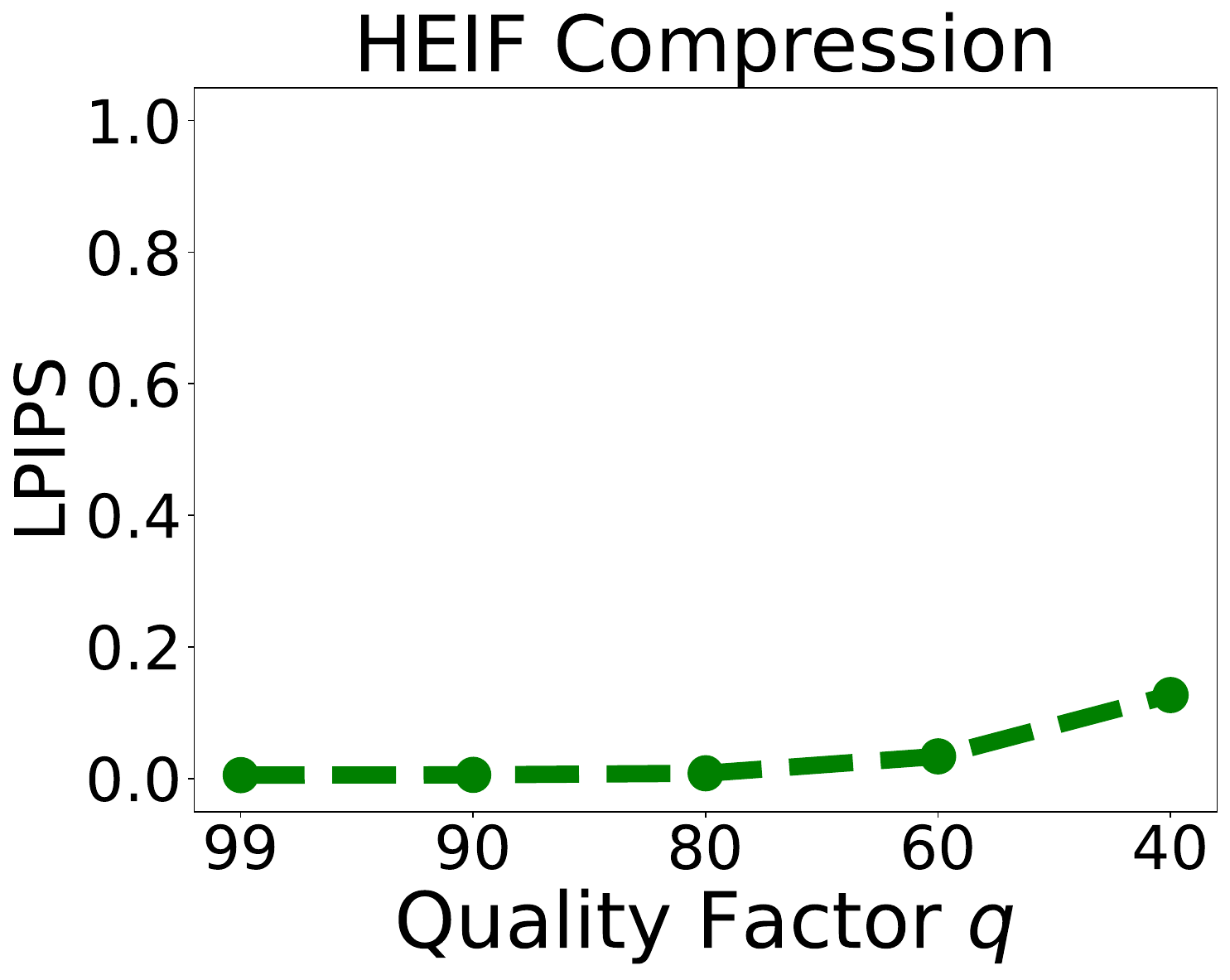}
        \includegraphics[width=0.24\textwidth]{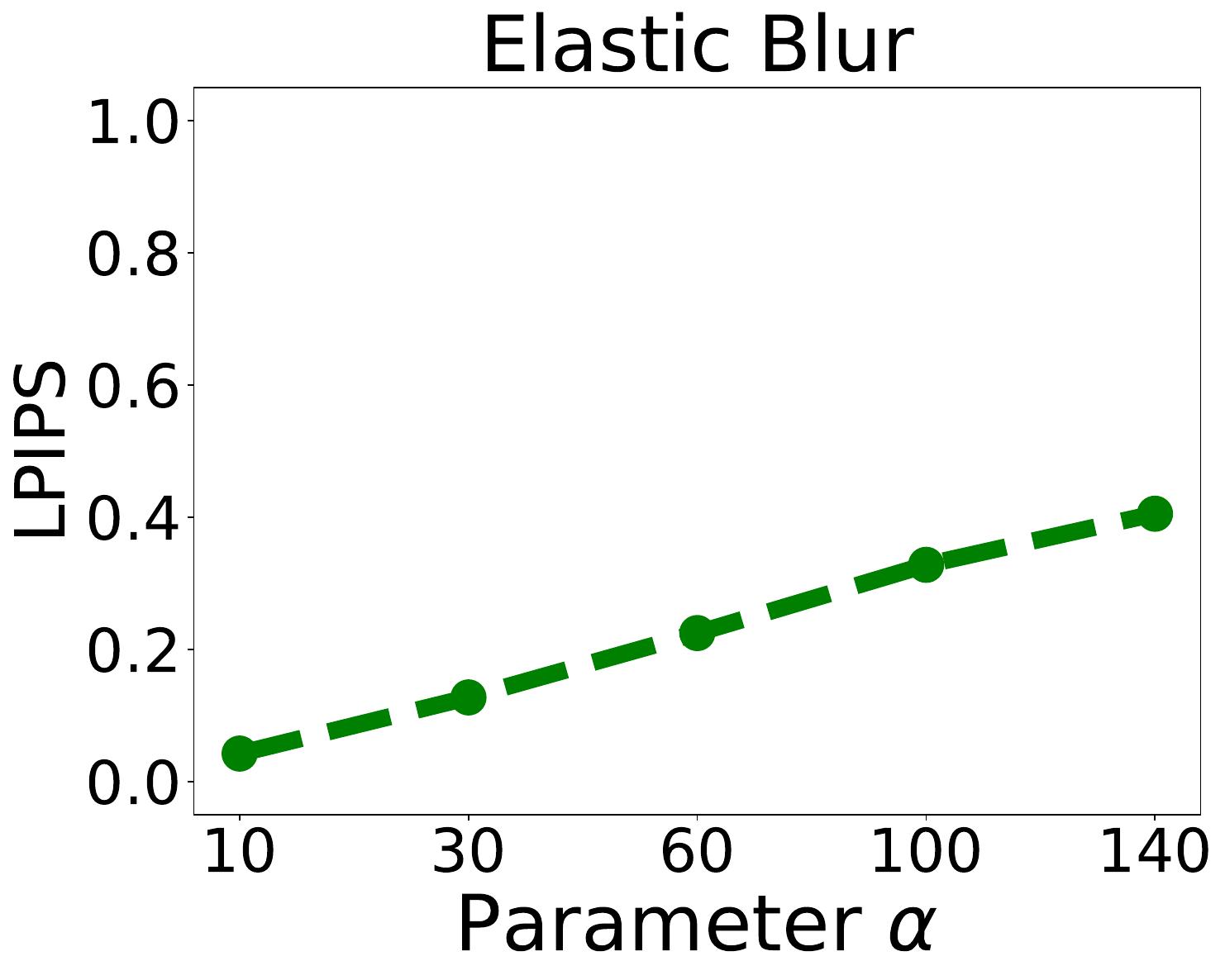}
        \includegraphics[width=0.24\textwidth]{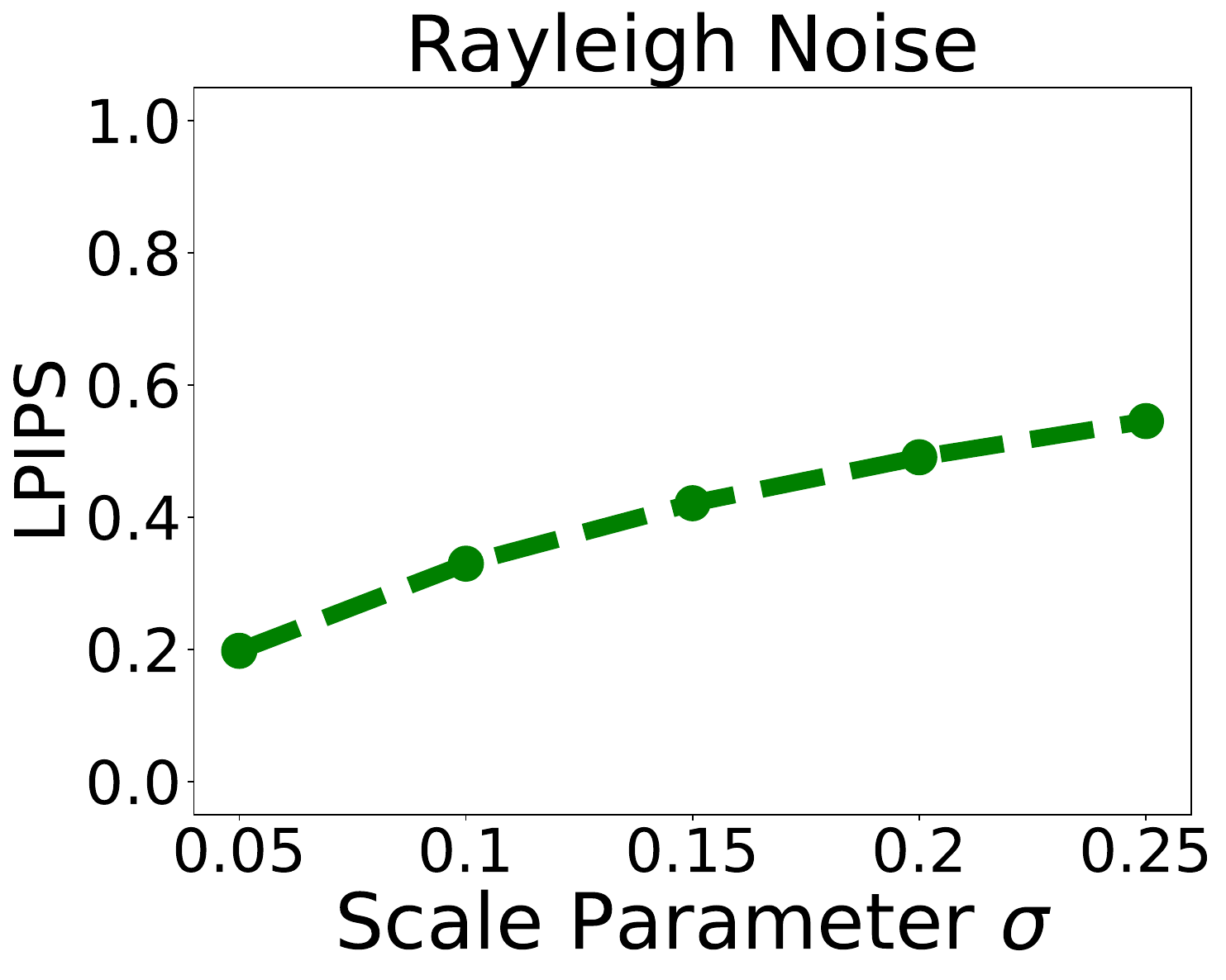}

    \caption{LPIPS under 8 types of common perturbations with varying parameter values.}
    \label{fig:common_perturbations}
\end{figure*}

\begin{figure*}[h!]
    \centering
    \begin{subfigure}[b]{0.22\textwidth}
        \includegraphics[width=\textwidth]{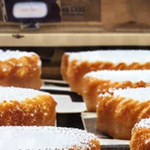}
        \caption{AI-generated}
    \end{subfigure}
    \hfill
    \begin{subfigure}[b]{0.22\textwidth}
        \includegraphics[width=\textwidth]{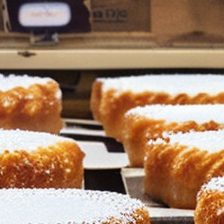}
        \caption{DE-FAKE}
    \end{subfigure}
    \hfill
    \begin{subfigure}[b]{0.22\textwidth}
        \includegraphics[width=\textwidth]{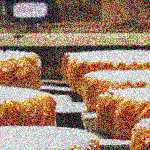}
        \caption{StegaStamp}
    \end{subfigure}
    \hfill
    \begin{subfigure}[b]{0.22\textwidth}
        \includegraphics[width=\textwidth]{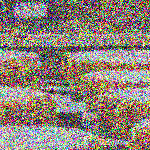}
        \caption{Smoothed StegaStamp}
    \end{subfigure}

    \vspace{0.5cm} 

    \begin{subfigure}[b]{0.22\textwidth}
        \includegraphics[width=\textwidth]{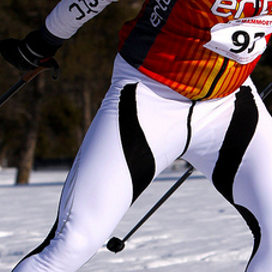}
        \caption{Non-AI-generated}
    \end{subfigure}
    \hfill
    \begin{subfigure}[b]{0.22\textwidth}
        \includegraphics[width=\textwidth]{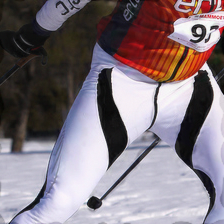}
        \caption{DE-FAKE}
    \end{subfigure}
    \hfill
    \begin{subfigure}[b]{0.22\textwidth}
        \includegraphics[width=\textwidth]{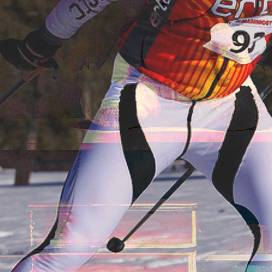}
        \caption{StegaStamp}
    \end{subfigure}
    \hfill
    \begin{subfigure}[b]{0.22\textwidth}
        \includegraphics[width=\textwidth]{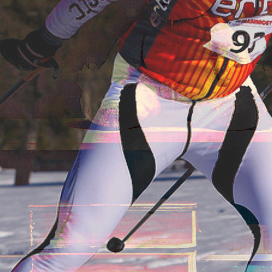}
        \caption{Smoothed StegaStamp}
    \end{subfigure}

    \caption{\emph{First row}: an AI-generated image and its adversarial versions found by the HopSkipJump attack to make the best passive and watermark-based detectors misclassify it as non-AI-generated. The watermarked version of the AI-generated image is omitted since it visually looks the same. \emph{Second row}: a non-AI-generated image and its adversarial versions found by the HopSkipJump attack to make the best passive and watermark-based detectors misclassify it as AI-generated. The HopSkipJump attack can deceive DE-FAKE in both removal and forgery attacks without sacrificing the image quality, while it substantially degrades the image quality when deceiving the watermark-based detectors especially Smoothed StegaStamp.}
    \label{fig:hsj_illus}
\end{figure*}